\documentclass[aip,amsmath,amssymb,reprint]{revtex4-1}

\usepackage{graphicx}
\usepackage{dcolumn}
\usepackage{bm}
\usepackage[utf8]{inputenc}
\usepackage[T1]{fontenc}
\usepackage{mathptmx}

\usepackage{units}

\begin{document}

\preprint{AIP/123-QED}
\title[]{FitzHugh-Nagumo oscillators on complex networks mimic epileptic-seizure-related synchronization phenomena}

\author{Moritz Gerster}
\affiliation{ 
Institut f\"ur Theoretische Physik, Technische Universit\"at Berlin, Hardenbergstr.\,36, 10623 Berlin
}%

\author{Rico Berner}
\affiliation{ Institut f\"ur Theoretische Physik, Technische Universit\"at Berlin, Hardenbergstr.\,36, 10623 Berlin
}%
\affiliation{Institut f\"ur Mathematik, Technische Universit\"at Berlin, Strasse des 17. Juni 136, 10623 Berlin}

\author{Jakub Sawicki}

\author{Anna Zakharova}
\affiliation{ Institut f\"ur Theoretische Physik, Technische Universit\"at Berlin, Hardenbergstr.\,36, 10623 Berlin
}%

\author{Anton\'in \v{S}koch}
\affiliation{ 
National Institute of Mental Health, Topolov\'{a} 748, 250 67 Klecany, Czech Republic 
}
\author{Jaroslav Hlinka}
\affiliation{ 
National Institute of Mental Health, Topolov\'{a} 748, 250 67 Klecany, Czech Republic 
}
\affiliation{ 
Institute of Computer Science of the Czech Academy of Sciences, Pod Vodarenskou vezi 2, 18207 Prague 8, Czech Republic
}

\author{Klaus Lehnertz}
\affiliation{ 
Department of Epileptology, University of Bonn Medical Centre, Venusberg Campus 1, 53127 Bonn, Germany}
\affiliation{Helmholtz Institute for Radiation and Nuclear Physics, University of Bonn, Nussallee 14--16, 53115 Bonn, Germany}
\affiliation{Interdisciplinary Center for Complex Systems, University of Bonn, Br{\"u}hler Stra\ss{}e 7, 53175 Bonn, Germany
}

\author{Eckehard Sch\"oll}
\email{schoell@physik.tu-berlin.de}
\affiliation{ 
Institut f\"ur Theoretische Physik, Technische Universit\"at Berlin, Hardenbergstr.\,36, 10623 Berlin
}
\affiliation{Bernstein Center for Computational Neuroscience Berlin, Humboldt-Universität, 10115 Berlin, Germany
}%

\date{\today}

\begin{abstract} 
We study patterns of partial synchronization in a network of FitzHugh-Nagumo oscillators with empirical structural connectivity measured in human subjects.
We report the spontaneous occurrence of synchronization phenomena that closely resemble the ones seen during epileptic seizures in humans. In order to obtain deeper insights into the interplay between dynamics and network topology, we perform long-term simulations of oscillatory dynamics on different paradigmatic network structures: random networks, regular nonlocally coupled ring networks, ring networks with fractal connectivities, and small-world networks with various rewiring probability. 
Among these networks, a small-world network with intermediate rewiring probability best mimics the findings achieved with the simulations using the empirical structural connectivity. 
For the other network topologies, either no spontaneously occurring epileptic-seizure-related synchronization phenomena can be observed in the simulated dynamics, or the overall degree of synchronization remains high throughout the simulation. 
This indicates that a topology with some balance between regularity and randomness favors the self-initiation and self-termination of episodes of seizure-like strong synchronization.
\end{abstract}

\maketitle

\begin{quotation}
\section*{}
Synchronization is a widespread natural phenomenon occurring in networks of oscillators~\cite{PIK01,BOC18}.
In the human brain, synchronization is essential for normal physiological functioning~\cite{SIN18}, but it is also strongly related to seizures,
which are the cardinal symptom of epilepsy\cite{LEH09,JIR13,JIR14}. This neurological disease is currently understood as a network disease~\cite{LEH14a}, and a better understanding of the role of the epileptic network's topology in seizure generation and termination is highly desirable.
Using complex networks of coupled oscillators, we simulate synchronization phenomena observed in the human brain.
We employ coupled oscillators of FitzHugh-Nagumo type since these are a paradigmatic model for neural dynamics~\cite{BAS18}.
With an empirical structural brain connectivity of human subjects as a coupling matrix, we observe spontaneously occurring periods of strong synchronization, which resemble the ones seen during epileptic seizures.
For a better insight into the network properties giving rise to such pathology-related events, we simulate the dynamics on various paradigmatic network topologies: we randomly rewire links in a small-world fashion, consider fractal connectivities, and exchange equal weights with empirical weights from diffusion-weighted magnetic resonance imaging. Moreover, we explore how global aspects of the networks --~as assessed with the average clustering coefficient and the mean shortest path length~-- impact on the dynamics of the epileptic-seizure-related synchronization phenomena. In order to strengthen our findings, we compare our model simulations to electroencephalographic (EEG) recordings of epileptic seizures. 
A better knowledge of the interplay between dynamics and network properties leading to complex synchronization phenomena is essential for understanding seizure dynamics.
\end{quotation}

\section{Introduction} 
Epilepsy is a neurological disorder that affects almost 70~million people worldwide~\cite{NGU11a}. 
People with epilepsy experience seizures characterized by a ``transient occurrence of signs and/or symptoms due to abnormal excessive or synchronous neuronal activity in the brain.''~\cite{FIS05}
Generalized seizures involve almost the entire brain~\cite{GAS70}, while focal seizures are confined to a circumscribed brain area. 
Generalized seizures are usually classified by symptoms such as muscle contractions, shaking of the limbs, muscle spasms, or rapid
loss of muscle activity.
There is only one kind of generalized seizure in which the gross muscular activity is unaffected, making it especially accessible to measurement using EEG by avoiding the problem of movement artifacts.
Such seizures are called \textit{absence seizures}.
If a person experiences an absence seizure while standing, he or she does not fall over.
Instead, the person may loose consciousness, stops any behavior engaged in before the seizure, keeps still, may blink his or her eyes, and picks up on the behavior right after the seizure terminates.
The person may have no memory of the seizure and is usually unaware that it happened, due to the possible loss of consciousness during
the seizure.

In epileptology, the development of the concept of an epileptic network~\cite{SPE02,RIC12a,KUH18} received a strong impetus from network-theoretical concepts.
An epileptic network comprises anatomically, and more importantly, functionally connected cortical and subcortical brain structures and regions. 
Seizures may emerge from, may spread via, and may be terminated by network constituents that generate and sustain normal, physiological brain dynamics during the seizure-free interval~\cite{SPE02}.

In order to advance the understanding of the epileptic network and its temporal evolution, research into seizure dynamics may benefit from research on the synchronization in complex networks. 
This topic is of great scientific interest due to its relevance for understanding synchronization phenomena in nature and technology\cite{PIK01,BOC06a,SCH16,BOC18,BIC20}. 
From this research, it is well known that the system's ability to synchronize depends on the local dynamics of the oscillators, their coupling, and their structural connectivity. 
In this article, we focus on the latter property and study the impact of the network structure on the emergence of seizures. 
To gain a better understanding of the dynamics of epileptic seizures, we are interested in synchronization events in neural networks which are (i) generalized (i.e., affect the entire system), (ii) have a long duration compared to the system dynamics, and (iii) are self-initiated and self-terminated. 
We aim at identifying network structures that provoke such events. 

Previous studies have identified characteristic changes in various properties of networks related to generalized seizures~\cite{PON09,CHA10c} and have highlighted the critical role of the coupling topology for their dynamics~\cite{BAI12,BEN12,TER12,PET14}.
For networks of neurons, modeled with the paradigmatic FitzHugh-Nagumo neuronal dynamics, epileptic-seizure-like dynamics has been investigated in the context of two topologies: an empirical structural brain connectivity (derived from diffusion-weighted magnetic resonance imaging) and a mathematically constructed network with modular fractal connectivity~\cite{CHO18}. 
Further, the role of partial synchronization phenomena~\cite{ROT14a,POE15,SAW20,SCH20b,ZAK20} for mechanisms of seizure initiation~\cite{AND16} and termination~\cite{ROT14} has been explored. 

The purpose of this work is to elucidate the role of the neural network coupling structure in causing epileptic-seizure-related synchronization phenomena. For better readability, we use the term seizure for \textit{epileptic-seizure-related synchronization phenomena} in the following.  
To this end, we compare various network topologies that are relevant in the neurosciences. 
Our goal is to conceive how the structure of a network facilitates events of spontaneous and prolonged synchronization in systems that are desynchronized most of the time. 

The paper is organized as follows: In Sec.~\ref{sect:model}, we introduce the dynamical model, which consists of coupled FitzHugh-Nagumo oscillators. 
In Sec.~\ref{sect:sim}, we present the results of our simulations for different network topologies that shed light on the role of the coupling structure for spontaneous synchronization.
Eventually, in Sec.~\ref{exp}, we compare our simulated seizures to those seen in electroencephalographic (EEG) recordings of generalized epileptic seizures in humans.

\section{The Model}
\label{sect:model}
\message{The column width is: \the\columnwidth}

We use the FitzHugh-Nagumo (FHN) model, which is a paradigmatic model for neuronal spiking~\cite{FIT61,NAG62, BAS18}. Note that while the FitzHugh-Nagumo model was originally developed as a simplified model of a single neuron, it is also often used as a generic model for excitable media on a coarse-grained level~\cite{CHE05e,CHE07a}. 
In this spirit, we model the 90 regions of the human brain labeled by the Automated Anatomical Labeling (AAL) atlas~\cite{TZO02} by a network of $N=90$ nodes, where each brain region is described by an FHN oscillator involving an activator variable (membrane potential) $u_k$ and an inhibitor variable (recovery variable) $v_k$.
We arrange the brain regions $k = 1, 2,\ldots, 90$ such that $k \in N_L = \{1, \ldots, 45\}$ corresponds to the left, and $k \in N_R = \{46, \ldots, 90\}$ corresponds to the right brain hemisphere. 
The dynamics of variables $u_k$ and $v_k$ is then given by: 
\begin{align}
    \varepsilon \dot{u}_k  = &u_k - \frac{u_k^3}{3} - v_k \notag\\
                    &+ \sigma \sum_{j=1}^{N} A_{kj} \left[ B_{uu}(u_j - u_k) + B_{uv}(v_j - v_k) \right] \notag\\
    \dot{v}_k  = &u_k + a \notag\\
 & + \sigma \sum_{j=1}^{N} A_{kj} \left[ B_{vu}(u_j - u_k) + B_{vv}(v_j - v_k) \right],   
\label{eq.1}
\end{align}
where $k=1,\ldots,N$, and $\varepsilon = 0.05$ describes the timescale separation between the fast activator variable $u_k$ and the slow inhibitor variable $v_k$~\cite{FIT61}. 
Depending on the threshold parameter $a$, each uncoupled node may exhibit excitable behavior ($\left| a \right| > 1$) or self-sustained limit cycle oscillations ($\left| a \right| < 1$), separated by a Hopf bifurcation at $\left| a \right| = 1$. 
We use the FHN model in the oscillatory regime and fix the threshold parameter at $a=0.5$ sufficiently far from the Hopf bifurcation point. 
The matrix elements $A_{kj}$ of the weighted adjacency matrix of size $90 \times 90$ determine the network topology. 
The overall coupling is determined by the coupling strength $\sigma$. 
The interaction scheme between activator and inhibitor variables is characterized by the $2 \times 2$ matrix $\mathbf{B}$. 
Employing a rotational matrix $\mathbf{B}$ is a simple way to parameterize the possibility of either diagonal coupling $(B_{uu},B_{vv})$ or activator-inhibitor cross-coupling $(B_{uv},B_{vu})$ by a single parameter $\varphi$:
\begin{align}
\mathbf{B} = 
\begin{pmatrix}
B_{uu} & B_{uv} \\
B_{vu} & B_{vv}
\end{pmatrix}
=
\begin{pmatrix}
\text{cos}\varphi & \text{sin}\varphi \\
-\text{sin}\varphi & \text{cos}\varphi 
\end{pmatrix} .
\end{align}
In the following we choose $\varphi = \frac{\pi}{2} - 0.1$, causing dominant activator-inhibitor cross-coupling~\cite{OME13}, which is a commonly employed mechanism in biology~\cite{KIS04}. In the neurosciences, the microscopic coupling schemes are very complex~\cite{PER14a}, but in our coarse-grained macroscopic description of a whole brain area by a pair of activator and inhibitor variables, activator-inhibitor coupling is a natural extension of pure activator-activator coupling. Mathematically, this means that signals of other neuronal areas are coupled via a coupling phase, which introduces a phase lag or time delay. The subtle interplay of excitatory and inhibitory interaction enables intermittent periods of either high or low synchronization. This is typical of the critical state at the edge of different dynamical regimes in which the brain operates~\cite{MAS15a}. In fact, in our simulations we have not found epileptic-seizure-related partial synchronization phenomena if we use pure activator-activator coupling.
The coupling phase $\varphi$ is similar to the phase-lag parameter of the paradigmatic Kuramoto phase oscillator model, which is widely used to describe synchronization phenomena in coupled oscillator networks. 
The coupling phase has been shown to be crucial for the modeling of nontrivial partial synchronization patterns in the Kuramoto model~\cite{OME10a} and in the FHN model\cite{OME13}.  

We use the global Kuramoto order parameter $r$ to measure the degree of synchronization of a network. 
It is calculated as
\begin{align}
{r(t) = \frac{1}{N} \left| \sum_{k=1}^N \text{exp}[i \phi_k(t)]\right|},
\label{GKOP}
\end{align}
utilizing an abstract dynamical phase $\phi_k$ obtained from the standard geometric phase ${\tilde{\phi}_k(t) = \text{arctan}(v_k/u_k)}$ by a transformation which yields constant phase velocity $\dot{\phi}_k$. 
For an uncoupled FHN oscillator, the function $t(\tilde{\phi}_k)$ is calculated numerically, assigning a value of time $0<t(\tilde{\phi}_k)<T$ for every value of the geometric phase, where $T$ is the oscillation period. 
The dynamical phase is then defined as $\phi_k=2 \pi t(\tilde{\phi}_k)/T$, which yields $\dot{\phi}_k = \text{const}$. 
Uncoupled oscillators have thus a constant phase velocity with respect to the dynamical phase. 
If the geometrical phase $\tilde{\phi}_k(t)$ were used instead, 
the slow-fast time scales of inhibitor and activator would result in a strongly inhomogeneous phase velocity, which would cause strong fluctuations of the order parameter $r$. 
Only by using the dynamical phase $\phi_k$, these fluctuations are suppressed, and a change in $r$ indeed reflects a change in the degree of synchronization. 
The Kuramoto order parameter may vary between 0 and 1, where $r=1$ corresponds to complete phase synchronization, small values characterize desynchronized states, and intermediate values correspond to partial synchronization.

\section{Simulation of spontaneous epileptic-seizure-related synchronization phenomena}
\label{sect:sim}
In the following, we present simulations of the dynamics described by Eq.~(\ref{eq.1}) for various network topologies $A_{kj}$.

\subsection{Empirical brain network}
\label{Human}

\begin{figure}
\centering
\includegraphics[width=\the\columnwidth]{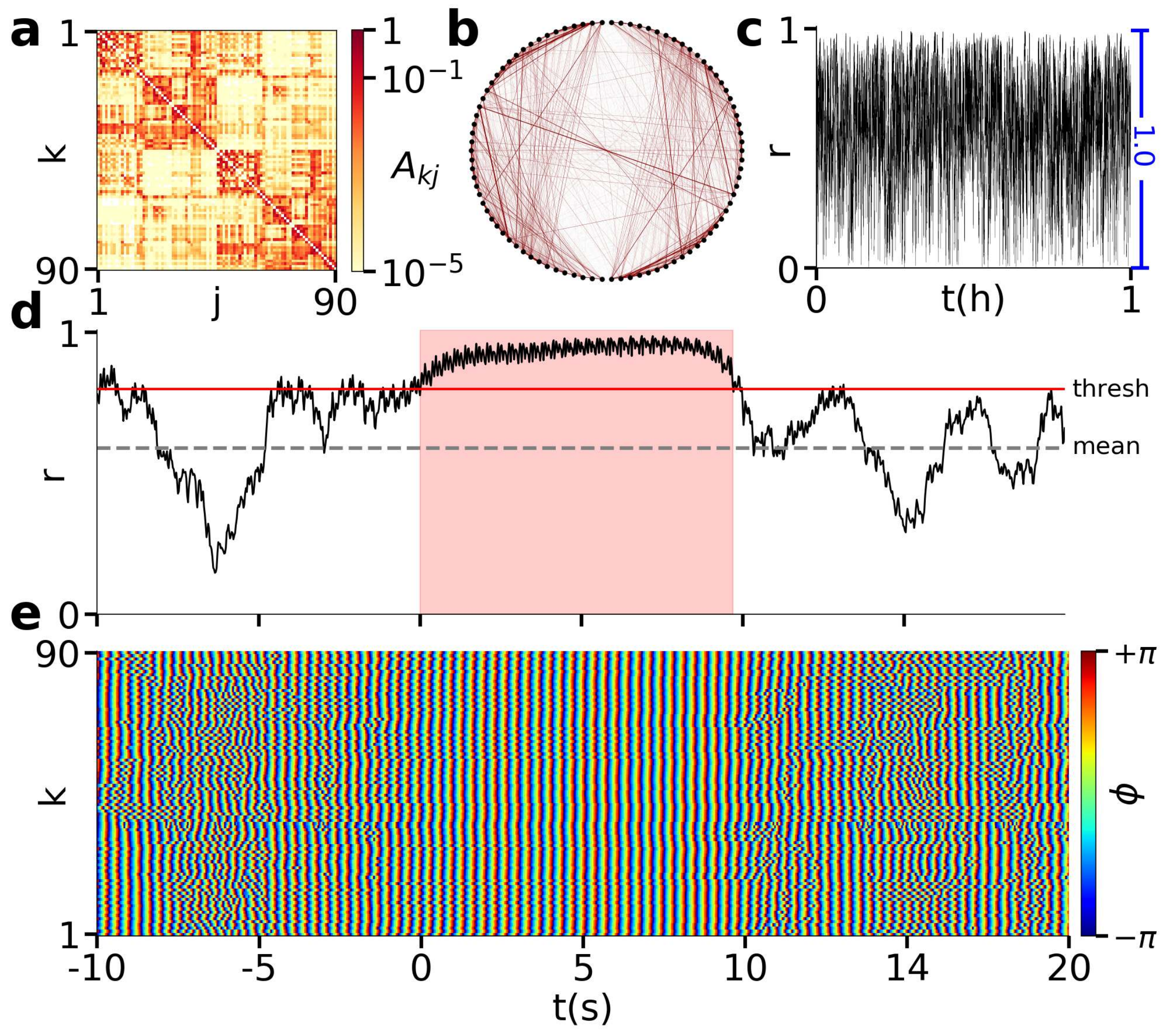}
\caption{Epileptic-seizure-like synchronization phenomena in a FitzHugh-Nagumo network with empirical connectivity. (a) Weighted adjacency matrix obtained from empirical human diffusion tensor imaging (DTI), averaged over 20 subjects\cite{MEL15}. Indices $k,j=1, \ldots, 45$, and $k,j=46,\ldots,90$ label the left and right hemisphere, respectively. The dense intra-hemispheric connectivities (first and second diagonal blocks) and the sparse inter-hemispheric connectivities can be clearly seen.
(b) Schematic plot of the network structure. The left (right) semicircle corresponds to the left (right) hemisphere, the nodes are numbered clockwise sequentially $1,\ldots,90$ starting from the bottom of the circle. The link thickness is proportional to the weights $A_{kj}$. 
(c) Global Kuramoto order parameter $r$ vs time (time interval one hour). It fluctuates strongly in the range from 0 to almost 1 (blue vertical bar). 
(d) Global Kuramoto order parameter $r$ vs time relative to the onset of a seizure (time interval 30 seconds). The horizontal dashed grey line denotes the time average over 3 hours. 
The horizontal full red line marks the threshold of $r=0.8$. If $r>0.8$ for more than 8 seconds, we define this as a seizure (pink shaded region). 
(e) Space-time plot of the dynamical phases corresponding to panel (d). The left (right) hemisphere is shown in the lower (upper) half. Simulation parameters: $a=0.5$, $\varepsilon=0.05$, $\varphi = \frac{\pi}{2} - 0.1$, $N=90$, $\sigma=0.6$.}
\label{DTI}
\end{figure}

First, we consider an empirical structural brain network. The brain network was obtained from diffusion-weighted magnetic resonance imaging data measured in healthy human subjects. For details regarding the experimental setup and data processing, see Ref.~\onlinecite{MEL15}, for previous utilization of the structural network to analyze partial synchronization phenomena, see Refs.~\onlinecite{CHO18,RAM19}, and for a short description of the Diffusion Tensor Imaging (DTI) data acquisition, we refer to the Appendix. The brains were segregated into 90 areas according to the Automated Anatomical Labeling (AAL) atlas~\cite{TZO02}. The 90 areas correspond to the 90 nodes of our network, and the connecting white-matter fibers between the areas correspond to the links. The anatomical names of the brain areas for each index $k$ are given in Tab.~SII 
of the Supplementary Material. To eliminate individual variation, the matrices of 20 subjects were averaged, giving rise to the topology of Fig.~\ref{DTI}(a), (b). In the present study, brain areas $k \in N_L = \{1,2,\ldots,45\}$ correspond to the left hemisphere and $k \in N_R = \{46,\ldots,90\}$ to the right hemisphere as in Ref.~\onlinecite{RAM19}. This contrasts the typical AAL indexing in which uneven $k$ are left, and even $k$ are right hemispheric areas. Using our labeling, the structure of the brain hemispheres can be easily distinguished: In the adjacency matrix in Fig.~\ref{DTI}(a), the connections within one hemisphere are much stronger than the connections between both hemispheres. In Fig.~\ref{DTI}(b), the network topology is schematically represented on a ring, where the left and right hemispheres correspond to the left and right half-circle, respectively. Most links are intra-hemispheric, and only very few inter-hemispheric connections can be seen. Note that the width of the links in Fig.~\ref{DTI}(b) is proportional to their weight.

In the simulations throughout this paper, we use the parameters $a=0.5$, $\varepsilon=0.05$, $\varphi = \frac{\pi}{2} - 0.1$. The coupling strength $\sigma$ is chosen such that it is as high as possible while still avoiding full synchronization for long simulations ($\approx 10 000$ time units). For the empirical connectivities, we choose $\sigma=0.6$. In order to compare our simulations with real data (EEG recordings of absence seizures; see Sec.~\ref{exp}), we transform the dimensionless time units of the FHN oscillator model to real time units by comparing the FHN oscillation period of a single FHN oscillator $T=2.56$ to the dominant frequency of an absence seizure at about $f=\unit[3] {Hz}$~\cite{GAS70, BAL00a, BLU00, SAD06}. Therefore, the simulation time is converted to real time by $\unit[1]{second} = 2.56/3 = 0.85$ simulation time units. 

The results of the simulation are shown in Fig.~\ref{DTI}(c), (d), (e). In panels (c) and (d), we show the global Kuramoto order parameter $r(t)$, which measures the degree of synchronization. Panels (c) and (d) also reveal periods of very high and of very low synchronization of the system as a function of time, varying in a range from 0 to almost 1 (panel (c)). The temporal average of the order parameter $\langle r \rangle$  (horizontal dashed grey line in (d)) and its standard deviation $\delta$ are given by $ \langle r \rangle \pm \delta = 0.59 \pm 0.21 $ for the full simulation of 164 minutes. We define a threshold of high synchrony as $r_{\text{th}} = \langle r \rangle + \delta = 0.8$ (horizontal red line in (d)). This threshold value is kept at 0.8 for all simulations in this article, even if the mean and the standard deviation differ for other topologies. In the simulation presented in Fig.~\ref{DTI}, the order parameter is found to be in high synchrony with $r>0.8$ during 17\% of the simulation time. Only if the synchronization remains above the threshold for at least 8 seconds, we define this time interval as a seizure.

In Fig.~\ref{DTI}(d), the order parameter is shown versus time for one exemplary seizure. Approximately 6 seconds prior to the start of the seizure, the order parameter drops to a low value of $r \approx 0.2$.
Such an apparent desynchronization can often be observed~\cite{MOR00,MOR03a,FEL07, AND16} prior to the onset of epileptic seizures. 
The order parameter then increases above $r>0.8$ (onset of seizure) and remains in high synchrony for almost 10 seconds. The seizure interval is shown as a pink shaded region; it marks the time of high synchronization without interruption. In the full simulation of 164 minutes, 11 seizures were detected, giving an average of 4 seizures per hour. Their average duration was 10.8 seconds, with a standard deviation of \unit[1.3]{s}. 
In Fig.~\ref{DTI}(e), the dynamic phases of the oscillators are shown as space-time plot for the same time interval as in (d). The lower half of the panel corresponds to the left hemisphere ($k \leq 45$), the upper half to the right hemisphere ($k\geq 46$). Since both hemispheres synchronize strongly, this resembles a generalized seizure.

\subsection{Random surrogate network}

\label{rand}

\begin{figure}
\centering
 \includegraphics[width=1\columnwidth]{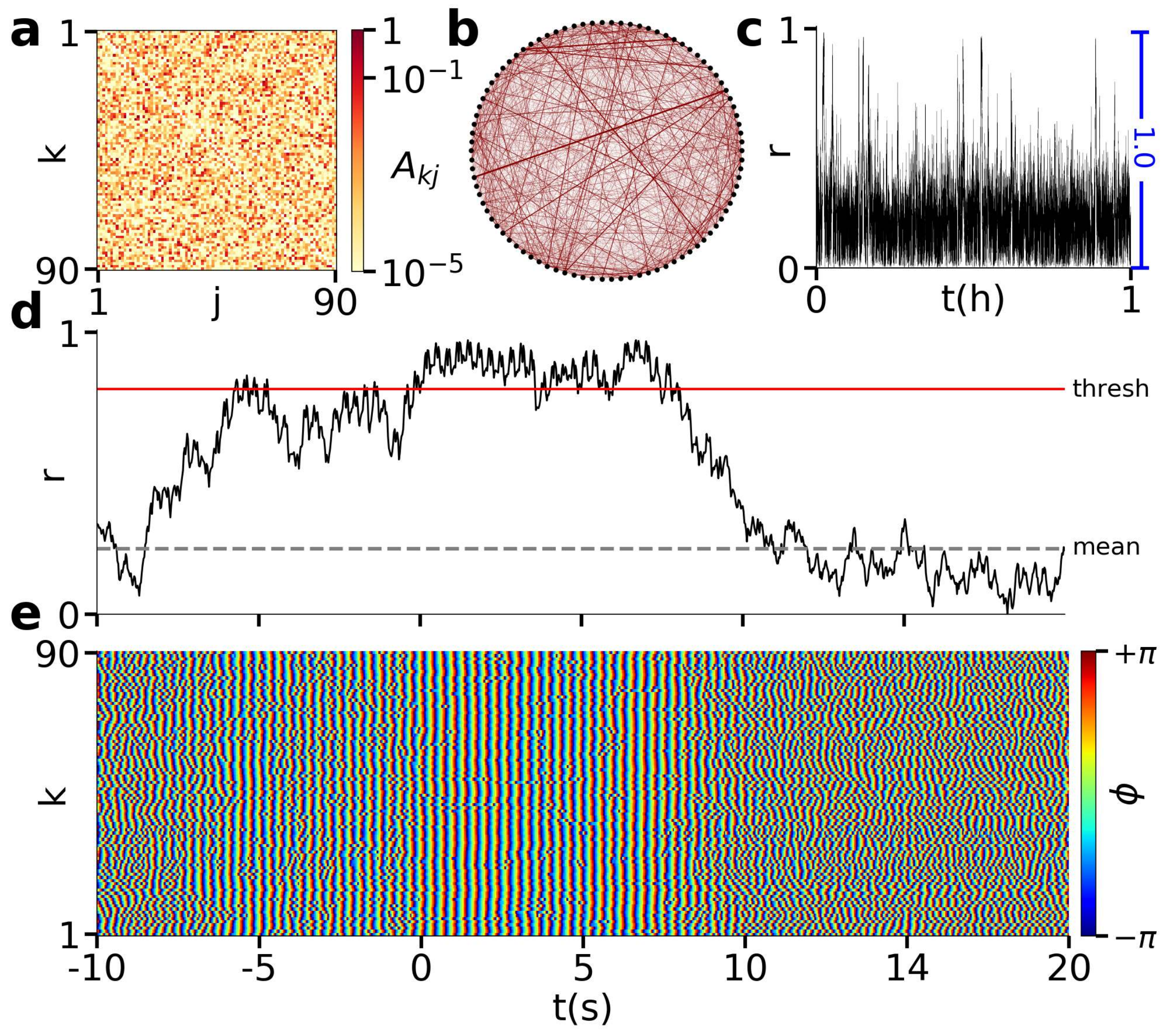}
\caption{Same as Fig.~\ref{DTI} with all links randomly rewired. Simulation parameters as in Fig.\ref{DTI}.}
\label{Random}
\end{figure}

In order to gain deeper insight into the interplay of dynamics and network topology, especially regarding the occurrence of seizures, we consider different artificially constructed networks. First, we study a surrogate network with all links of the empirical connectivity matrix randomly rewired (Fig.~\ref{Random}). Note that the set of weights of all links is the same as in Fig.~\ref{DTI}. However, the graph in panel (b) looks much denser due to the larger number of inter-hemispheric long-range connections. The simulation of the global Kuramoto order parameter $r(t)$, for one hour, shows that, on average, the system is less synchronized, see Fig.~\ref{Random}(c). Note that there are very short intervals of strong synchronization despite the very low average degree of synchronization. However, high synchrony $r>0.8$ is observed only in 1\% of the simulation time, and one such event is shown in Fig.~\ref{Random}(d). The global synchronization at $t\approx \unit[7]{s}$ is similar to the global synchronization in Fig.~\ref{DTI}(d) at $t\approx \unit[7]{s}$. However, the dynamic phases of the space-time plot in Fig.~\ref{Random}(e) appear less coherent since the connectivity of neighboring nodes is, on average, much smaller in the random network, preventing local synchronization. Since $r>0.8$ never holds for more than 8 seconds, according to our definition, not a single seizure is found in the simulation. The average degree of synchronization $\langle r \rangle= 0.23$ decreases significantly as compared to the one seen for the empirical connectivity (Fig.~\ref{DTI}). The links in this random network are the same as in the empirical network. Therefore, also the average node strength is equal, see Tab.~\ref{tabSUM}. However, the weighted clustering coefficient~\cite{BAR04b} and the average weighted shortest path length~\cite{BOC06} decrease by 41\% and 38\%, respectively. It is difficult to assess which network measure is an appropriate characteristic quantity related to the decrease in the average degree of synchronization. We address this question later in Sec.~\ref{Watts}.

Interestingly, by increasing $\sigma$ to 0.7 in the random surrogate network, the system attains an average Kuramoto order parameter of $\langle r \rangle = 0.60$, similar to the empirical connectivity. However, even if high synchrony $r>0.8$ is observed 47\% of the time (see Fig.~S1  
in the Supplementary Material), only 4.7 seizures per hour were detected, compared to 4 seizures per hour and 17\% high synchrony with the empirical connectivity, see Tab.~\ref{seizures}.

\subsection{Fractal connectivity}
\label{fract}

\begin{figure}
\centering
\includegraphics[width=8cm]{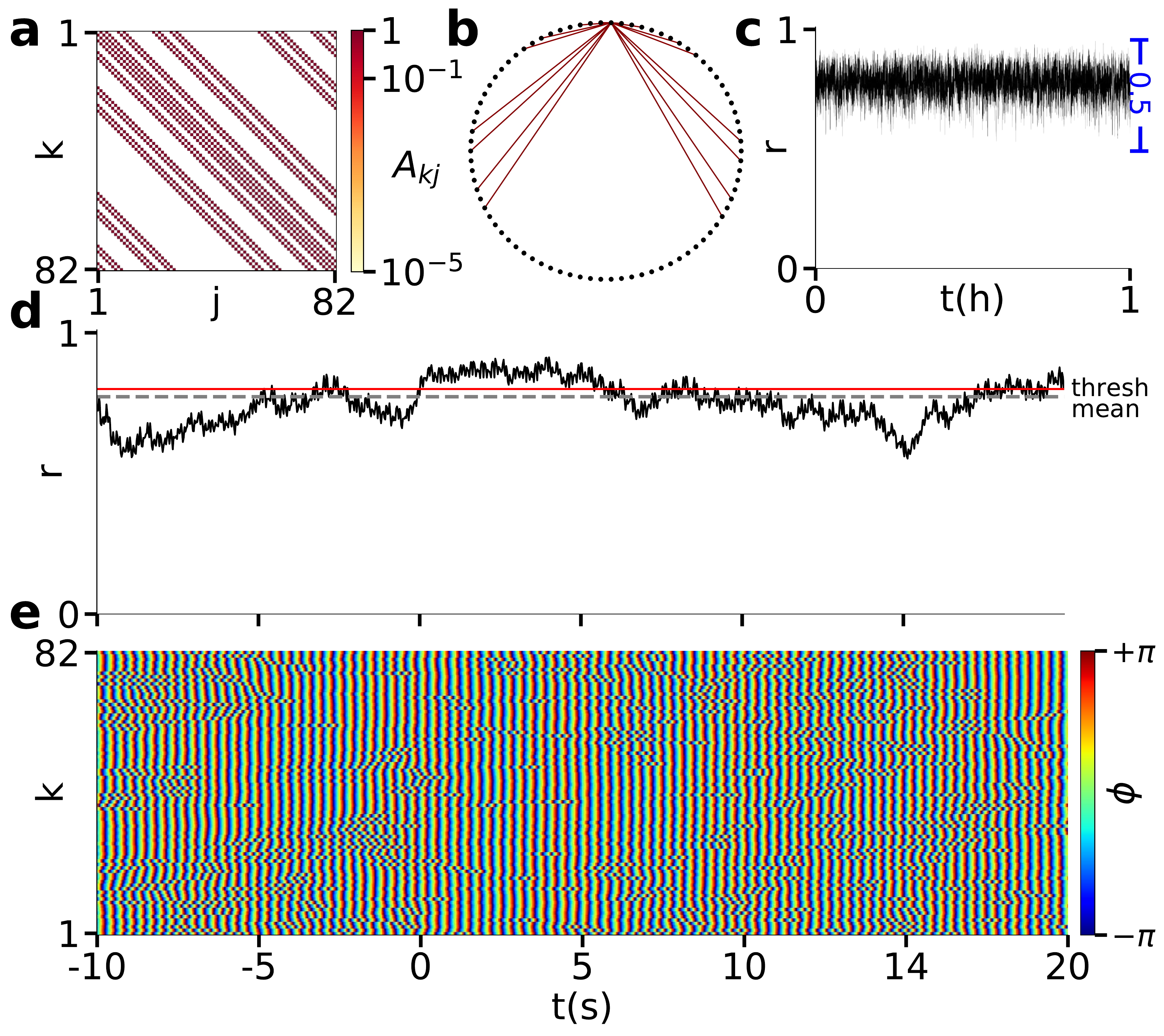}
\caption{Same as Fig.~\ref{DTI} for a ring network with fractal connectivity. 
Simulation parameters as in Fig.\ref{DTI}, except for $\sigma=0.01$ and $N=82$. In (b) for clarity, only links emanating from one representative node are shown. All other nodes have the same coupling topology.}
\label{Fract_Unw} 
\end{figure}

In mathematics, a \textit{fractal} is a self-similar structure  with a non-integer Hausdorff dimension. In nature, structures similar to fractals appear frequently\cite{ENR05, HAH05}. 
The white matter tracts in the human brain were reported to have a quasi-fractal structure~\cite{KAT09,KAT12}. This inspired simulations of networks of FHN oscillators with a one-dimensional~\cite{OME15,PLO16a,SAW19,NIK19} or two-dimensional~\cite{KRI17,CHO18} fractal coupling structure, and also for other dynamical models~\cite{HIZ15,ULO16,TSI16,TSI17,SAW17,BON18}. 
To create a (one-dimensional) ring network with fractal connectivity, we follow the procedure described in Ref. \onlinecite{KRI17}. Choose a base pattern $b_{\text{init}}$ that consists of a string of ones and zeros. In this article, $b_{\text{init}}=(1 0 1)$ is used. Then iterate this base $n$ times: For each $1$, substitute the initial base pattern $b_{\text{init}}$, for each $0$, substitute a string of zeros of size $b$ with $b=3$ corresponding to the length of the initial base pattern. The $n$th hierarchy level is reached after $n-1$ iterations. A mathematical fractal is obtained in the limit of an infinite number of iterations $n\rightarrow \infty$. Since a network is finite, only a finite number of iterations can be performed, and the resulting string is a \textit{quasi-fractal}. Put 0 in front of the string to exclude self-coupling~\cite{ULO16}. The resulting string of binary digits becomes the first row of the fractal ring adjacency matrix, where $1$ represents a link, $0$ represents no link. For every following row, the string is shifted by one element to the right. Via this procedure, a circulant matrix is obtained. After $n$ hierarchical steps, the obtained network consists of $N=b^n+1$ nodes. In this paper, $n=4$, so the network has 82 nodes. The adjacency matrix is shown in Fig.~\ref{Fract_Unw} (a) and the connections for one exemplary node are sketched in Fig.~\ref{Fract_Unw} (b).

The mean node strength $S=\sum_i s_i/N$, where $s_i=\sum_j a_{ij}$ is the $i$-th node strength or node degree, is very large ($S=16$) for this network. Consequently, it synchronizes completely at a relatively small coupling strength $\sigma$.  
Therefore, in order to avoid permanent complete synchronization, we reduce $\sigma$ to 0.01. Figure~\ref{Fract_Unw} (c) shows the order parameter for the (unweighted) fractal ring.
The average order parameter is $\langle r \rangle=0.77$, high synchrony $r>0.8$ is observed during 32\% of the time, but the range of $r$-values in (c) is comparatively small (0.5). Overall, the synchronization varies little, and no clearly defined seizures were found (d). The space-time plot (e) shows an overall moderately synchronized pattern without distinct bursts of synchrony. Thus it seems that fractal connectivities are not appropriate to model realistic epileptic seizures. Either different fractal connectivities or a larger network size or a particular weight distribution is necessary. The latter point is addressed next.

\subsection{Weighted fractal connectivity}
\label{fract_weight}

\begin{figure}
\centering
\includegraphics[width=8.5cm]{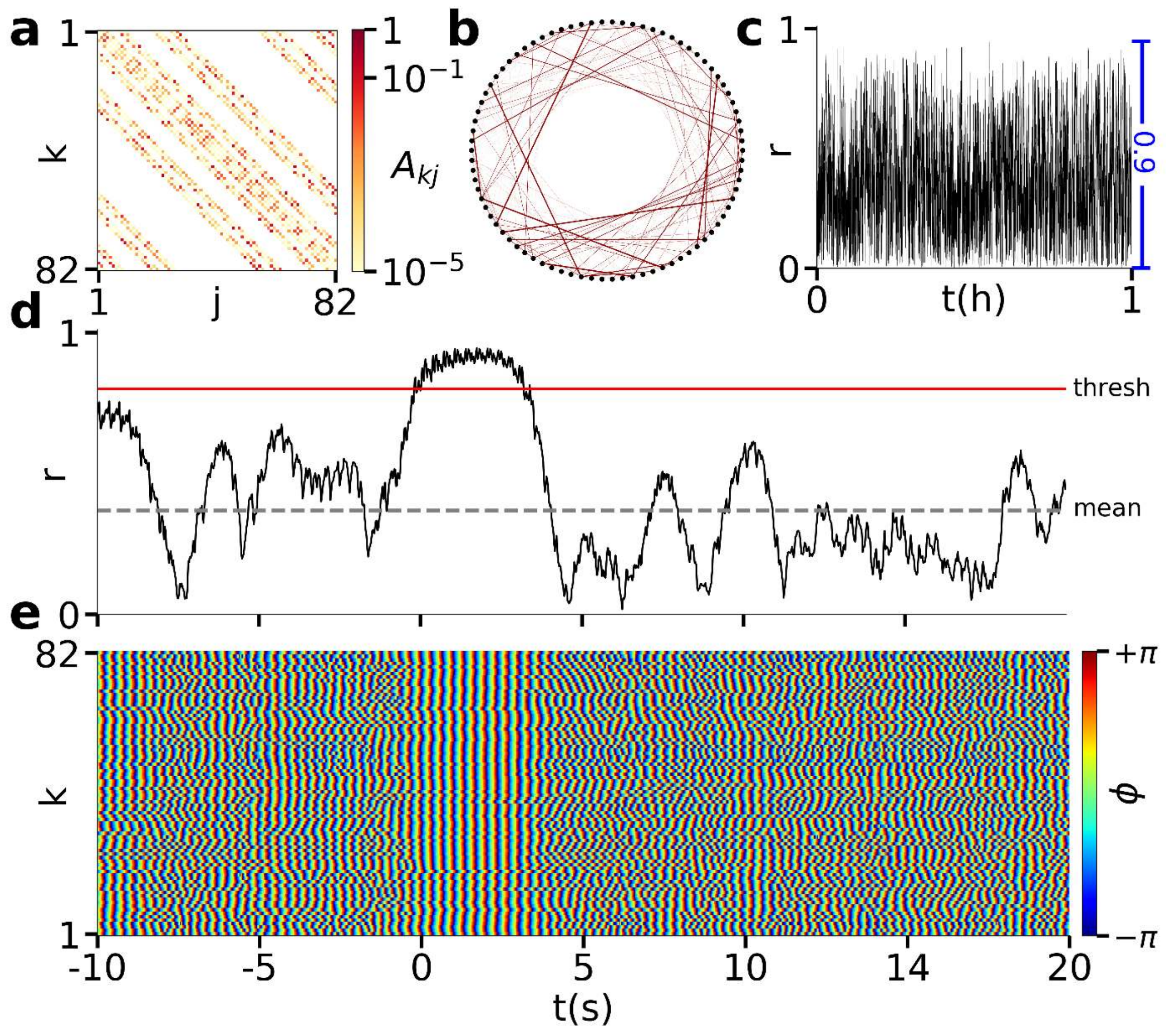}
\caption{Same as Fig.~\ref{DTI} for a ring network with fractal connectivity and weights selected randomly from the empirical connectivity matrix in Fig.~\ref{DTI}. 
Simulation parameters as in Fig.\ref{DTI}, except for $\sigma=6.1$ and $N=82$.} 
\label{Fract_DTI}
\end{figure} 

To achieve a more realistic weight distribution than just 1 and 0, we replace all 1312 non-zero links of the fractal ring in Fig.~\ref{Fract_Unw} by randomly chosen weights of the 7793 links of the empirical connectivity matrix in Fig.~\ref{Fract_DTI}(a), (b). Due to the much smaller weights, we have to increase $\sigma$ from 0.01 to $\sigma=6.1$ in order to obtain partial synchrony. Now the dynamics resembles the empirical one, as shown in Fig.~\ref{Fract_DTI}(c). 
The order parameter varies strongly in time. However, despite a few short events of high synchrony $r>0.8$ during 2\% of the total time, not a single seizure was detected, Fig.~\ref{Fract_Unw}(d), (e).
For higher values of $\sigma$, still, no seizure can be found because the system starts to stabilize at a fixed value of the order parameter of about $r \approx 0.7$.
Overall, the fractal connectivity with empirical weights shows rich, brain-like synchronization behavior. However, despite short phases with high synchrony, no well-defined seizures were found.

\subsection{Small-world networks}
\label{Watts}

\begin{figure}
\centering
\includegraphics[width=8.5cm]{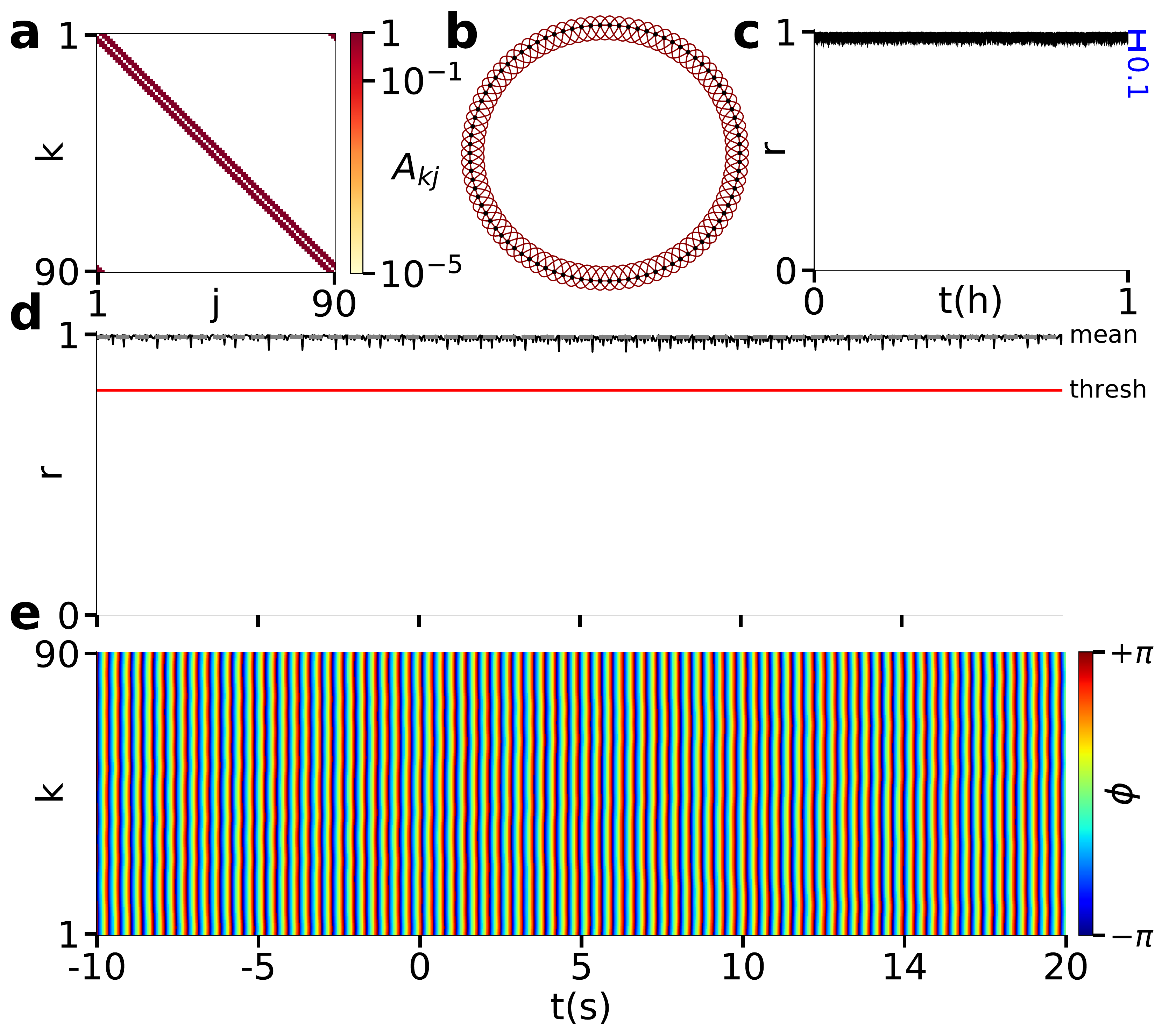}
\caption{Same as Fig.~\ref{DTI} for a nonlocally coupled ring network, corresponding to the Watts-Strogatz model with rewiring probability $p=0$ and $N=90$, $S=6$ (every node is coupled to 3 neighbors on each side). 
Average clustering coefficient $C=0.60$,  mean shortest path length $L=7.92$. Simulation parameters as in Fig.\ref{DTI}, except for $\sigma=0.0506$.} \label{p0}
\end{figure} 

Next, we consider small-world-like networks, which can be constructed according to the Watts-Strogatz algorithm~\cite{WAT98} by starting from a nonlocally coupled ring and randomly rewiring links with a probability $p$. With increasing $p$, these networks are characterized by decreasing average clustering coefficient (which quantifies the strongly coupled neighborhoods) and decreasing mean shortest path length. In some intermediate regime of $p$ and for large enough average node degree $S$, the small-world property of large clustering coefficient and short path length is found. For the largest probability $p=1$, we obtain an Erd{\H o}s-R{\'e}nyi random network\cite{ERD60}. 
The simulations are repeated 10 times for each probability $p$ with different random initial conditions. The coupling strength $\sigma$ is chosen for each $p$ such that the simulations give a sufficiently high degree of synchronization while avoiding complete synchrony during the whole simulation.

In the nonlocally coupled ring at $p=0$, each of the 90 nodes is connected to its three nearest neighbors on each side; thus, the node degree (strength) is $S=6$. The links are nondirected and nonweighted. This nonlocally coupled ring is shown in Fig.~\ref{p0}(a),(b). 
It is necessary to tune the coupling parameter $\sigma=0.0506$ very carefully. For slightly larger values of $\sigma$, the system would fully synchronize during the whole simulation time, while for lower $\sigma$, it would not synchronize at all. This underlines the system's sensitivity to small parameter changes~\cite{KAR14b,ANS16}.
For a long shortest path length and maximum clustering in the nonlocally coupled ring with $p=0$, the system fully synchronizes ($\langle r \rangle=0.99$) in 8 simulations. In 2 simulations, it is completely desynchronized ($\langle r \rangle=0.01$), leading to an arithmetic mean of $\overline{\langle r \rangle}=0.79$ for all simulations. One typical simulation, with $\langle r \rangle=0.99$, is shown in Fig.~\ref{p0} (c)-(e).

\begin{figure}
\centering
\includegraphics[width=8.5cm]{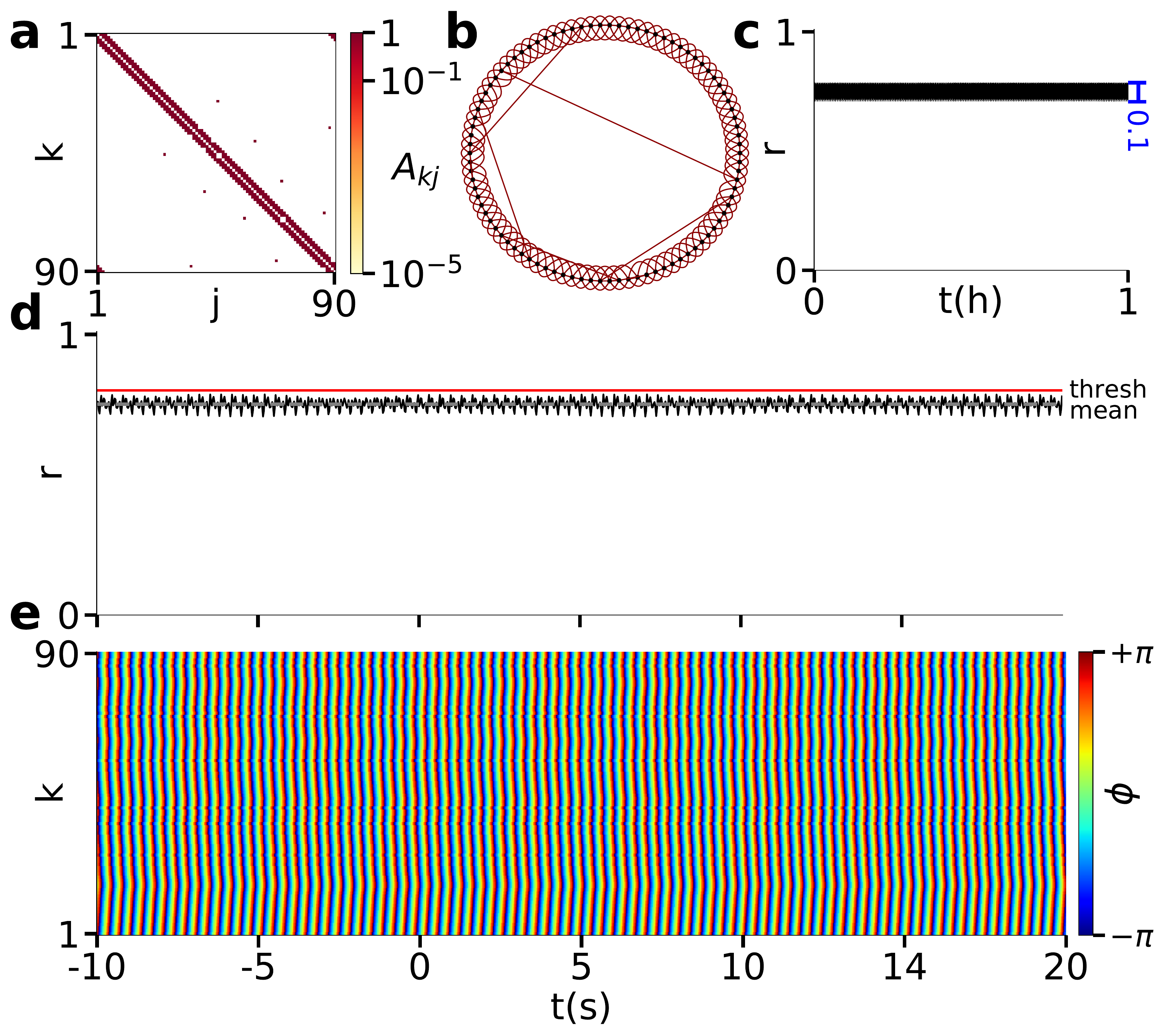}
\caption{Same as Fig.~\ref{DTI} for a Watts-Strogatz network with rewiring probability $p=0.006$. 
Average clustering coefficient $C=0.57$, mean shortest path length $L= 5.07$. Simulation parameters as in Fig.\ref{DTI}, except for $\sigma=0.0506$, mean node degree $S=6$.}
\label{p3}
\end{figure}

If the rewiring probability is increased slightly to $p=0.006$, we obtain a small-world network. It still has 95\% of the clustering coefficient of the nonlocally coupled ring, but its mean shortest path length is reduced by 36\%. The topology is shown in Fig.~\ref{p3} (a) and (b).
With an increase of $p$ and the resulting reduction of the average shortest path length, the synchronization decreases, see Fig.~\ref{p3}(c). While all other parameters are kept constant, and the clustering coefficient remains high, $ \overline{\langle r \rangle}=0.80$ is reduced to $\overline{\langle r \rangle}= 0.47$ averaged over all 10 simulations. For $p=0.006$, $\langle r \rangle$ is for some simulations as low as $\langle r \rangle =0.03$ and for some simulations as high as $\langle r \rangle=0.76$, showing a higher sensitivity to initial conditions. In both scenarios, $p=0$, and $p=0.006$, and for all simulations, the instantaneous degree of synchronization is approximately constant in time and varies in a very small range of $0.1$. In fact, by inspecting Figs.~\ref{p0} (d) and \ref{p3} (d) closely, one can recognize small periodic amplitude fluctuations of $r(t)$. This seems to be related to the slow-fast nature of the FHN system. Since the range of these fluctuations is very small, when full synchronization does not occur, no seizures are detected. The average degree of synchronization, however, depends sensitively upon the random initial conditions. All values are listed in Tab.~SI 
of the Supplementary Material. 

By comparing Figs.~\ref{p0} and \ref{p3}, the impact of the shortest path length upon the network dynamics becomes apparent. In general, after decreasing the network's shortest path length, the synchronization reduces significantly. In theory, this should reduce the risk of epileptic seizures; it contradicts earlier findings of an, on average, shorter path lengths in the functional networks of subjects with epilepsy~\cite{CHA10c}. However, our results should be taken with caution: The range of the order parameter is smaller than 0.1 in both cases and does not allow for realistic brain modeling. This indicates that brain networks must not have too large clustering coefficients, which questions the hypothesis that the human brain is a small-world network~\cite{HIL16,PAP16,GAS16,BIA10,HLI12a,HLI17}.

\begin{figure}
\centering
\includegraphics[width=8.5cm]{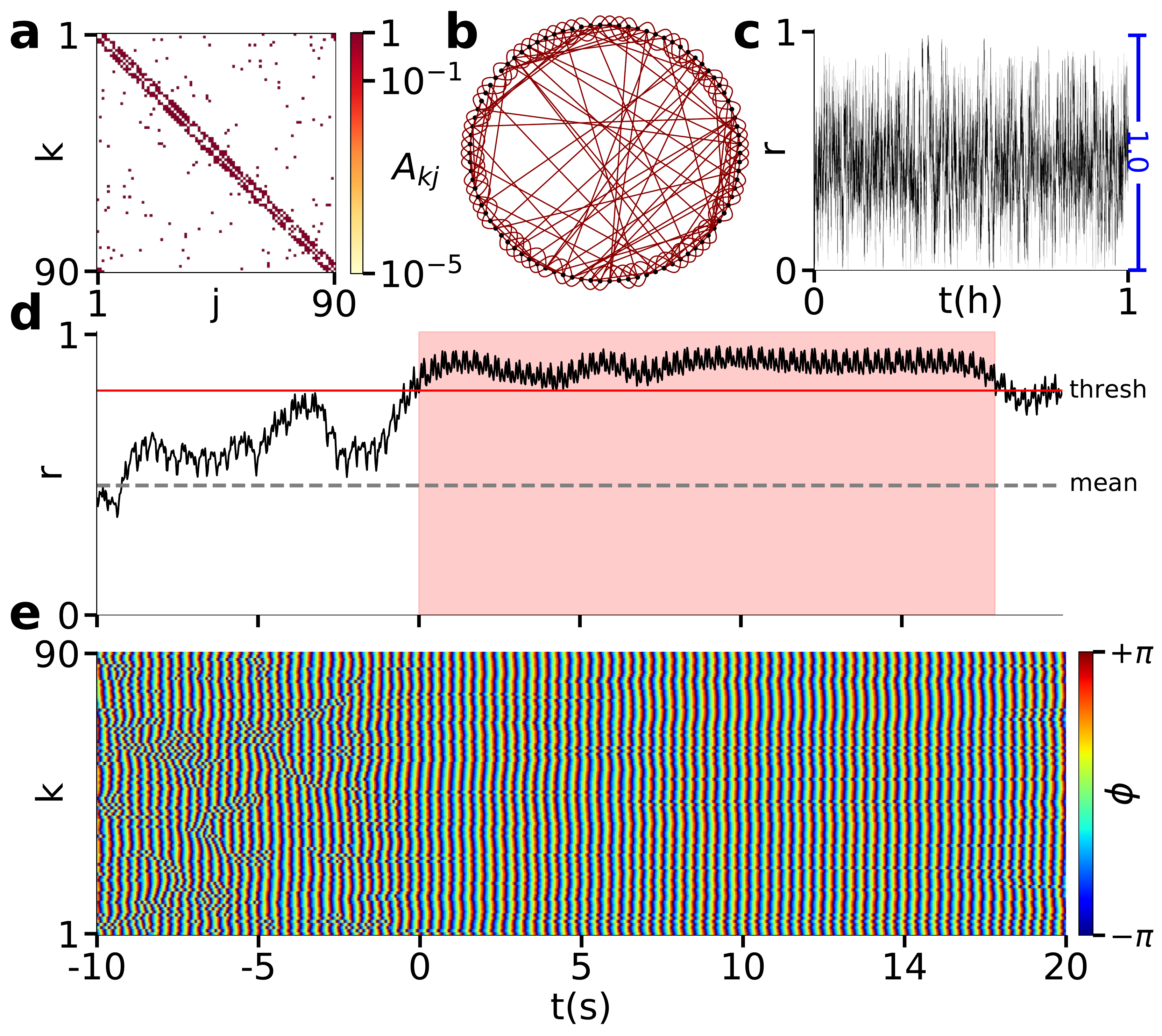}
\caption{Same as Fig.~\ref{DTI} for a Watts-Strogatz network with rewiring probability $p=0.232$. Average clustering coefficient $C= 0.25$, mean shortest path length $L=2.97$. Simulation parameters as in Fig.\ref{DTI}, except for $\sigma=0.0506$, mean node degree $S=6$.}
\label{p8}
\end{figure}

Next, we increase the rewiring probability to $p=0.232$, see Fig.~\ref{p8}. The clustering and shortest path are reduced to 41\% and  38\% of their original values. At $p=0.232$, the mean shortest path length is very close to its minimum, and the clustering is still significant. The order parameter is $\langle r \rangle \approx 0.52$ for all simulations, proving its independence of initial conditions. We find high synchrony during 14\% of the time, and 0.6 seizures per hour with an average duration of 14.7 seconds with a standard deviation of \unit[3.5]{s}. One seizure is shown in Fig.~\ref{p8} (d) and (e). The dynamics is similar to the one for the empirical connectivity. 

\begin{figure}
\centering
\includegraphics[width=8.5cm]{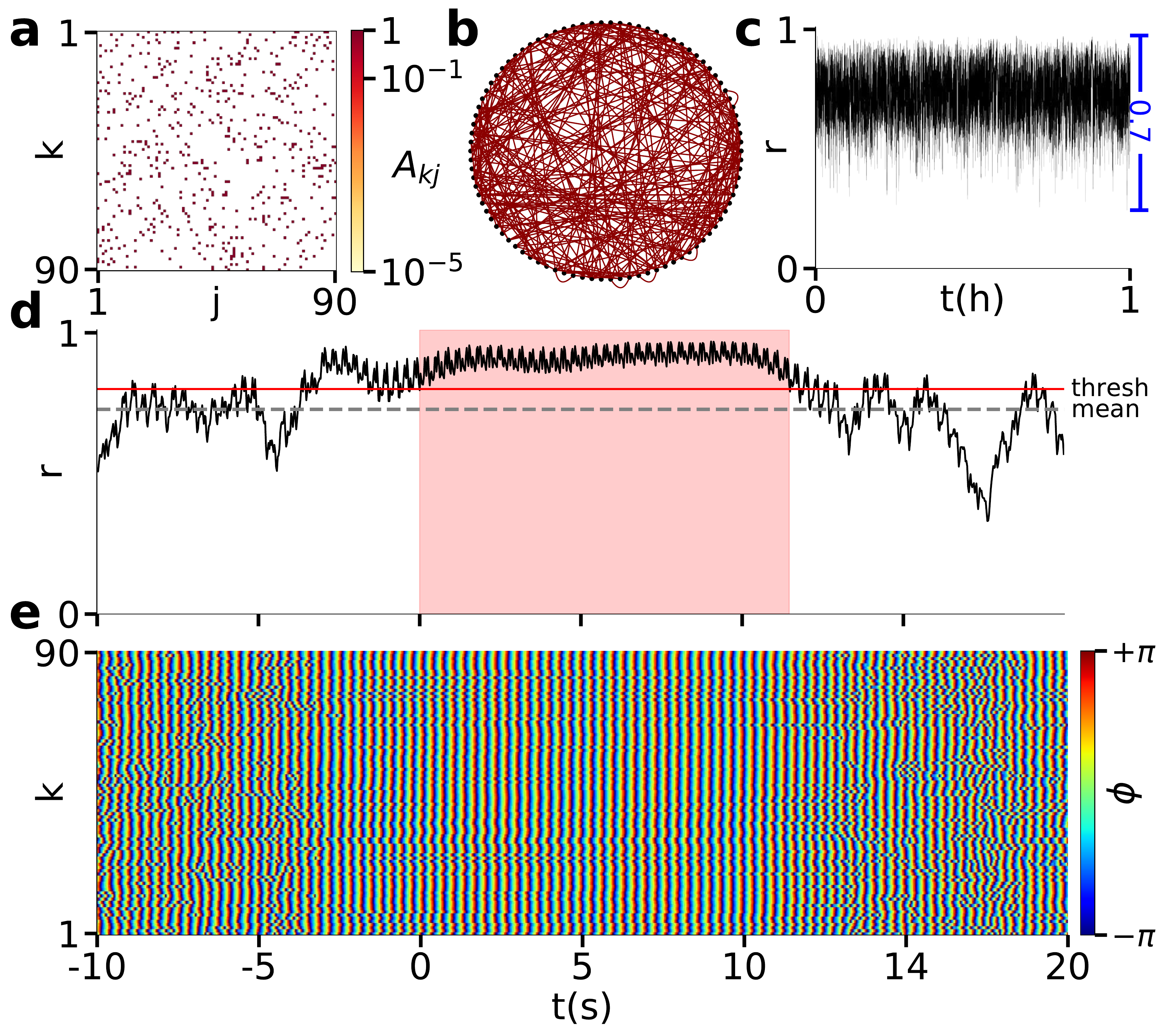}
\caption{Same as Fig.~\ref{DTI} for a random network (Watts-Strogatz network with rewiring probability $p=1$). Average clustering coefficient $C=0.053$, mean shortest path length $L=2.67$. Simulation parameters as in Fig.\ref{DTI}, except for $\sigma=0.0506$, mean node degree $S=6$.}
\label{p10}
\end{figure} 
For $p=1$, corresponding to a random network (Fig.~\ref{p10}), the mean shortest path length reduces further by only 11\%, whereas the clustering reduces further by 79\%. Now we find a very high Kuramoto order parameter of $\langle r \rangle=0.73$ and a high degree of synchrony during 25\% of the time. Note the very high average of $r$. Strikingly though, despite the strong synchronization in general, fewer seizures (only 0.5 per hour) are observed as compared to our findings reported in Fig.~\ref{p8}. One such seizure is shown in Fig.~\ref{p10}(d),(e). The values of the network measures for all topologies are summarized in Tab.~\ref{tabSUM}; and the simulation results are summarized in Tab.~\ref{seizures}.

By comparing Figs.~\ref{p8} and \ref{p10}, it becomes apparent that seizures are more likely for a higher average clustering coefficient of the network than for a smaller clustering coefficient.  This is in line with Ref.~\onlinecite{CHA10c}. 
Remarkably, this is the case even though the average degree of synchronization decreases significantly for a larger average clustering coefficient.  This shows that the occurrence of seizures is not proportional to the average global Kuramoto order parameter of the network, as one might expect. For a reduction of the clustering coefficient, the synchronization might increase while the probability of seizures decreases. 

When comparing Fig.~\ref{p8}(c) with Fig.~\ref{p10}(c), the impact of clustering on seizure probability shows up (since the mean shortest path length changes little). For rich synchronization dynamics, including both very low and very high synchronization, the clustering coefficient needs to be sufficiently high:  For $p\leq0.006$, the difference between maximum and minimum of $r(t)$ is $\Delta r = 0.1$, whereas for $p \geq 0.232$ it is $\Delta r  \geq 0.8$.

In conclusion, our simulations indicate that the human brain can only effectively function in a specific window of medium clustering. If the clustering is too large, the neural synchronization is approximately constant in time ($\Delta r \approx 0.1$). The brain, however, shows both low and high synchronization values on the EEG during different tasks and mental states such as sleep. Moreover, epileptic brains, which function normally most of the time, appear to synchronize during generalized seizures fully. This shows that the brain is capable of sustaining both very coherent and very incoherent oscillatory states, which is not possible for too large clustering coefficients. 

On the other hand, if the clustering coefficient is too small, the synchronization fluctuates rapidly in time and does not resemble the dynamics of simulations with an empirical brain network. Furthermore, the range of the degree of synchronization ($\Delta r \approx 0.7$) is decreased compared to medium clustering ($\Delta r \approx 1$). One might speculate, based on these simulations, that the difference between healthy and epileptic brains might show up in the network's slightly altered clustering coefficient~\cite{HOR10,ANS12}.

\begin{table}[]
\begin{tabular}{|l|l|l|l|l|l|l|l|l|}
\hline
Network & Nodes & Links & Weights & $S$ & $C$ & $L$ \\ \hline\hline
Empirical &	90 &	7793 & weight. & 1.3 & $1.7e^{-3}$ & $2.9e^{-7}$ \\ \hline
Random Surr.&	90 &	7793 & weight. & 1.3 & $1.0e^{-3}$ & $1.8e^{-7}$      \\ \hline
Fractal Unw. & 82	 &1312&unw.&  16	& 0 & 2.1\\ \hline
Fractal DTI & 82	 &1312&weight.& 0.11& 0&$1.7e^{-6}$\\ \hline
SW $p=0$ &90 &270  &		unw.& 6& 0.60 &7.9 \\ \hline
SW $p=0.006$  &90& 270 &unw.& 6& 0.57 &5.1 \\ \hline
SW $p=0.232$ &90 & 270  &unw.& 6& 0.25&3.0 \\ \hline
SW $p=1$ &90& 270 &unw.&  6&0.05 & 2.7 \\ \hline
\end{tabular}
 \caption{Network measures for various topologies. Empirical: empirical Diffusion Tensor Imaging network, Random: Random surrogate network with DTI weights, Fractal Unw: Fractal connectivity (unweighted), Fractal DTI: Fractal connectivity with randomly selected DTI weights, SW: Small-world Watts-Strogatz model with rewiring probabilities $p$; Nodes: number of nodes; Links: number of (non-zero) links; weights: weighted/unweighted; $\bar w_{ij}$: mean weight (median weight) of all links; $S$: average node strength; $C$: average (weighted) clustering coefficient; $L$: mean (weighted) shortest path length.}
\label{tabSUM}
\end{table}

\begin{table}[]
\begin{tabular}{|l|l|l|l|l|l|l|l|l|}
\hline
Topology & $\sigma$ & $\overline{\langle r \rangle}$ & $\Delta r$ & $r>0.8$ & Num & Duration  \\ \hline\hline
Empirical  &0.6 &0.59  &0.99& 17 \%& 4.0 & $\unit[10.8\pm1.3]{s}$\\ \hline 
Random   & 0.6& 0.23 &0.99&   1\% &   0   &   - \\ \hline 
Random   & 0.7& 0.60 &0.99&   47\% &   4.7   &   $\unit[10.2\pm2.6]{s}$ \\ \hline 
Fractal Unw   & 0.01 & 0.77 & 0.47 &   32\% &   0   &   -   \\ \hline
Fractal DTI   & 6.1 & 0.37 &0.95 &   2\% &   0   &   -    \\ \hline
$p=0$   & 0.05 & 0.79 &0.07 &   80\% &   0  &   -    \\ \hline 
$p=0.006$   & 0.05 & 0.47 &0.10 &   0\% &   0  &   -    \\ \hline 
$p=0.232$  & 0.05 & 0.52 &0.98 &   14\% &   0.6  &   $\unit[14.7\pm3.5]{s}$ \\ \hline 
$p=1$   & 0.05 & 0.73 &0.81 &   25\% &   0.5  &   $\unit[9.0\pm0.1]{s}$ \\ \hline 
 \end{tabular}
 \caption{Comparison of seizures for different networks. $\sigma$: coupling strength; $\overline{\langle r \rangle}$: average 
 order parameter, $\Delta r=r_{\text{max}}-r_{\text{min}}$: range of $r$; $r>0.8$: percentage of time with high synchrony; Num: number of seizures per 1 simulation hour; Duration: average duration of  seizures. For each of the last four topologies, 10 simulations with an average of 2.9 hours were performed. For $p=0$, $r>0.8$ was true 0\% of the time for 2 simulations and 100\% for 8 simulations.  Minimum duration of seizures 8 s.}
\label{seizures}
\end{table}

\section{Comparison with EEG-recorded absence seizures}
\label{exp}

\begin{table*}[]
\centering
\begin{tabular}{|c|c|c|c|c|c|c|c|c|c|c|c|}
\hline
left/right & Fp1-F3/ & Fp1-F7/& F3-C3/& C3-Cz/& F7-T3/& FT9-T3/& T3-T5/&T3-C3/  & C3-P3/ & T5-O1/ & P3-O1/ \\ 
electrodes & Fp2-F4 &Fp2-F8& F4-C4& Cz-C4& F8-T4& T4-FT10& T4-T6&C4-T4  & C4-P4 & T6-O2 & P4-O2 \\ \hline
AAL number & 2/47	&4/49&10/55&	1/46&7/52&	41/86&	43/88&	29/74&	30/75&	26/71&	25/70\\ \hline
Brain Area &Front Sup	&Front Mid	&Supp Mot Area	&Precentral	&Front Inf Tri	&Temp Sup	&Temp Mid&	Postcentral	&Par Sup	&Occ Mid&	Occ Sup \\ \hline
\end{tabular}
 \caption{Mapping of the EEG electrodes to the AAL. Electrodes were referenced in bipolar montage. This means that the EEG time series for AAL area 7 in Fig.~\ref{Exp} (d), for example, shows the measured voltage difference of electrodes F7 and T3. The AAL area covered by their locations were approximately identified and assigned. The electrode locations are shown in Fig.~\ref{Exp} (b).}
\label{tabAreas}
\end{table*}

\begin{figure*}
\centering
\includegraphics[width=\the \textwidth]{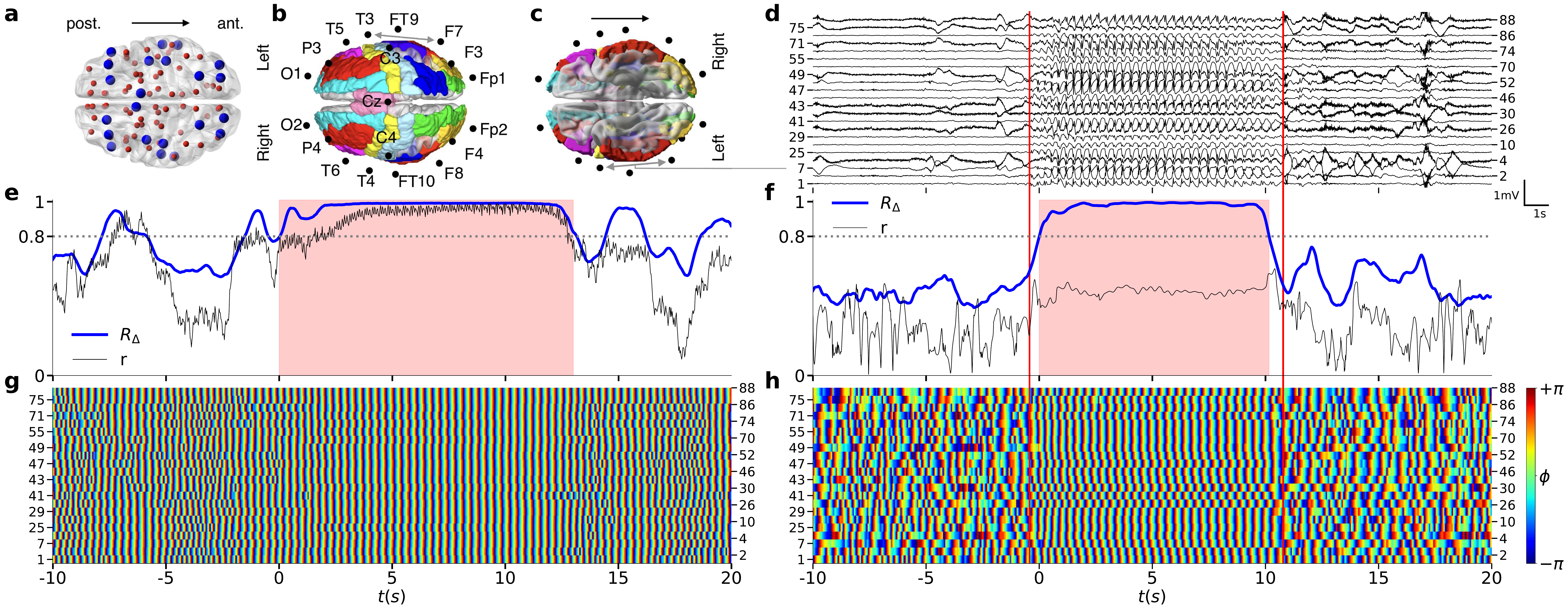}
\caption{Comparison of simulated and measured epileptic-seizure-related synchronization phenomena. (a) The 90 AAL brain areas marked by red and blue dots. The simulations of the network shown in panels (e), (g) are based on all areas (red and blue). However, the global phase coherence and global Kuramoto order parameter are calculated, for better comparison with the EEG-recorded seizure, for the areas shown in blue only. (b) Top view on brain: Electrode locations on the brain (black dots). 
All AAL brain areas corresponding to the blue dots in panel (a) are assigned to the EEG locations and highlighted in color. (c) Bottom view on brain: 
the grey line connects the exemplary electrode pair F7-T3 to their EEG time series shown in panel (d). 
(d) EEG recording of an exemplary absence seizure. Red vertical lines mark the onset and termination. (e) Simulated global Kuramoto order parameter $r$ vs time (time interval 30 seconds) in black during one simulated seizure. Global phase coherence $R_{\Delta}$ vs time in blue. The horizontal grey dashed line marks the threshold of $r=0.8$. In contrast to the previous figures, here the threshold $r>0.8$ is applied to $R_{\Delta}$ instead of $r$ to enable a comparison between simulated and measured data. The resulting duration of the simulated seizure is shown as a pink shaded region. (f) Same as in panel (e) for the EEG recording in (d). (g) Simulated space-time plot of the dynamic phases corresponding to panel (e). The dynamical phases of the left (right) hemisphere are shown in the lower (upper) half. The dynamics results from all 90 brain areas, but only the phases of the 22 areas sampled with the EEG are shown. Simulation parameters: $a=0.5$, $\varepsilon=0.05$, $\varphi = \frac{\pi}{2} - 0.1$, $N=90$, $\sigma=0.6$.  (h) Same as panel (g) for EEG recording in panels (d) and (f). (a)-(c) were created using BrainNet Viewer~\cite{XIA13a}.}
\label{Exp}
\end{figure*} 

We have obtained our EEG recordings from the Department of Epileptology of the University of Bonn from a 12 years old subject who suffered from absence seizures. The study was approved by the ethics committee of the University of Bonn, and a parent gave written informed consent that the clinical data might be used and published for research purposes. 
EEG data were acquired at a sampling rate of \unit[256]{Hz} (16 bit A/D conversion) within a bandwidth of \unit[0.3~--~70]{Hz} from 19 electrodes in bipolar montage. Locations and nomenclature of these electrodes are standardized by the American Electroencephalographic Society\cite{SHA91}. 

In order to facilitate a comparison between EEG-recorded and simulated epileptic-seizure-related synchronization phenomena, we estimated which AAL brain area was recorded by the different EEG electrode pairs. We chose those AAL regions that are located between the electrode pairs right below the skull (see Tab.~\ref{tabAreas}) and considered the electrode-to-brain area assignment by Ref.~\onlinecite{OKA04}. We emphasize that a perfect assignment between EEG electrodes and AAL brain areas is not possible. 

We apply a Morlet wavelet to filter the EEG signals for frequencies in the range $f_i \in \{\unit[1]{Hz}, ..., \unit[32]{Hz}\}$ 
and extract the corresponding phases. From the phase data, we calculate the global Kuramoto order parameter $r$ according to equation (\ref{GKOP}). Additionally, we introduce a new measure that we call global phase coherence $R_{\Delta}$. It generalizes the mean phase coherence between two oscillators $i, j$, studied earlier,~\cite{HOK89,MOR00,CHE07a,CHE09a,KAS17} to its arithmetic mean over all pairs $(i, j)$: 
\begin{equation}
R_{\Delta} = \frac{1}{N(N-1)/2} \sum^{N}_{\substack{i,j=1,\\ i > j}}  \left | \frac{1}{T} \sum^T_{t=1}e^{i\Delta \Phi_{ij}(t)}\right |,
\end{equation}
where $\Delta \Phi_{ij}(t) = \phi_i(t)-\phi_j(t)$ is the dynamic phase difference corresponding to two electrode pairs $(i, j)$ at time $t$. The complex number $e^{i\Delta \Phi_{ij}(t)}$ is averaged over the time window $T$: If the phase relationship of the two pairs is constant throughout the time window $T$ (frequency-locking), the modulus of the time-averaged $e^{i\Delta \Phi_{ij}(t)}$ becomes 1. If the phases are decoupled, the modulus of the averaged $e^{i\Delta \Phi_{ij}(t)}$ decreases. For small $T$, the time resolution of the mean phase coherence is large. However, for very small $T$ in the limit $T \to 0$, the global phase coherence $R_{\Delta} \to 1$ because any two pairs are perfectly phase-locked for an infinitely small time interval which gives no information about the temporal evolution of the phase coherence. For large $T$, smaller temporal fluctuations are averaged out, and significant changes in the coherence versus time can be observed. However, in the limit $T \to \infty$, the time resolution of $R_{\Delta}$ is lost. We choose a time window of 3 oscillation periods for each frequency band, focusing on high time resolution while allowing for strong temporal fluctuations of $R_{\Delta}$. The global phase coherence is defined in such a way that the normalization factor makes it 1 for perfect coherence. We have calculated the global phase coherence during epileptic seizures for different frequencies and found the best seizure identification at  $f=\unit[3]{Hz}$, which is typical for absence seizures\cite{SAD06,GAS70}. For this frequency, our time window of 3 periods corresponds to $T=\unit[1]{s}$. Note that this coherence measure is 1 for phase-locked states with a fixed phase difference, in contrast to the Kuramoto order parameter, which is 1 only for complete in-phase synchronization of all oscillators.

Figure~\ref{Exp} presents a comparison of our simulations for an empirical connectivity with an EEG recording of an absence seizure. Electrodes and brain areas are depicted in Fig.~\ref{Exp}(a)-(c), measured EEG data in Fig.~\ref{Exp}(d), simulated and EEG-related global phase coherence and global Kuramoto order parameter are shown in Fig.~\ref{Exp}(e) and (f), respectively, and the simulated and EEG-related space-time plot of the phases in Fig.~\ref{Exp}(g) and (h), respectively. 

Figure~\ref{Exp}(a) shows the 90 AAL brain areas underlying the simulations in (e) and (g) as red and blue dots. In (b), the approximated electrode locations are shown as black dots. The EEG time series are obtained by subtracting the voltages from two electrode pairs. For each of the 22 bipolar recording channels, we assign the closest AAL areas, which are colored in (b). These colored brain areas are marked blue in (a), and the ones which are not associated with electrodes in red. Fig.~\ref{Exp}(c) provides the bottom view of (b) to clarify that no subcortical brain areas are assigned to the comparison. The EEG recording is presented in (d). For three exemplary electrode pairs, denoted by grey arrows in (c), grey lines connect them to the data in (d). The seizure start and end are marked by vertical red lines and are determined by visual inspection of the EEG by an expert. 

In Figures~\ref{Exp}(e) and (f), both the global Kuramoto order parameter $r(t)$ (black line) and the global phase coherence $R_{\Delta}(t)$ (blue curve) are shown for simulation and EEG recording, respectively. In the simulation, the coupled dynamics of all 90 brain areas is computed. However, only the 22 areas that were accessible with the EEG are used to calculate the simulated measures $R_{\Delta}$ and $r$ for better comparison. 

The phases underlying the computation of  $R_{\Delta}$ and $r$ are shown in the space-time plots of Figs.~\ref{Exp}(g) and (h). We extracted phases from the EEG data shown in Fig.~\ref{Exp}(d) after filtering for $f=\unit[3]{Hz}$. The frequency synchronization during the seizure can easily be distinguished from the pre- and post-seizure activity. The left and the right y-axis show the label of each of the 22 assigned AAL brain areas in Tab.~\ref{tabAreas}.

Let us examine the EEG-recorded seizure in more detail. The value of the global phase coherence in panel (f) before and after the seizure is $R_{\Delta} \approx 0.5$. At the (electrical) onset of the seizure, $R_{\Delta}$ increases quickly and reaches almost $R_{\Delta} \approx 1$ within one second. Applying our threshold condition $R_{\Delta} > 0.8$ to the global phase coherence, we mark the onset and termination of the seizure, shown as a pink shaded region. Note that the start of the seizure (vertical red line) slightly precedes the pink shaded region. This is expected since we obtain $R_{\Delta}$ by averaging over the time window $T=\unit[1]{s}$, resulting in a lower time resolution of $R_{\Delta}$. For the same reason, the end of the seizure, according to our threshold definition, slightly precedes the marked seizure termination. However, in both cases, the measured absence seizure lasts for approximately \unit[10]{s}. Overall, 4 of the 5 seizures from the data set (all shown in the Supplementary Material) were identified correctly by our seizure detection criterion.

Note that during the recorded seizure temporal fluctuations of the Kuramoto order parameter $r(t)$ are stabilized, in accordance with our simulations.
However, contrasting our simulations, $r(t)$ remains at a low value of $r \approx 0.5$. Therefore, if we apply our previous seizure detection criterion $r > 0.8$ to the Kuramoto order parameter, we would not detect a seizure from the EEG recording. 
This highlights the difference between both synchronization measures. Due to time delays in the real system (brain), we cannot expect complete phase synchronization but only synchronization with respect to relative phases, which is measured by $R_{\Delta}$. For this reason, $R_{\Delta}$ is more appropriate than $r$ for quantifying brain synchrony. 

When examining the simulated data in Fig.~\ref{Exp}(e), we notice that $R_{\Delta}$ and $r$ correlate strongly, since delay is not included in the simulated equations (\ref{eq.1}). 
Due to the small averaging window of $T=\unit[1]{s}$, the signal fluctuates strongly before and after the simulated seizure. We show the same plot with $T=\unit[3]{s}$ (Fig.~S2) and with $T=\unit[5]{s}$ (Fig.~S3) in the Supplementary Material. For these longer averaging windows, the fluctuations are successively averaged out, and substantial similarity between EEG recording and simulation can be seen. However, the time resolution decreases, prohibiting the precise determination of the start and end of the simulated seizure. For a better comparison of our simulation with the EEG recording, we have also applied our seizure detection criterion to $R_{\Delta}$ in the simulation.   The simulated seizure lasts \unit[13]{s} and is shown as a pink shaded region. In general, $R_{\Delta}$ and $r$ are similar for the simulation, but due to the weaker definition of synchronization, the global phase coherence is, on average, larger than the order parameter.

\section{Conclusion}

We have shown that FitzHugh-Nagumo oscillators, coupled via empirical structural connectivities measured in human subjects, exhibit synchronization phenomena that resemble the ones seen during epileptic seizures. Comparing our long-term simulations using empirical connectivities to EEG-recorded epileptic seizures, we have found that the simulations show striking similarities to the real data.
By simulating FitzHugh-Nagumo oscillators on a variety of networks, we have gained insight into the interplay of network structure and synchronization. Our work highlights that both the weight distribution and the clustering coefficient of the network are critical components in the synchronization behavior. For a better understanding of epilepsy, it might be useful to compare these and other network measures in both structural and functional brain networks of people with epilepsy and healthy subjects~\cite{LEH14a}.
In order to obtain insight into the interplay of dynamics and network topology, we have performed simulations for different artificial network structures: random networks, regular nonlocally coupled ring networks, ring networks with fractal connectivities, and small-world networks with various rewiring probabilities. 
In more detail, by randomly rewiring its links, we have artificially destroyed the highly organized structure of the empirical connectivity matrix, while keeping the weight distribution and average node strength constant. Next, we have used a quasi-fractal connectivity on a ring network. To enable a better comparison with the empirical network, we have transformed the quasi-fractal ring into a weighted network with an empirical weight distribution. Finally, we have evaluated the impact of the average clustering coefficient and the average shortest path length on the number of observed epileptic-seizure-related synchronization episodes. We have  examined both network measures by employing the Watts-Strogatz small-world algorithm with specific rewiring probabilities. Among the artificial networks, a small-world network with intermediate rewiring probability results in the best agreement with the simulations for empirical structural connectivity. For the other network topologies, either no spontaneously occurring epileptic-seizure-related synchronization phenomena are found in the simulated dynamics, or the overall degree of synchronization remains high throughout the simulation. This indicates that a topology with some balance of regularity and randomness favors the self-initiation and self-termination of episodes of high, seizure-like synchronization. In particular, the value of the clustering coefficient should not be too high (as for regular ring networks) and not too low (as for pure random networks), and thus the rewiring probability should assume intermediate values between 0 and 1.

It is known that epilepsy can be caused by macroscopic changes in the network structure, such as brain lesions caused, e.g., by stroke~\cite{KIL90}. Furthermore, in epilepsy surgery, the brain's network structure is purposely changed to treat certain types of epilepsy~\cite{GOO16, SIN16a, OLM19}. Therefore, future perspectives of our work might be directed towards the question whether potential differences in the network structure of the brains of people with epilepsy compared to the ones of healthy subjects might perhaps exist.

\begin{acknowledgments}
This work was supported by the Deutsche Forschungsgemeinschaft (DFG, German Research Foundation) - Project Nos. 163436311 - SFB 910, 440145547, 411803875 and 308748074. ES acknowledges the Bernstein Center for Computational Neuroscience Berlin.
KL acknowledges support from the Deutsche Forschungsgemeinschaft (Grant No: LE 660/7-1). JH acknowledges the project Nr. LO1611 with financial support from the MEYS under the NPU I program, and the Czech Health Research Council Project No. NV17-28427A.
\end{acknowledgments}

\section*{Data Availability Statement}
The simulation data that support the findings of this study are available within the article. The EEG data that support the findings of this study are available from the corresponding author upon reasonable request. The EEG data are not publicly available due to privacy restrictions.

\section*{Appendix: Data acquisition for the empirical network}
\label{sec:Appendix}
The anatomical network of the cortex and subcortex was measured using Diffusion Tensor Imaging (DTI). The \textit{Functional Magnetic Resonance Imaging of the Brain (FMRIB) Software Library} (www.fmrib.ox.ac.uk/fsl/) was employed to apply probabilistic tractography to the data, enabling segregation of the brain into 90 areas according to the Automated Anatomical Labeling (AAL) atlas~\cite{TZO02}.  The anatomical names of the brain areas for each index $k$ are given in Tab.~SII of the Supplementary Material. The connecting white-matter fibers between the areas, which correspond to links in our network, were estimated by measurement of the preferred diffusion directions: For each voxel, through probabilistic tractography, a set of $n_s=5000$ streamlines was obtained which are hypothesized to correlate with the white-matter tracts. The proportion of streamlines connecting from area $j$ to all other areas $k$ is given by the probability coefficient $P_{jk}$ from which the adjacency matrix $A_{kj}$ is constructed. To eliminate individual variation, the matrices of 20 subjects (mean age 33 years, standard deviation 5.7 years, 10 females, 2 left-handed) are averaged, giving rise to the topology of Fig.~\ref{DTI}(a),(b). The pipeline for processing the DTI data has been adopted from a previous study of differences in connectivity patterns between healthy subjects and schizophrenia patients~\cite{CAB13b}. Obtaining such connectivity information using diffusion tractography is known to face a range of challenges~\cite{SCH19d,HLI12}.

\section*{Supplementary Material}
See the supplementary material for figures of simulations with other network parameters, additional EEG recorded seizure data, long-time simulations, and a table of cortical and subcortical regions according to the Automated Anatomical Labeling atlas (AAL).

%

\newpage

\begin{center}
	\textbf{\large }
\end{center}
\pagebreak

\onecolumngrid
\begin{center}
	\textbf{\large Supplementary Material for:\\ [.2cm]
		FitzHugh-Nagumo oscillators on complex networks mimic epileptic-seizure-related synchronization phenomena}\\[.2cm]
	Moritz Gerster$^{1}$, Rico Berner$^{1, 2}$, Jakub Sawicki$^{1}$, Anna Zakharova$^{1}$, Anton\'in \v{S}koch$^{3}$, Jaroslav Hlinka$^{3,4}$, Klaus Lehnertz$^{5,6,7}$, and Eckehard Sch\"oll$^{1,8}$\\[.1cm]
	{\itshape $^{1}$Institut f\"ur Theoretische Physik, Technische Universit\"at Berlin, Hardenbergstr.\,36, 10623 Berlin, Germany\\
		$^{2}$Institut f\"ur Mathematik, Technische Universit\"at Berlin, Stra\ss e des 17. Juni\,136, 10623 Berlin, Germany\\
		$^{3}$National Institute of Mental Health, Topolov\'{a} 748, 250 67 Klecany, Czech Republic \\
		$^{4}$Institute of Computer Science of the Czech Academy of Sciences, Pod Vodarenskou vezi 2, 18207 Prague 8, Czech Republic\\
		$^{5}$Department of Epileptology, University of Bonn Medical Centre, Venusberg Campus 1, 53127 Bonn, Germany\\
		$^{6}$Helmholtz Institute for Radiation and Nuclear Physics, University of Bonn, Nussallee 14--16, 53115 Bonn, Germany\\
		$^{7}$Interdisciplinary Center for Complex Systems, University of Bonn, Br{\"u}hler Stra\ss{}e 7, 53175 Bonn, Germany\\
		$^{8}$Bernstein Center for Computational Neuroscience Berlin, Humboldt-Universität, 10115 Berlin, Germany}\\
\end{center}

\setcounter{equation}{0}
\setcounter{figure}{0}
\setcounter{table}{0}
\setcounter{page}{1}
\setcounter{secnumdepth}{1}
\renewcommand{\theequation}{S\arabic{equation}}
\renewcommand{\thefigure}{S\arabic{figure}}
\renewcommand{\thetable}{S\Roman{table}}
\renewcommand{\bibnumfmt}[1]{[S#1]}
\renewcommand{\citenumfont}[1]{#1}

\section*{Random Surrogate Networks}

\begin{figure}[h]
	\centering
	\includegraphics[width=8.4cm]{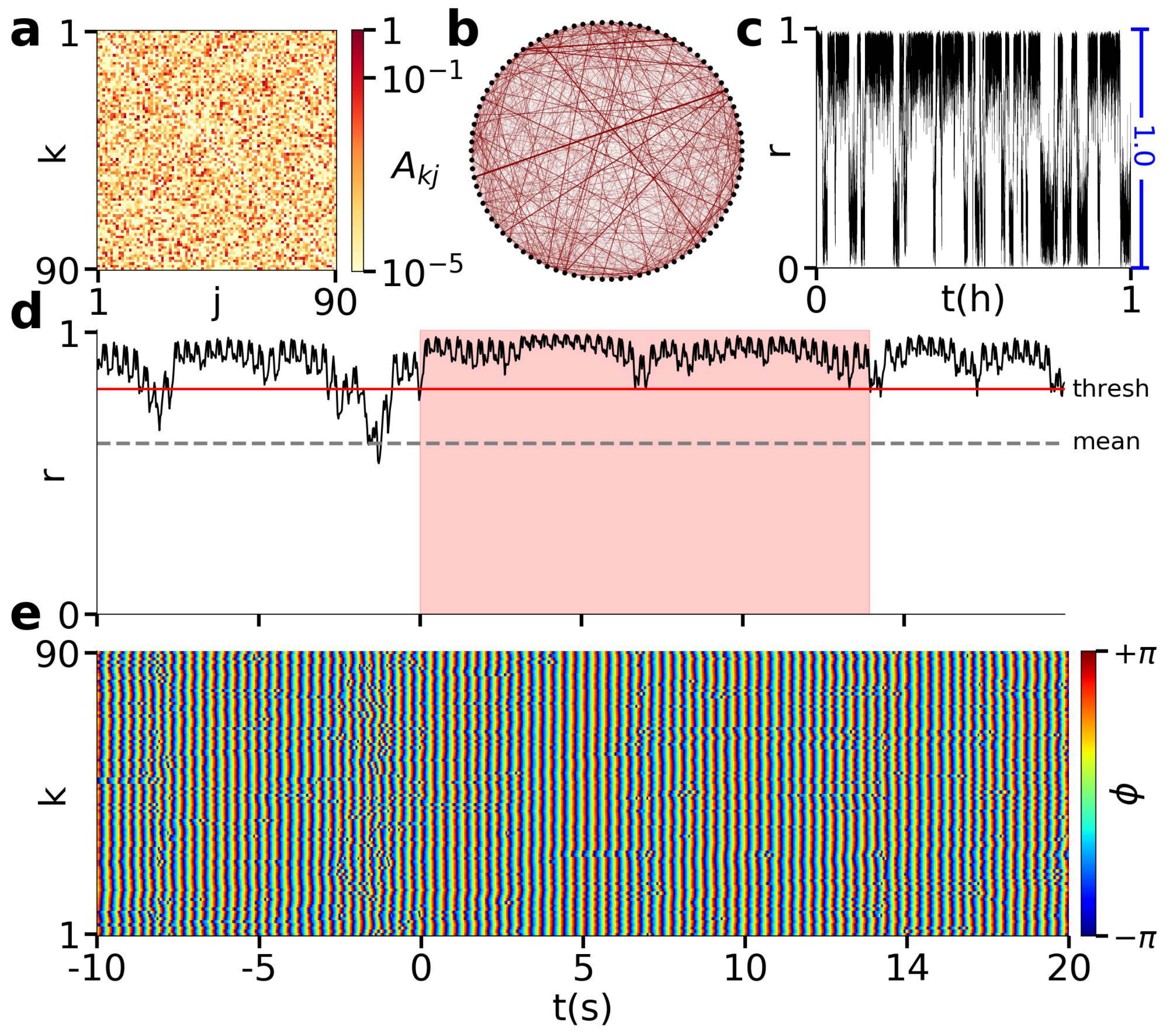}
	\caption{Same as Fig.~2 (random surrogate network) with a coupling strength of $\sigma=0.7$. Note that $\langle r \rangle = 0.60$ both for the random surrogate network and the empirical network. However, high synchrony $ r > 0.8$ makes up 47\% of the time in the random network and only  17\% of the time in the empirical network. Interestingly, the number of seizures seen on the random network (4.7 per hour) is not much larger than the number of seizures seen on the empirical network (4 per hour).} 
	\label{R07}
\end{figure}

The random surrogate network of Fig.~2 shows smaller synchronization values compared to the empirical network, even though it consists of the same set of weights. Therefore, no epileptic-seizure-related synchronization phenomena (seizures) were found. If the coupling strength is increased to $\sigma=0.7$, the average order parameter $\langle r \rangle$ becomes 0.6, which corresponds to the average order parameter of the empirical network, see Fig.~\ref{R07}. Notably, despite high synchrony $r>0.8$ during 47\% of the time, only 4.7 seizures per hour are found (compared to 17\% high synchrony and 4 seizures for the empirical network). In addition, the time series in Fig.~\ref{R07}(c) shows time intervals of both extreme synchronization (shown in (d) and (e)) and extreme desynchronization. Medium synchronization, at about $r\approx 0.5$, which we consider as corresponding to a healthy state, is rarely apparent in the time series. This supports our claim that the clustering coefficient in brain networks must not be too small. 

\section*{Small-world networks}

For Figs.~5 to 8, we repeated all simulations 10 times to average out dynamic differences that might be caused by different random initial conditions. Tab.~\ref{tabSW2} summarizes the mean global Kuramoto order parameter $\langle r \rangle$ for each simulation. Note that for $p=0.232$ and $p=1$, the dynamics is independent of the initial conditions. For $p=0$ and $p=0.006$, on the other hand, the initial conditions determine the entire time series.

\begin{table}[h!]
	\begin{tabular}{|l|l|l|l|l|l|l|l|l|l|l|l|}
		\hline
		Simulation &1&2&3&4&5&6&7&8&9&10 \\ \hline\hline
		$p=0$&0.99& 0.99& 0.99& 0.99& 0.99& 0.99& 0.02& 0.99& 0.01& 0.99 \\ \hline
		$p=0.006$ &0.13& 0.46& 0.03& 0.76& 0.75& 0.21& 0.34& 0.46& 0.75& 0.75 \\ \hline
		$p=0.232$ &0.46&0.47& 0.46& 0.46& 0.47& 0.47& 0.47& 0.47& 0.47& 0.46 \\ \hline
		$p=1$ &0.73& 0.72& 0.73& 0.73& 0.73& 0.73& 0.73& 0.73& 0.73& 0.73 \\ \hline
	\end{tabular}
	\caption{For the small-world networks, we repeat the simulation 10 times with different random initial conditions for each rewiring probability $p$. The table shows the global Kuramoto order parameter $\langle r \rangle$, averaged over the simulation time. Each simulation has a duration of approximately 3 hours in real time.}
	\label{tabSW2}
\end{table}

\section*{Comparison with EEG-recorded absence seizures}

\begin{figure}
	\includegraphics[width=8.4cm]{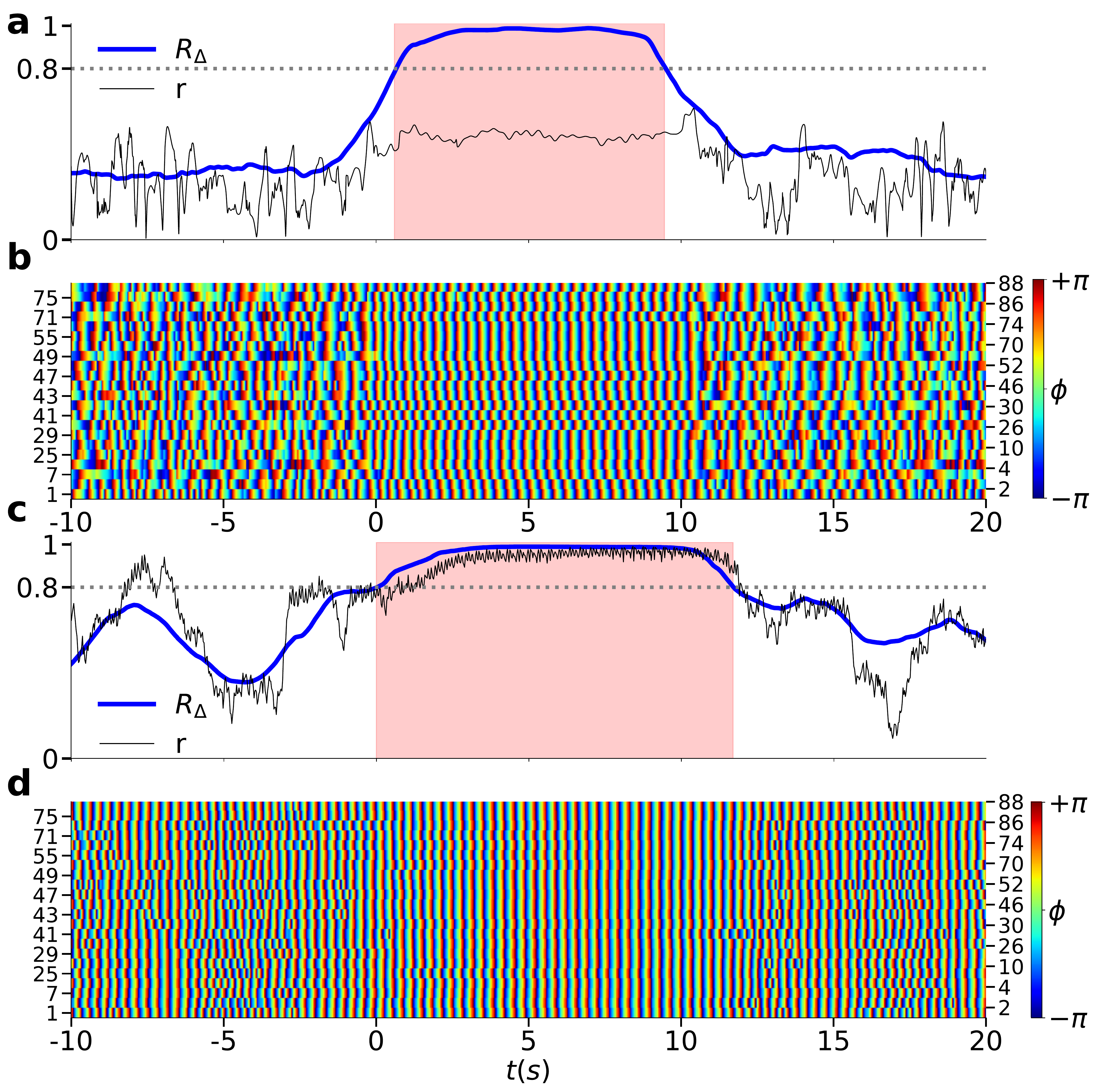}
	\caption{Comparison of EEG-recorded data (a),(b) with simulated data (c),(d): Same as Fig.~9f,h,e,g with an increased averaging window of $T=\unit[3]{s}$.}
	\label{Exp3s}
\end{figure}

\begin{figure}
	\includegraphics[width=8.4cm]{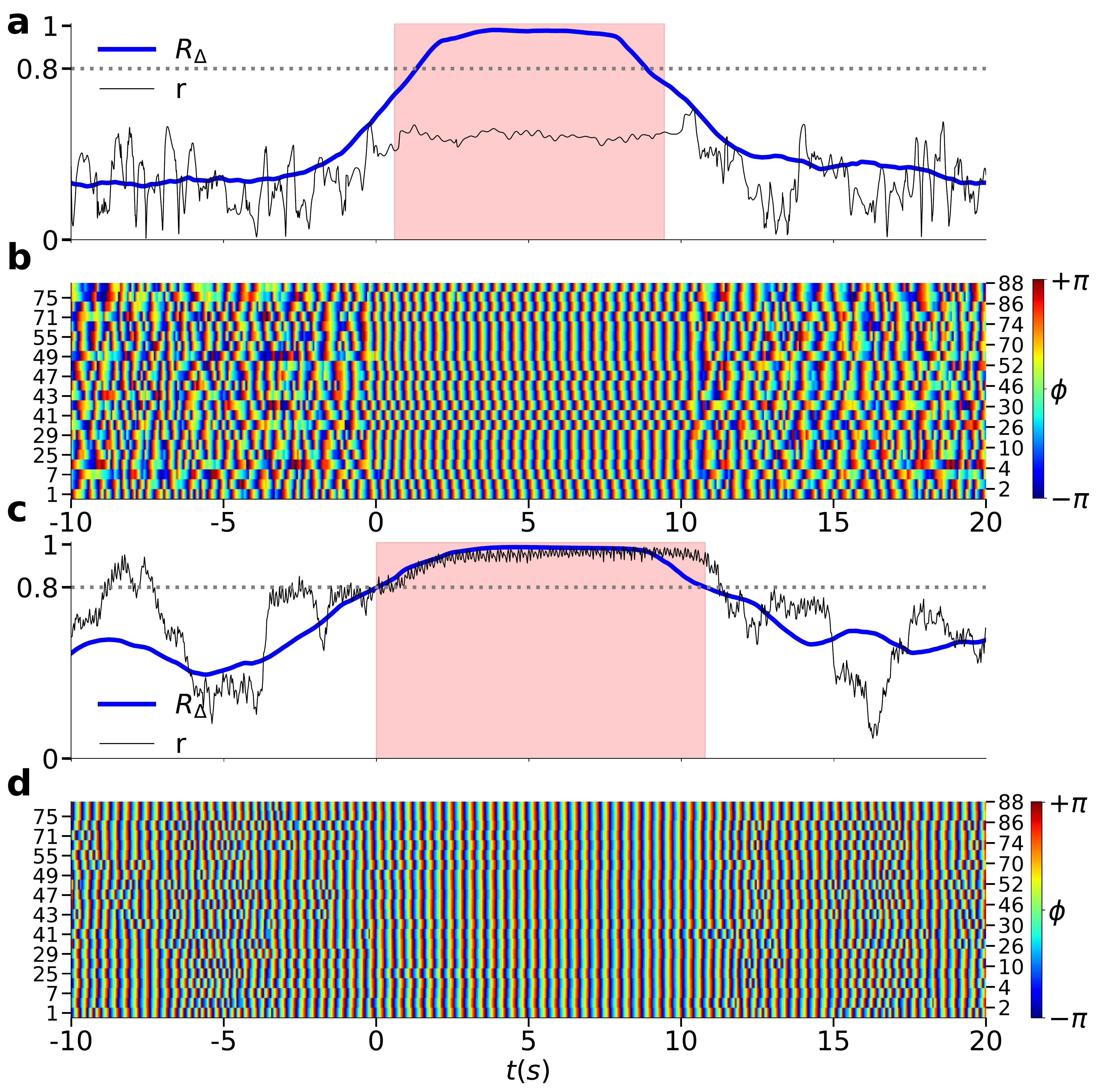}
	\caption{Same as Fig.~9 with an increased averaging window of $T=\unit[5]{s}$.}
	\label{Exp5s}
\end{figure}

\begin{figure*}
	\centering
	\includegraphics[width=8.4cm]{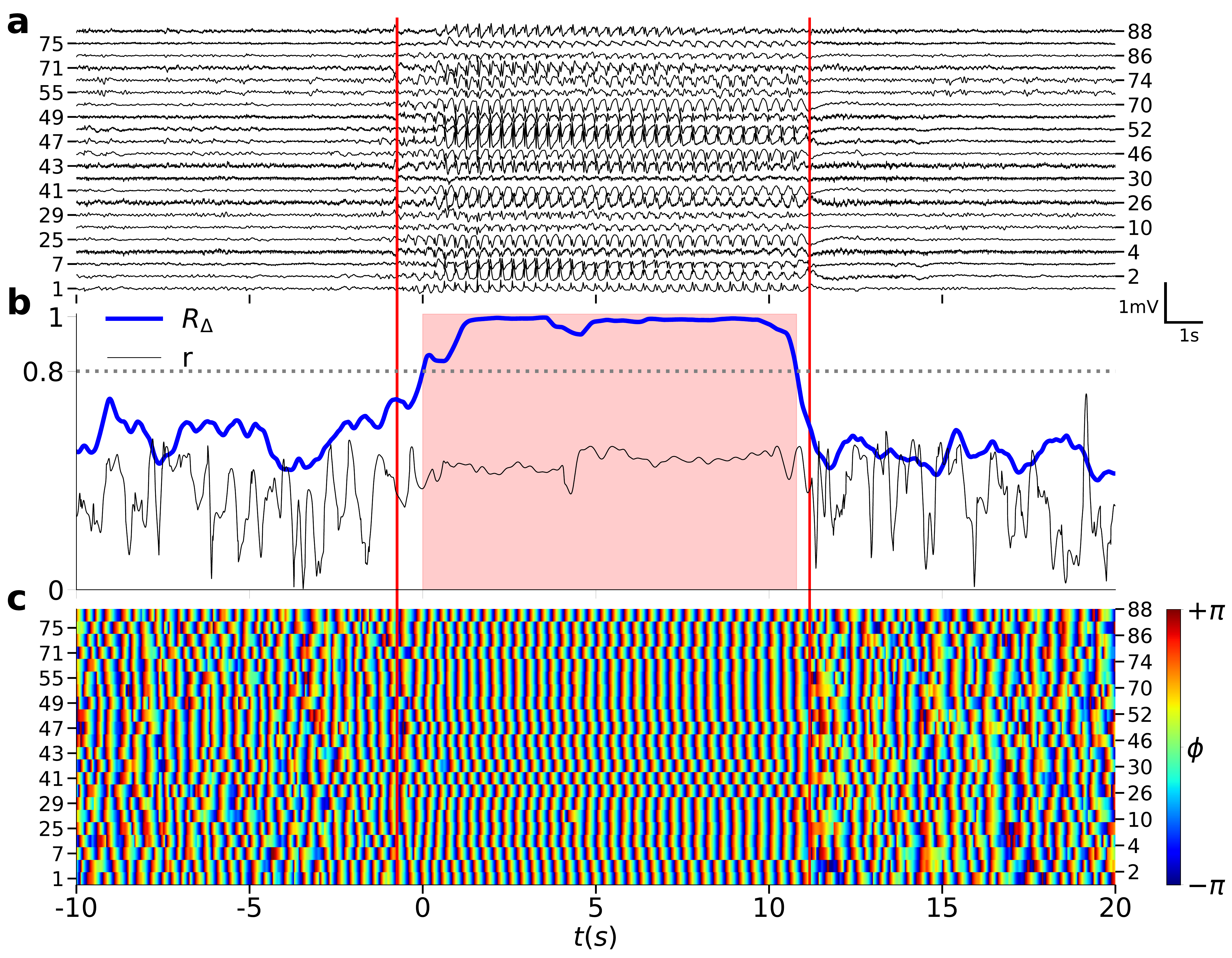}\caption{EEG recorded data: Same as Fig.~9 (d), (f), (h), but for another seizure of the same subject with epilepsy.}\label{Seiz1}
\end{figure*}

\begin{figure*}
	\centering
	\includegraphics[width=8.4cm]{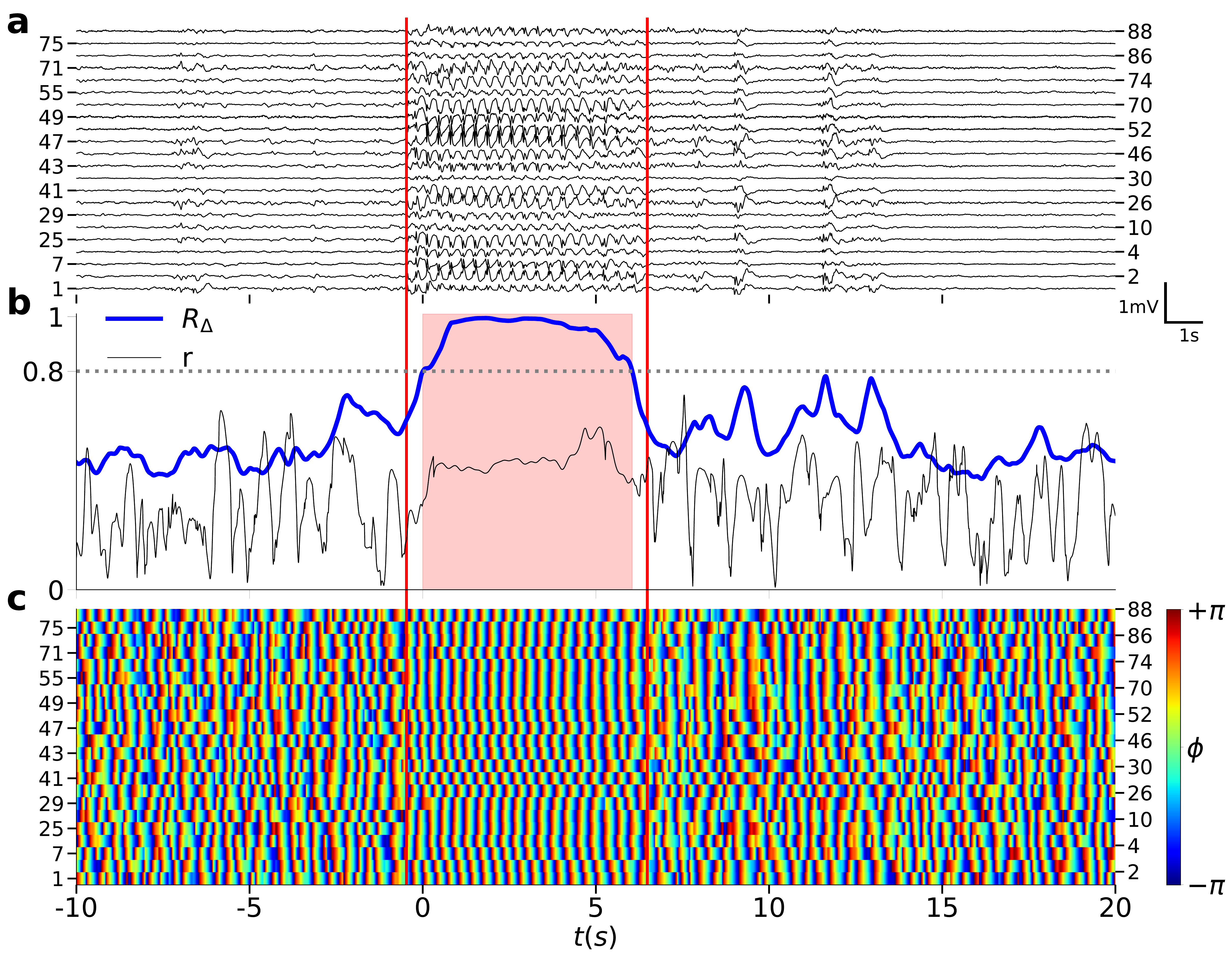}
	\caption{EEG recorded data: Same as Fig.~9 (d), (f), (h), but for another seizure of the same subject.}\label{Seiz2}
\end{figure*}

\begin{figure*}
	\centering
	\includegraphics[width=8.4cm]{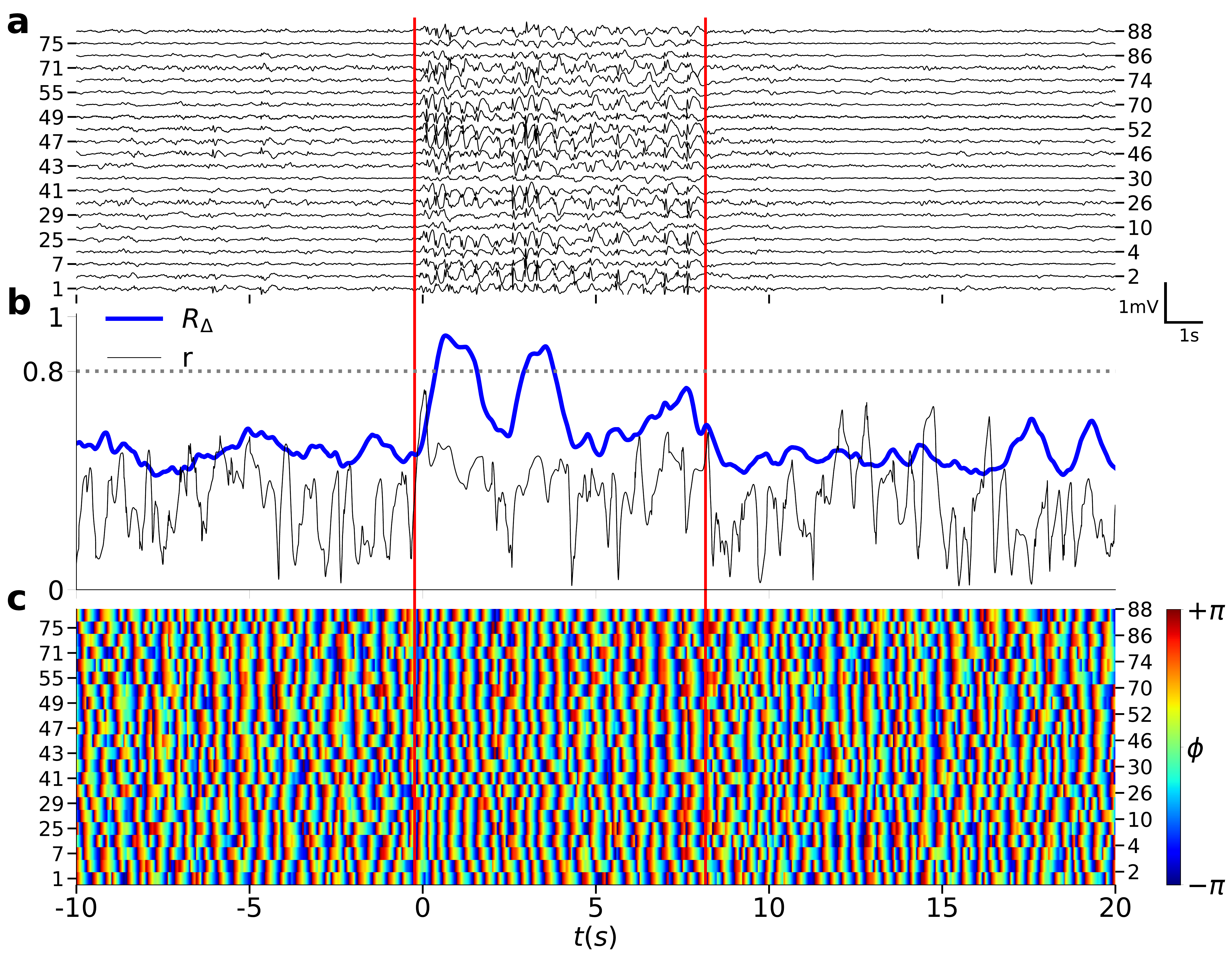}\caption{EEG recorded data: Same as Fig.~9 (d), (f), (h), but for another seizure of the same subject. For this seizure, our threshold definition 
		did not identify the seizure correctly.}\label{Seiz3}
\end{figure*}

\begin{figure*}
	\centering
	\includegraphics[width=8.4cm]{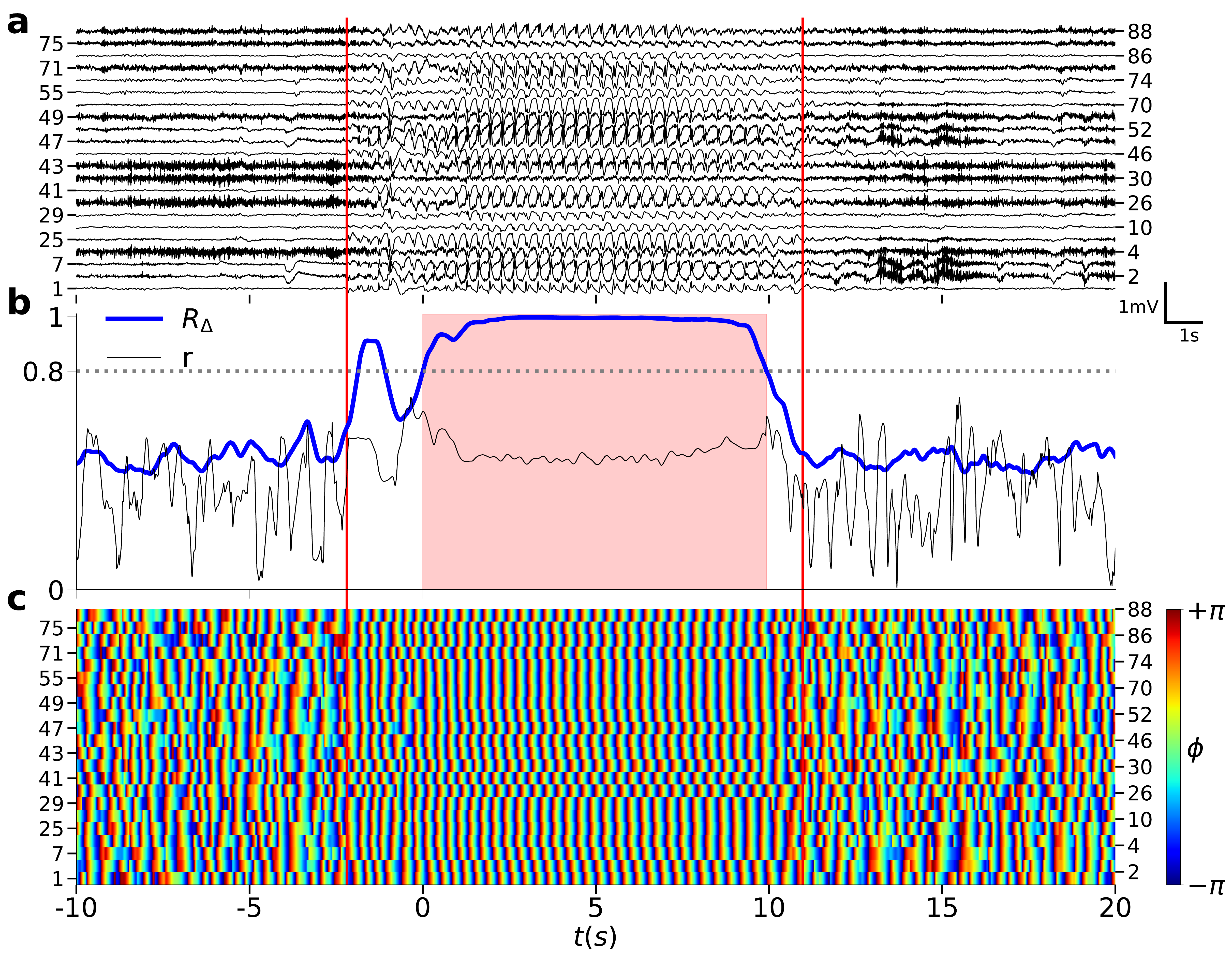}
	\caption{EEG recorded data: Same as Fig.~9 (d), (f), (h), but for another seizure of the same subject.}\label{Seiz4}
\end{figure*}

\begin{figure*}
	\centering
	\includegraphics[width=\the \textwidth]{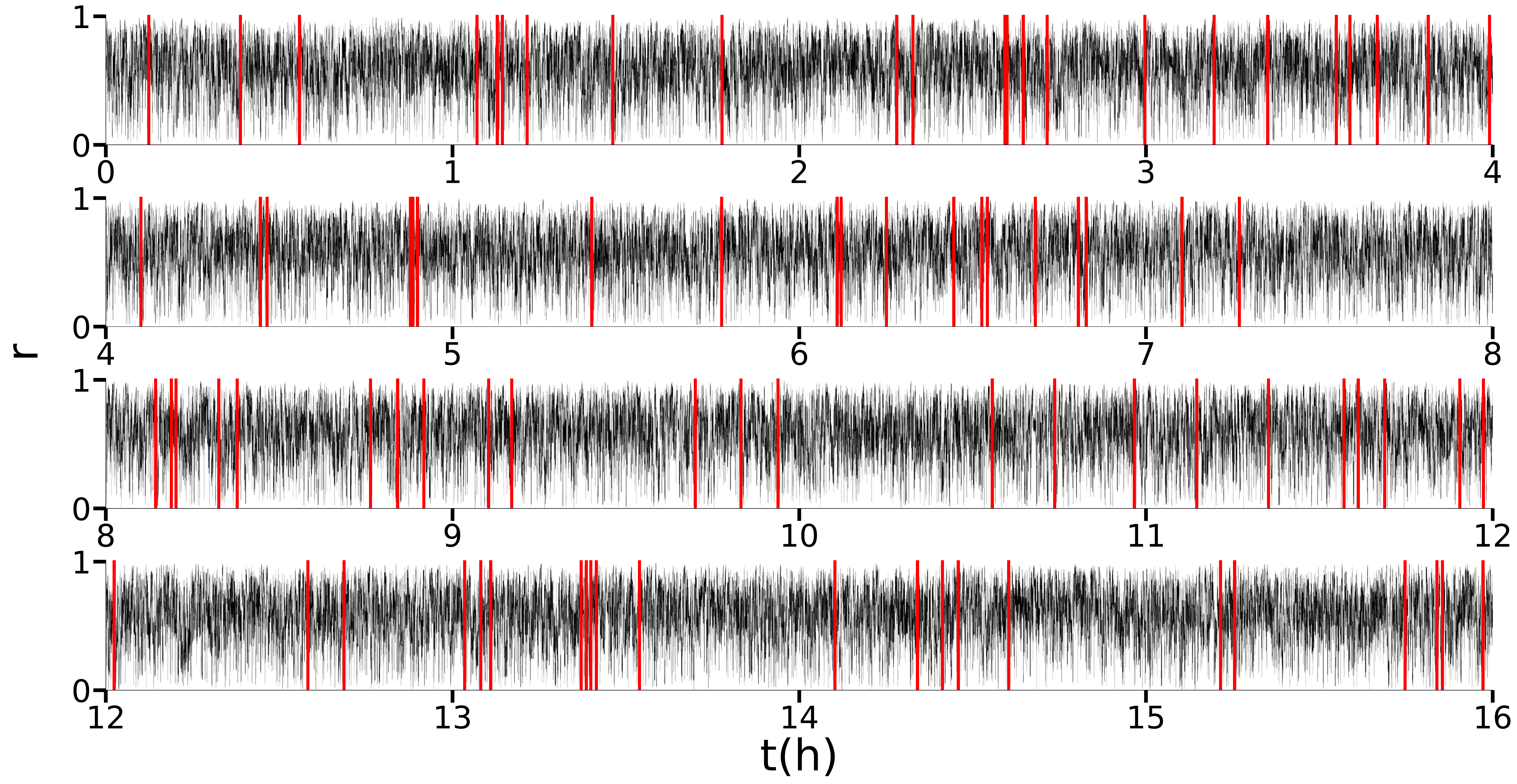}
	\caption{Global Kuramoto order parameter $r(t)$ for a very long simulation of 16 hours with the empirical connectivity. Vertical red lines indicate seizures.}\label{long_DTI}
\end{figure*}

\begin{figure*}
	\centering
	\includegraphics[width=12cm]{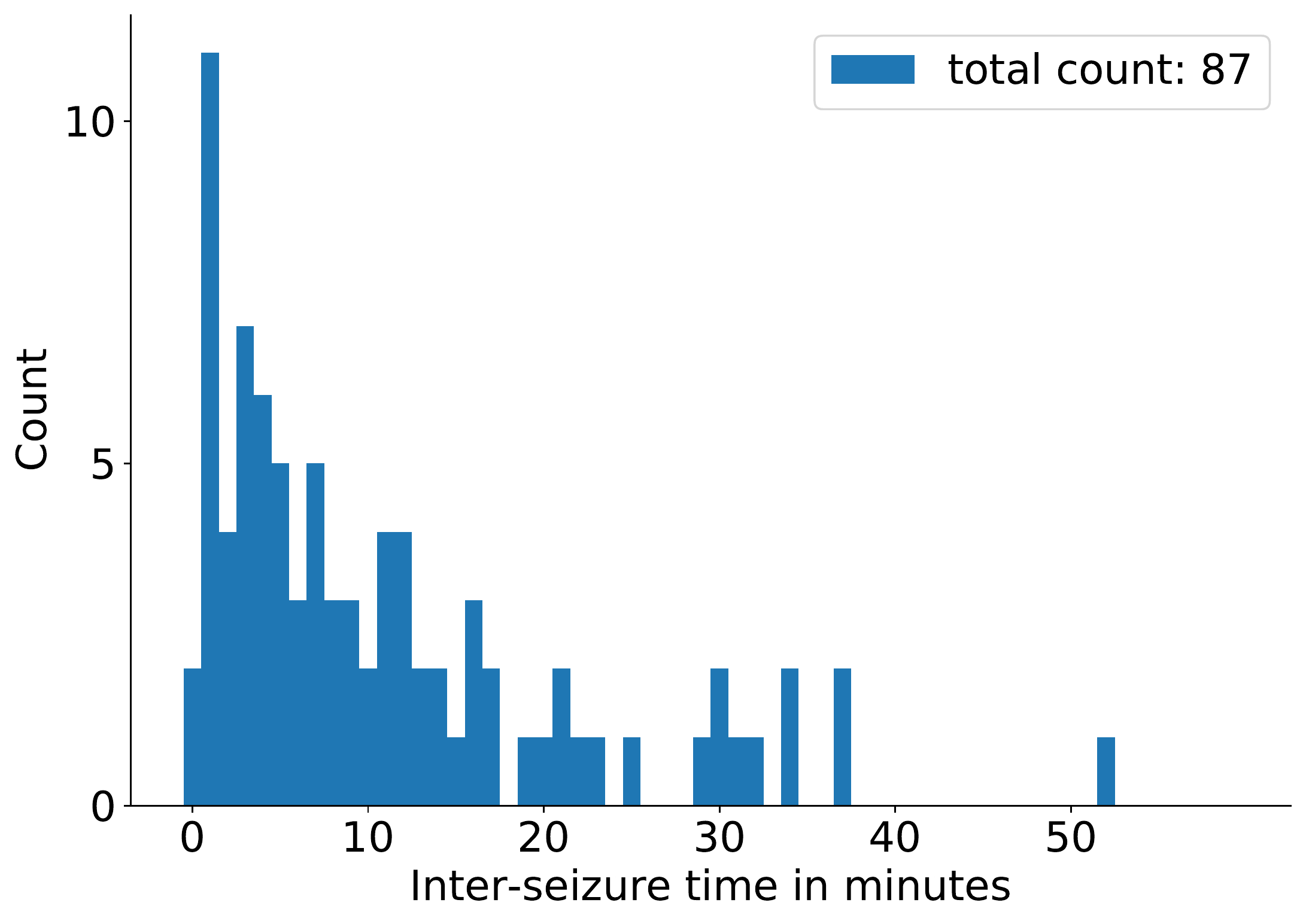}
	\caption{Histogram of the inter-seizure intervals of Fig.~\ref{long_DTI}. Bin size = 1 minute.}\label{hist_DTI}
\end{figure*}

\begin{figure*}
	\centering
	\includegraphics[width=\the \textwidth]{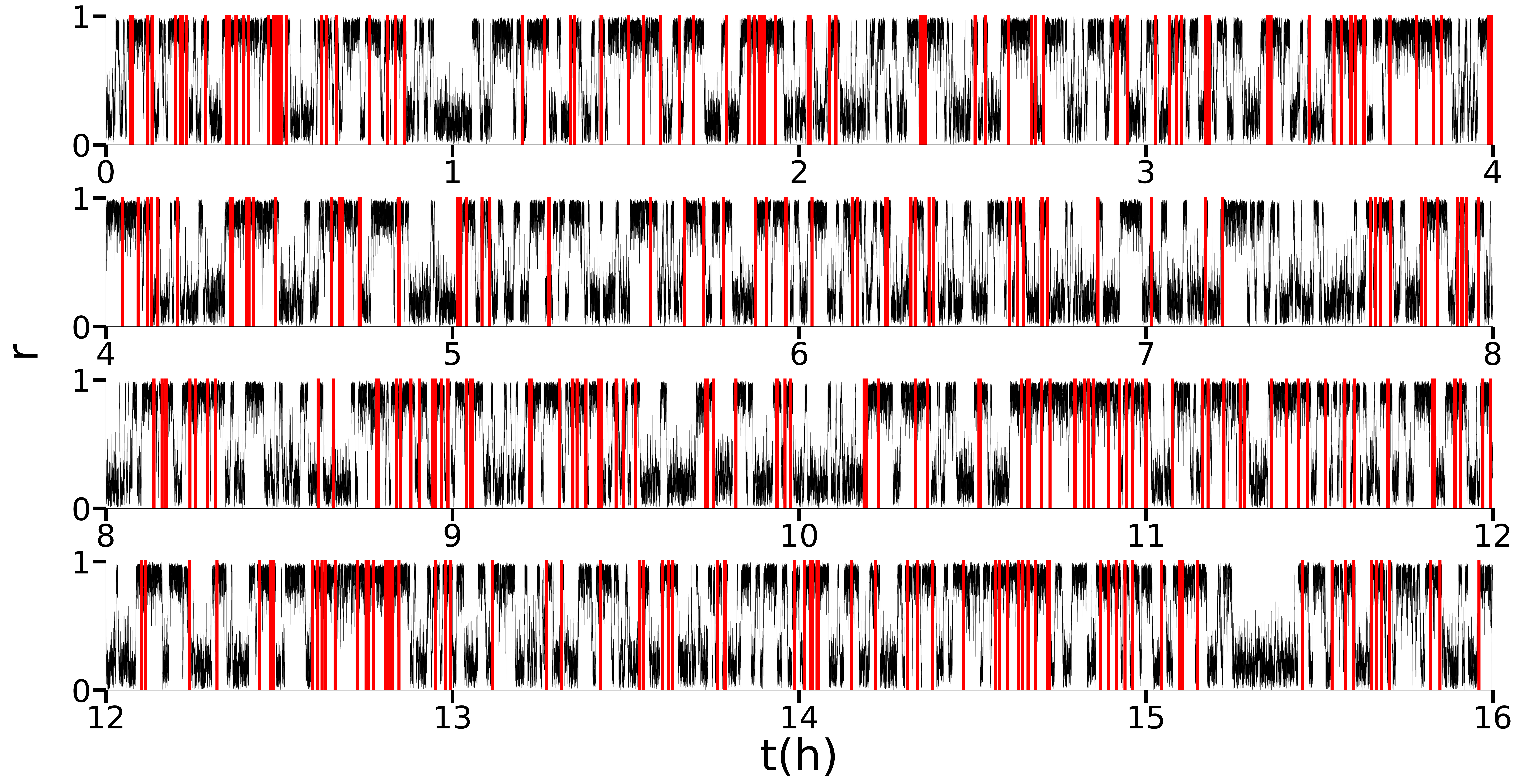}
	\caption{Global Kuramoto order parameter $r(t)$ for a very long simulation of 16 hours with the random surrogate connectivity. Vertical red lines indicate seizures.}\label{long_Rand}
\end{figure*}

\begin{figure*}
	\centering
	\includegraphics[width=12cm]{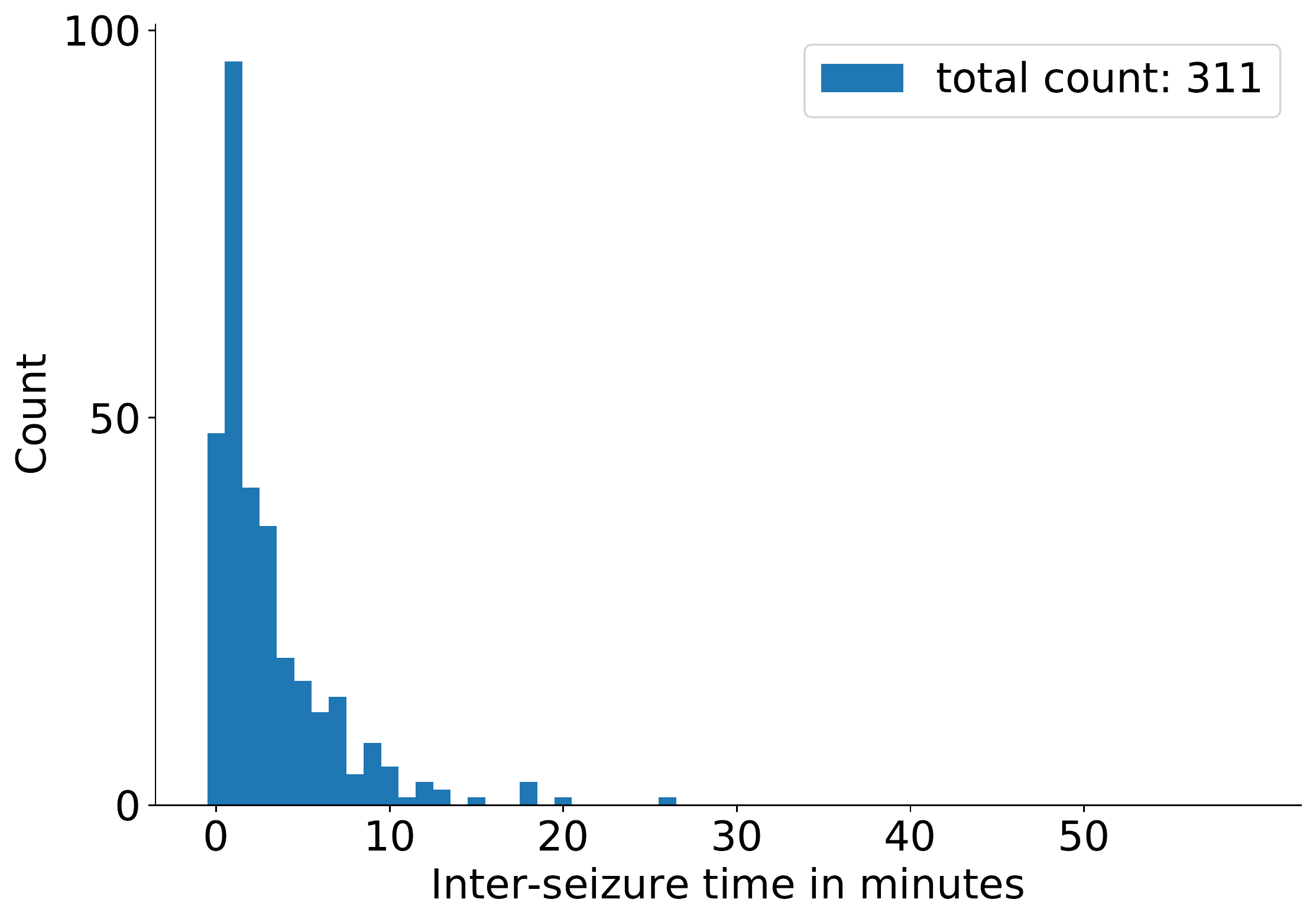}
	\caption{Histogram of the inter-seizure intervals of Fig.~\ref{long_Rand}. Bin size = 1 minute.}\label{hist_Rand}
\end{figure*}

For the calculation of the global phase coherence $R_{\Delta}$, the average phase-locking is calculated over a time window $T$. For visualizing $R_{\Delta}$, it is of great importance whether the averaging window $T$ is chosen rather small or rather large. In Figs.~\ref{Exp3s} and \ref{Exp5s}, we re-plot data from Fig.~9 from the main text for different values of the averaging window $T$. Note that for longer averaging windows, the simulation data reproduces the data from the EEG recording better. However, due to the lower time resolution, the precise onset and end of the seizure cannot be resolved. Figs.~\ref{Seiz1}-\ref{Seiz4} show additional EEG recorded data for other seizures of the same subject with epilepsy. Figs.~\ref{long_DTI} and \ref{long_Rand} show very long simulated time series for the empirical connectivity and the random surrogate connectivity, respectively, and the corresponding histograms of the distribution of inter-seizure intervals (Figs.~\ref{hist_DTI}, \ref{hist_Rand}). Table \ref{tab:appA_aal} lists the cortical and subcortical brain regions according to the Automated Anatomical Labeling atlas (AAL).


\begin{thebibliography}{91}%
\makeatletter
\providecommand \@ifxundefined [1]{%
 \@ifx{#1\undefined}
}%
\providecommand \@ifnum [1]{%
 \ifnum #1\expandafter \@firstoftwo
 \else \expandafter \@secondoftwo
 \fi
}%
\providecommand \@ifx [1]{%
 \ifx #1\expandafter \@firstoftwo
 \else \expandafter \@secondoftwo
 \fi
}%
\providecommand \natexlab [1]{#1}%
\providecommand \enquote  [1]{``#1''}%
\providecommand \bibnamefont  [1]{#1}%
\providecommand \bibfnamefont [1]{#1}%
\providecommand \citenamefont [1]{#1}%
\providecommand \href@noop [0]{\@secondoftwo}%
\providecommand \href [0]{\begingroup \@sanitize@url \@href}%
\providecommand \@href[1]{\@@startlink{#1}\@@href}%
\providecommand \@@href[1]{\endgroup#1\@@endlink}%
\providecommand \@sanitize@url [0]{\catcode `\\12\catcode `\$12\catcode
  `\&12\catcode `\#12\catcode `\^12\catcode `\_12\catcode `\%12\relax}%
\providecommand \@@startlink[1]{}%
\providecommand \@@endlink[0]{}%
\providecommand \url  [0]{\begingroup\@sanitize@url \@url }%
\providecommand \@url [1]{\endgroup\@href {#1}{\urlprefix }}%
\providecommand \urlprefix  [0]{URL }%
\providecommand \Eprint [0]{\href }%
\providecommand \doibase [0]{http://dx.doi.org/}%
\providecommand \selectlanguage [0]{\@gobble}%
\providecommand \bibinfo  [0]{\@secondoftwo}%
\providecommand \bibfield  [0]{\@secondoftwo}%
\providecommand \translation [1]{[#1]}%
\providecommand \BibitemOpen [0]{}%
\providecommand \bibitemStop [0]{}%
\providecommand \bibitemNoStop [0]{.\EOS\space}%
\providecommand \EOS [0]{\spacefactor3000\relax}%
\providecommand \BibitemShut  [1]{\csname bibitem#1\endcsname}%
\let\auto@bib@innerbib\@empty
\bibitem [{\citenamefont {Pikovsky}\ \emph {et~al.}(2001)\citenamefont
  {Pikovsky}, \citenamefont {Rosenblum},\ and\ \citenamefont {Kurths}}]{PIK01}%
  \BibitemOpen
  \bibfield  {author} {\bibinfo {author} {\bibfnamefont {A.}~\bibnamefont
  {Pikovsky}}, \bibinfo {author} {\bibfnamefont {M.}~\bibnamefont {Rosenblum}},
  \ and\ \bibinfo {author} {\bibfnamefont {J.}~\bibnamefont {Kurths}},\
  }\href@noop {} {\emph {\bibinfo {title} {Synchronization: a universal concept
  in nonlinear sciences}}}\ (\bibinfo  {publisher} {Cambridge University
  Press},\ \bibinfo {address} {Cambridge},\ \bibinfo {year} {2001})\BibitemShut
  {NoStop}%
\bibitem [{\citenamefont {Boccaletti}\ \emph {et~al.}(2018)\citenamefont
  {Boccaletti}, \citenamefont {Pisarchik}, \citenamefont {del Genio},\ and\
  \citenamefont {Amann}}]{BOC18}%
  \BibitemOpen
  \bibfield  {author} {\bibinfo {author} {\bibfnamefont {S.}~\bibnamefont
  {Boccaletti}}, \bibinfo {author} {\bibfnamefont {A.~N.}\ \bibnamefont
  {Pisarchik}}, \bibinfo {author} {\bibfnamefont {C.~I.}\ \bibnamefont {del
  Genio}}, \ and\ \bibinfo {author} {\bibfnamefont {A.}~\bibnamefont {Amann}},\
  }\href@noop {} {\emph {\bibinfo {title} {Synchronization: From Coupled
  Systems to Complex Networks}}}\ (\bibinfo  {publisher} {Cambridge University
  Press},\ \bibinfo {address} {Cambridge},\ \bibinfo {year} {2018})\BibitemShut
  {NoStop}%
\bibitem [{\citenamefont {Singer}(2018)}]{SIN18}%
  \BibitemOpen
  \bibfield  {author} {\bibinfo {author} {\bibfnamefont {W.}~\bibnamefont
  {Singer}},\ }\href@noop {} {\bibfield  {journal} {\bibinfo  {journal} {Eur.
  J. Neurosci.}\ }\textbf {\bibinfo {volume} {48}},\ \bibinfo {pages} {2389}
  (\bibinfo {year} {2018})}\BibitemShut {NoStop}%
\bibitem [{\citenamefont {Lehnertz}\ \emph {et~al.}(2009)\citenamefont
  {Lehnertz}, \citenamefont {Bialonski}, \citenamefont {Horstmann},
  \citenamefont {Krug}, \citenamefont {Rothkegel}, \citenamefont {Staniek},\
  and\ \citenamefont {Wagner}}]{LEH09}%
  \BibitemOpen
  \bibfield  {author} {\bibinfo {author} {\bibfnamefont {K.}~\bibnamefont
  {Lehnertz}}, \bibinfo {author} {\bibfnamefont {S.}~\bibnamefont {Bialonski}},
  \bibinfo {author} {\bibfnamefont {M.-T.}\ \bibnamefont {Horstmann}}, \bibinfo
  {author} {\bibfnamefont {D.}~\bibnamefont {Krug}}, \bibinfo {author}
  {\bibfnamefont {A.}~\bibnamefont {Rothkegel}}, \bibinfo {author}
  {\bibfnamefont {M.}~\bibnamefont {Staniek}}, \ and\ \bibinfo {author}
  {\bibfnamefont {T.}~\bibnamefont {Wagner}},\ }\href@noop {} {\bibfield
  {journal} {\bibinfo  {journal} {J. Neurosci. Methods}\ }\textbf {\bibinfo
  {volume} {183}},\ \bibinfo {pages} {42} (\bibinfo {year} {2009})}\BibitemShut
  {NoStop}%
\bibitem [{\citenamefont {Jiruska}\ \emph {et~al.}(2013)\citenamefont
  {Jiruska}, \citenamefont {de~Curtis}, \citenamefont {Jefferys}, \citenamefont
  {Schevon}, \citenamefont {Schiff},\ and\ \citenamefont {Schindler}}]{JIR13}%
  \BibitemOpen
  \bibfield  {author} {\bibinfo {author} {\bibfnamefont {P.}~\bibnamefont
  {Jiruska}}, \bibinfo {author} {\bibfnamefont {M.}~\bibnamefont {de~Curtis}},
  \bibinfo {author} {\bibfnamefont {J.~G.~R.}\ \bibnamefont {Jefferys}},
  \bibinfo {author} {\bibfnamefont {C.~A.}\ \bibnamefont {Schevon}}, \bibinfo
  {author} {\bibfnamefont {S.~J.}\ \bibnamefont {Schiff}}, \ and\ \bibinfo
  {author} {\bibfnamefont {K.}~\bibnamefont {Schindler}},\ }\href@noop {}
  {\bibfield  {journal} {\bibinfo  {journal} {J. Physiol.}\ }\textbf {\bibinfo
  {volume} {591.4}},\ \bibinfo {pages} {787} (\bibinfo {year}
  {2013})}\BibitemShut {NoStop}%
\bibitem [{\citenamefont {Jirsa}\ \emph {et~al.}(2014)\citenamefont {Jirsa},
  \citenamefont {Stacey}, \citenamefont {Quilichini}, \citenamefont {Ivanov},\
  and\ \citenamefont {Bernard}}]{JIR14}%
  \BibitemOpen
  \bibfield  {author} {\bibinfo {author} {\bibfnamefont {V.~K.}\ \bibnamefont
  {Jirsa}}, \bibinfo {author} {\bibfnamefont {W.~C.}\ \bibnamefont {Stacey}},
  \bibinfo {author} {\bibfnamefont {P.~P.}\ \bibnamefont {Quilichini}},
  \bibinfo {author} {\bibfnamefont {A.~I.}\ \bibnamefont {Ivanov}}, \ and\
  \bibinfo {author} {\bibfnamefont {C.}~\bibnamefont {Bernard}},\ }\href@noop
  {} {\bibfield  {journal} {\bibinfo  {journal} {Brain}\ }\textbf {\bibinfo
  {volume} {137}},\ \bibinfo {pages} {2210} (\bibinfo {year}
  {2014})}\BibitemShut {NoStop}%
\bibitem [{\citenamefont {Lehnertz}\ \emph {et~al.}(2014)\citenamefont
  {Lehnertz}, \citenamefont {Ansmann}, \citenamefont {Bialonski}, \citenamefont
  {Dickten}, \citenamefont {Geier},\ and\ \citenamefont {Porz}}]{LEH14a}%
  \BibitemOpen
  \bibfield  {author} {\bibinfo {author} {\bibfnamefont {K.}~\bibnamefont
  {Lehnertz}}, \bibinfo {author} {\bibfnamefont {G.}~\bibnamefont {Ansmann}},
  \bibinfo {author} {\bibfnamefont {S.}~\bibnamefont {Bialonski}}, \bibinfo
  {author} {\bibfnamefont {H.}~\bibnamefont {Dickten}}, \bibinfo {author}
  {\bibfnamefont {C.}~\bibnamefont {Geier}}, \ and\ \bibinfo {author}
  {\bibfnamefont {S.}~\bibnamefont {Porz}},\ }\href@noop {} {\bibfield
  {journal} {\bibinfo  {journal} {Physica D}\ }\textbf {\bibinfo {volume}
  {267}},\ \bibinfo {pages} {7} (\bibinfo {year} {2014})}\BibitemShut {NoStop}%
\bibitem [{\citenamefont {Bassett}\ \emph {et~al.}(2018)\citenamefont
  {Bassett}, \citenamefont {Zurn},\ and\ \citenamefont {Gold}}]{BAS18}%
  \BibitemOpen
  \bibfield  {author} {\bibinfo {author} {\bibfnamefont {D.~S.}\ \bibnamefont
  {Bassett}}, \bibinfo {author} {\bibfnamefont {P.}~\bibnamefont {Zurn}}, \
  and\ \bibinfo {author} {\bibfnamefont {J.~I.}\ \bibnamefont {Gold}},\ }\href
  {\doibase 10.1038/s41583-018-0038-8} {\bibfield  {journal} {\bibinfo
  {journal} {Nat. Rev. Neurosci.}\ }\textbf {\bibinfo {volume} {19}},\ \bibinfo
  {pages} {566} (\bibinfo {year} {2018})}\BibitemShut {NoStop}%
\bibitem [{\citenamefont {Ngugi}\ \emph {et~al.}(2011)\citenamefont {Ngugi},
  \citenamefont {Kariuki}, \citenamefont {Bottomley}, \citenamefont
  {Kleinschmidt}, \citenamefont {Sander},\ and\ \citenamefont
  {Newton}}]{NGU11a}%
  \BibitemOpen
  \bibfield  {author} {\bibinfo {author} {\bibfnamefont {A.~K.}\ \bibnamefont
  {Ngugi}}, \bibinfo {author} {\bibfnamefont {S.}~\bibnamefont {Kariuki}},
  \bibinfo {author} {\bibfnamefont {C.}~\bibnamefont {Bottomley}}, \bibinfo
  {author} {\bibfnamefont {I.}~\bibnamefont {Kleinschmidt}}, \bibinfo {author}
  {\bibfnamefont {J.~W.}\ \bibnamefont {Sander}}, \ and\ \bibinfo {author}
  {\bibfnamefont {C.~R.}\ \bibnamefont {Newton}},\ }\href@noop {} {\bibfield
  {journal} {\bibinfo  {journal} {Neurology}\ }\textbf {\bibinfo {volume}
  {77}},\ \bibinfo {pages} {1005} (\bibinfo {year} {2011})}\BibitemShut
  {NoStop}%
\bibitem [{\citenamefont {Fisher}\ \emph {et~al.}(2005)\citenamefont {Fisher},
  \citenamefont {van Emde~Boas}, \citenamefont {Blume}, \citenamefont {Elger},
  \citenamefont {Genton}, \citenamefont {Lee},\ and\ \citenamefont {{Engel
  Jr}}}]{FIS05}%
  \BibitemOpen
  \bibfield  {author} {\bibinfo {author} {\bibfnamefont {R.~S.}\ \bibnamefont
  {Fisher}}, \bibinfo {author} {\bibfnamefont {W.}~\bibnamefont {van
  Emde~Boas}}, \bibinfo {author} {\bibfnamefont {W.}~\bibnamefont {Blume}},
  \bibinfo {author} {\bibfnamefont {C.~E.}\ \bibnamefont {Elger}}, \bibinfo
  {author} {\bibfnamefont {P.}~\bibnamefont {Genton}}, \bibinfo {author}
  {\bibfnamefont {P.}~\bibnamefont {Lee}}, \ and\ \bibinfo {author}
  {\bibfnamefont {J.}~\bibnamefont {{Engel Jr}}},\ }\href {\doibase
  10.1111/j.0013-9580.2005.66104.x} {\bibfield  {journal} {\bibinfo  {journal}
  {Epilepsia}\ }\textbf {\bibinfo {volume} {46}},\ \bibinfo {pages} {470}
  (\bibinfo {year} {2005})}\BibitemShut {NoStop}%
\bibitem [{\citenamefont {Gastaut}(1970)}]{GAS70}%
  \BibitemOpen
  \bibfield  {author} {\bibinfo {author} {\bibfnamefont {H.}~\bibnamefont
  {Gastaut}},\ }\href@noop {} {\bibfield  {journal} {\bibinfo  {journal}
  {Epilepsia}\ }\textbf {\bibinfo {volume} {11}},\ \bibinfo {pages} {102}
  (\bibinfo {year} {1970})}\BibitemShut {NoStop}%
\bibitem [{\citenamefont {Spencer}(2002)}]{SPE02}%
  \BibitemOpen
  \bibfield  {author} {\bibinfo {author} {\bibfnamefont {S.~S.}\ \bibnamefont
  {Spencer}},\ }\href {\doibase 10.1046/j.1528-1157.2002.26901.x} {\bibfield
  {journal} {\bibinfo  {journal} {Epilepsia}\ }\textbf {\bibinfo {volume}
  {43}},\ \bibinfo {pages} {219} (\bibinfo {year} {2002})}\BibitemShut
  {NoStop}%
\bibitem [{\citenamefont {Richardson}(2012)}]{RIC12a}%
  \BibitemOpen
  \bibfield  {author} {\bibinfo {author} {\bibfnamefont {M.~P.}\ \bibnamefont
  {Richardson}},\ }\href@noop {} {\bibfield  {journal} {\bibinfo  {journal} {J.
  Neurol. Neurosurg. Psychiatry}\ }\textbf {\bibinfo {volume} {83}},\ \bibinfo
  {pages} {1238} (\bibinfo {year} {2012})}\BibitemShut {NoStop}%
\bibitem [{\citenamefont {Kuhlmann}\ \emph {et~al.}(2018)\citenamefont
  {Kuhlmann}, \citenamefont {Lehnertz}, \citenamefont {Richardson},
  \citenamefont {Schelter},\ and\ \citenamefont {Zaveri}}]{KUH18}%
  \BibitemOpen
  \bibfield  {author} {\bibinfo {author} {\bibfnamefont {L.}~\bibnamefont
  {Kuhlmann}}, \bibinfo {author} {\bibfnamefont {K.}~\bibnamefont {Lehnertz}},
  \bibinfo {author} {\bibfnamefont {M.~P.}\ \bibnamefont {Richardson}},
  \bibinfo {author} {\bibfnamefont {B.}~\bibnamefont {Schelter}}, \ and\
  \bibinfo {author} {\bibfnamefont {H.~P.}\ \bibnamefont {Zaveri}},\
  }\href@noop {} {\bibfield  {journal} {\bibinfo  {journal} {Nat. Rev.
  Neurol.}\ }\textbf {\bibinfo {volume} {14}},\ \bibinfo {pages} {618}
  (\bibinfo {year} {2018})}\BibitemShut {NoStop}%
\bibitem [{\citenamefont {Boccaletti}\ \emph {et~al.}(2006)\citenamefont
  {Boccaletti}, \citenamefont {Latora}, \citenamefont {Moreno}, \citenamefont
  {Chavez},\ and\ \citenamefont {Hwang}}]{BOC06a}%
  \BibitemOpen
  \bibfield  {author} {\bibinfo {author} {\bibfnamefont {S.}~\bibnamefont
  {Boccaletti}}, \bibinfo {author} {\bibfnamefont {V.}~\bibnamefont {Latora}},
  \bibinfo {author} {\bibfnamefont {Y.}~\bibnamefont {Moreno}}, \bibinfo
  {author} {\bibfnamefont {M.}~\bibnamefont {Chavez}}, \ and\ \bibinfo {author}
  {\bibfnamefont {D.~U.}\ \bibnamefont {Hwang}},\ }\href {\doibase doi:
  10.1016/j.physrep.2005.10.009} {\bibfield  {journal} {\bibinfo  {journal}
  {Phys. Rep.}\ }\textbf {\bibinfo {volume} {424}},\ \bibinfo {pages} {175}
  (\bibinfo {year} {2006})}\BibitemShut {NoStop}%
\bibitem [{\citenamefont {Sch{\"o}ll}\ \emph {et~al.}(2016)\citenamefont
  {Sch{\"o}ll}, \citenamefont {Klapp},\ and\ \citenamefont
  {H{\"o}vel}}]{SCH16}%
  \BibitemOpen
  \bibfield  {author} {\bibinfo {author} {\bibfnamefont {E.}~\bibnamefont
  {Sch{\"o}ll}}, \bibinfo {author} {\bibfnamefont {S.~H.~L.}\ \bibnamefont
  {Klapp}}, \ and\ \bibinfo {author} {\bibfnamefont {P.}~\bibnamefont
  {H{\"o}vel}},\ }\href@noop {} {\emph {\bibinfo {title} {Control of
  self-organizing nonlinear systems}}},\ edited by\ \bibinfo {editor}
  {\bibfnamefont {E.}~\bibnamefont {Sch{\"o}ll}}, \bibinfo {editor}
  {\bibfnamefont {S.~H.~L.}\ \bibnamefont {Klapp}}, \ and\ \bibinfo {editor}
  {\bibfnamefont {P.}~\bibnamefont {H{\"o}vel}}\ (\bibinfo  {publisher}
  {Springer},\ \bibinfo {address} {Berlin},\ \bibinfo {year}
  {2016})\BibitemShut {NoStop}%
\bibitem [{\citenamefont {Bick}\ \emph {et~al.}(2020)\citenamefont {Bick},
  \citenamefont {Goodfellow}, \citenamefont {Laing},\ and\ \citenamefont
  {Martens}}]{BIC20}%
  \BibitemOpen
  \bibfield  {author} {\bibinfo {author} {\bibfnamefont {C.}~\bibnamefont
  {Bick}}, \bibinfo {author} {\bibfnamefont {M.}~\bibnamefont {Goodfellow}},
  \bibinfo {author} {\bibfnamefont {C.~R.}\ \bibnamefont {Laing}}, \ and\
  \bibinfo {author} {\bibfnamefont {E.~A.}\ \bibnamefont {Martens}},\ }\href
  {\doibase https://doi.org/10.1186/s13408-020-00086-9} {\bibfield  {journal}
  {\bibinfo  {journal} {J. Math. Neurosci.}\ }\textbf {\bibinfo {volume}
  {10}},\ \bibinfo {pages} {9} (\bibinfo {year} {2020})}\BibitemShut {NoStop}%
\bibitem [{\citenamefont {Ponten}\ \emph {et~al.}(2009)\citenamefont {Ponten},
  \citenamefont {Douw}, \citenamefont {Bartolomei}, \citenamefont
  {Re\ij{}neveld},\ and\ \citenamefont {Stam}}]{PON09}%
  \BibitemOpen
  \bibfield  {author} {\bibinfo {author} {\bibfnamefont {S.~C.}\ \bibnamefont
  {Ponten}}, \bibinfo {author} {\bibfnamefont {L.}~\bibnamefont {Douw}},
  \bibinfo {author} {\bibfnamefont {F.}~\bibnamefont {Bartolomei}}, \bibinfo
  {author} {\bibfnamefont {J.~C.}\ \bibnamefont {Re\ij{}neveld}}, \ and\
  \bibinfo {author} {\bibfnamefont {C.~J.}\ \bibnamefont {Stam}},\ }\href
  {\doibase 10.1016/j.expneurol.2009.02.001} {\bibfield  {journal} {\bibinfo
  {journal} {Exp. Neurol.}\ }\textbf {\bibinfo {volume} {217}},\ \bibinfo
  {pages} {197} (\bibinfo {year} {2009})}\BibitemShut {NoStop}%
\bibitem [{\citenamefont {Chavez}\ \emph {et~al.}(2010)\citenamefont {Chavez},
  \citenamefont {Valencia}, \citenamefont {Navarro}, \citenamefont {Latora},\
  and\ \citenamefont {Martinerie}}]{CHA10c}%
  \BibitemOpen
  \bibfield  {author} {\bibinfo {author} {\bibfnamefont {M.}~\bibnamefont
  {Chavez}}, \bibinfo {author} {\bibfnamefont {M.}~\bibnamefont {Valencia}},
  \bibinfo {author} {\bibfnamefont {V.}~\bibnamefont {Navarro}}, \bibinfo
  {author} {\bibfnamefont {V.}~\bibnamefont {Latora}}, \ and\ \bibinfo {author}
  {\bibfnamefont {J.}~\bibnamefont {Martinerie}},\ }\href@noop {} {\bibfield
  {journal} {\bibinfo  {journal} {Phys. Rev. Lett.}\ }\textbf {\bibinfo
  {volume} {104}},\ \bibinfo {pages} {118701} (\bibinfo {year}
  {2010})}\BibitemShut {NoStop}%
\bibitem [{\citenamefont {Baier}\ \emph {et~al.}(2012)\citenamefont {Baier},
  \citenamefont {Goodfellow}, \citenamefont {Taylor}, \citenamefont {Wang},\
  and\ \citenamefont {Garry}}]{BAI12}%
  \BibitemOpen
  \bibfield  {author} {\bibinfo {author} {\bibfnamefont {G.}~\bibnamefont
  {Baier}}, \bibinfo {author} {\bibfnamefont {M.}~\bibnamefont {Goodfellow}},
  \bibinfo {author} {\bibfnamefont {P.~N.}\ \bibnamefont {Taylor}}, \bibinfo
  {author} {\bibfnamefont {Y.}~\bibnamefont {Wang}}, \ and\ \bibinfo {author}
  {\bibfnamefont {D.~J.}\ \bibnamefont {Garry}},\ }\href@noop {} {\bibfield
  {journal} {\bibinfo  {journal} {Front. Physiol.}\ }\textbf {\bibinfo {volume}
  {3}},\ \bibinfo {pages} {281} (\bibinfo {year} {2012})}\BibitemShut {NoStop}%
\bibitem [{\citenamefont {Benjamin}\ \emph {et~al.}(2012)\citenamefont
  {Benjamin}, \citenamefont {Fitzgerald}, \citenamefont {Ashwin}, \citenamefont
  {Tsaneva-Atanasova}, \citenamefont {Chowdhury}, \citenamefont {Richardson},\
  and\ \citenamefont {Terry}}]{BEN12}%
  \BibitemOpen
  \bibfield  {author} {\bibinfo {author} {\bibfnamefont {O.}~\bibnamefont
  {Benjamin}}, \bibinfo {author} {\bibfnamefont {T.~H.}\ \bibnamefont
  {Fitzgerald}}, \bibinfo {author} {\bibfnamefont {P.}~\bibnamefont {Ashwin}},
  \bibinfo {author} {\bibfnamefont {K.}~\bibnamefont {Tsaneva-Atanasova}},
  \bibinfo {author} {\bibfnamefont {F.}~\bibnamefont {Chowdhury}}, \bibinfo
  {author} {\bibfnamefont {M.~P.}\ \bibnamefont {Richardson}}, \ and\ \bibinfo
  {author} {\bibfnamefont {J.~R.}\ \bibnamefont {Terry}},\ }\href@noop {}
  {\bibfield  {journal} {\bibinfo  {journal} {J. Math. Neurosci.}\ }\textbf
  {\bibinfo {volume} {2}},\ \bibinfo {pages} {1} (\bibinfo {year}
  {2012})}\BibitemShut {NoStop}%
\bibitem [{\citenamefont {Terry}\ \emph {et~al.}(2012)\citenamefont {Terry},
  \citenamefont {Benjamin},\ and\ \citenamefont {Richardson}}]{TER12}%
  \BibitemOpen
  \bibfield  {author} {\bibinfo {author} {\bibfnamefont {J.~R.}\ \bibnamefont
  {Terry}}, \bibinfo {author} {\bibfnamefont {O.}~\bibnamefont {Benjamin}}, \
  and\ \bibinfo {author} {\bibfnamefont {M.~P.}\ \bibnamefont {Richardson}},\
  }\href {\doibase 10.1111/j.1528-1167.2012.03560.x} {\bibfield  {journal}
  {\bibinfo  {journal} {Epilepsia}\ }\textbf {\bibinfo {volume} {{53}}},\
  \bibinfo {pages} {e166} (\bibinfo {year} {2012})}\BibitemShut {NoStop}%
\bibitem [{\citenamefont {Petkov}\ \emph {et~al.}(2014)\citenamefont {Petkov},
  \citenamefont {Goodfellow}, \citenamefont {Richardson},\ and\ \citenamefont
  {Terry}}]{PET14}%
  \BibitemOpen
  \bibfield  {author} {\bibinfo {author} {\bibfnamefont {G.}~\bibnamefont
  {Petkov}}, \bibinfo {author} {\bibfnamefont {M.}~\bibnamefont {Goodfellow}},
  \bibinfo {author} {\bibfnamefont {M.~P.}\ \bibnamefont {Richardson}}, \ and\
  \bibinfo {author} {\bibfnamefont {J.~R.}\ \bibnamefont {Terry}},\ }\href
  {\doibase 10.3389/fneur.2014.00261} {\bibfield  {journal} {\bibinfo
  {journal} {Front. Neurol.}\ }\textbf {\bibinfo {volume} {5}},\ \bibinfo
  {pages} {261} (\bibinfo {year} {2014})}\BibitemShut {NoStop}%
\bibitem [{\citenamefont {Chouzouris}\ \emph {et~al.}(2018)\citenamefont
  {Chouzouris}, \citenamefont {Omelchenko}, \citenamefont {Zakharova},
  \citenamefont {Hlinka}, \citenamefont {Jiruska},\ and\ \citenamefont
  {Sch{\"o}ll}}]{CHO18}%
  \BibitemOpen
  \bibfield  {author} {\bibinfo {author} {\bibfnamefont {T.}~\bibnamefont
  {Chouzouris}}, \bibinfo {author} {\bibfnamefont {I.}~\bibnamefont
  {Omelchenko}}, \bibinfo {author} {\bibfnamefont {A.}~\bibnamefont
  {Zakharova}}, \bibinfo {author} {\bibfnamefont {J.}~\bibnamefont {Hlinka}},
  \bibinfo {author} {\bibfnamefont {P.}~\bibnamefont {Jiruska}}, \ and\
  \bibinfo {author} {\bibfnamefont {E.}~\bibnamefont {Sch{\"o}ll}},\ }\href
  {\doibase https://doi.org/10.1063/1.5009812} {\bibfield  {journal} {\bibinfo
  {journal} {Chaos}\ }\textbf {\bibinfo {volume} {28}},\ \bibinfo {pages}
  {045112} (\bibinfo {year} {2018})}\BibitemShut {NoStop}%
\bibitem [{\citenamefont {Rothkegel}\ and\ \citenamefont
  {Lehnertz}(2014{\natexlab{a}})}]{ROT14a}%
  \BibitemOpen
  \bibfield  {author} {\bibinfo {author} {\bibfnamefont {A.}~\bibnamefont
  {Rothkegel}}\ and\ \bibinfo {author} {\bibfnamefont {K.}~\bibnamefont
  {Lehnertz}},\ }\href@noop {} {\bibfield  {journal} {\bibinfo  {journal}
  {Europhys. Lett.}\ }\textbf {\bibinfo {volume} {105}},\ \bibinfo {pages}
  {30003} (\bibinfo {year} {2014}{\natexlab{a}})}\BibitemShut {NoStop}%
\bibitem [{\citenamefont {Poel}\ \emph {et~al.}(2015)\citenamefont {Poel},
  \citenamefont {Zakharova},\ and\ \citenamefont {Sch{\"o}ll}}]{POE15}%
  \BibitemOpen
  \bibfield  {author} {\bibinfo {author} {\bibfnamefont {W.}~\bibnamefont
  {Poel}}, \bibinfo {author} {\bibfnamefont {A.}~\bibnamefont {Zakharova}}, \
  and\ \bibinfo {author} {\bibfnamefont {E.}~\bibnamefont {Sch{\"o}ll}},\
  }\href {\doibase 10.1103/physreve.91.022915} {\bibfield  {journal} {\bibinfo
  {journal} {Phys. Rev. E}\ }\textbf {\bibinfo {volume} {91}},\ \bibinfo
  {pages} {022915} (\bibinfo {year} {2015})}\BibitemShut {NoStop}%
\bibitem [{\citenamefont {Sawicki}(2019)}]{SAW20}%
  \BibitemOpen
  \bibfield  {author} {\bibinfo {author} {\bibfnamefont {J.}~\bibnamefont
  {Sawicki}},\ }\href {\doibase 10.1007/978-3-030-34076-6_5} {\emph {\bibinfo
  {title} {Delay controlled partial synchronization in complex networks}}},\
  Springer Theses\ (\bibinfo  {publisher} {Springer},\ \bibinfo {address}
  {Heidelberg},\ \bibinfo {year} {2019})\BibitemShut {NoStop}%
\bibitem [{\citenamefont {Sch{\"o}ll}\ \emph {et~al.}(2020)\citenamefont
  {Sch{\"o}ll}, \citenamefont {Zakharova},\ and\ \citenamefont
  {Andrzejak}}]{SCH20b}%
  \BibitemOpen
  \bibfield  {author} {\bibinfo {author} {\bibfnamefont {E.}~\bibnamefont
  {Sch{\"o}ll}}, \bibinfo {author} {\bibfnamefont {A.}~\bibnamefont
  {Zakharova}}, \ and\ \bibinfo {author} {\bibfnamefont {R.~G.}\ \bibnamefont
  {Andrzejak}},\ }\href {\doibase 10.3389/978-2-88963-311-1} {\emph {\bibinfo
  {title} {Chimera States in Complex Networks}}},\ Research Topics, Front.
  Appl. Math. Stat.\ (\bibinfo  {publisher} {Lausanne: Frontiers Media SA},\
  \bibinfo {year} {2020})\ \bibinfo {note} {ebook}\BibitemShut {NoStop}%
\bibitem [{\citenamefont {Zakharova}(2020)}]{ZAK20}%
  \BibitemOpen
  \bibfield  {author} {\bibinfo {author} {\bibfnamefont {A.}~\bibnamefont
  {Zakharova}},\ }\href {\doibase 10.1007/978-3-030-21714-3} {\emph {\bibinfo
  {title} {Chimera Patterns in Networks: Interplay between Dynamics, Structure,
  Noise, and Delay}}},\ Understanding Complex Systems\ (\bibinfo  {publisher}
  {Springer},\ \bibinfo {year} {2020})\BibitemShut {NoStop}%
\bibitem [{\citenamefont {Andrzejak}\ \emph {et~al.}(2016)\citenamefont
  {Andrzejak}, \citenamefont {Rummel}, \citenamefont {Mormann},\ and\
  \citenamefont {Schindler}}]{AND16}%
  \BibitemOpen
  \bibfield  {author} {\bibinfo {author} {\bibfnamefont {R.~G.}\ \bibnamefont
  {Andrzejak}}, \bibinfo {author} {\bibfnamefont {C.}~\bibnamefont {Rummel}},
  \bibinfo {author} {\bibfnamefont {F.}~\bibnamefont {Mormann}}, \ and\
  \bibinfo {author} {\bibfnamefont {K.}~\bibnamefont {Schindler}},\ }\href
  {\doibase 10.1038/srep23000} {\bibfield  {journal} {\bibinfo  {journal} {Sci.
  Rep.}\ }\textbf {\bibinfo {volume} {6}},\ \bibinfo {pages} {23000} (\bibinfo
  {year} {2016})}\BibitemShut {NoStop}%
\bibitem [{\citenamefont {Rothkegel}\ and\ \citenamefont
  {Lehnertz}(2014{\natexlab{b}})}]{ROT14}%
  \BibitemOpen
  \bibfield  {author} {\bibinfo {author} {\bibfnamefont {A.}~\bibnamefont
  {Rothkegel}}\ and\ \bibinfo {author} {\bibfnamefont {K.}~\bibnamefont
  {Lehnertz}},\ }\href@noop {} {\bibfield  {journal} {\bibinfo  {journal} {New
  J. Phys.}\ }\textbf {\bibinfo {volume} {16}},\ \bibinfo {pages} {055006}
  (\bibinfo {year} {2014}{\natexlab{b}})}\BibitemShut {NoStop}%
\bibitem [{\citenamefont {FitzHugh}(1961)}]{FIT61}%
  \BibitemOpen
  \bibfield  {author} {\bibinfo {author} {\bibfnamefont {R.}~\bibnamefont
  {FitzHugh}},\ }\href@noop {} {\bibfield  {journal} {\bibinfo  {journal}
  {Biophys. J.}\ }\textbf {\bibinfo {volume} {1}},\ \bibinfo {pages} {445}
  (\bibinfo {year} {1961})}\BibitemShut {NoStop}%
\bibitem [{\citenamefont {Nagumo}\ \emph {et~al.}(1962)\citenamefont {Nagumo},
  \citenamefont {Arimoto},\ and\ \citenamefont {Yoshizawa.}}]{NAG62}%
  \BibitemOpen
  \bibfield  {author} {\bibinfo {author} {\bibfnamefont {J.}~\bibnamefont
  {Nagumo}}, \bibinfo {author} {\bibfnamefont {S.}~\bibnamefont {Arimoto}}, \
  and\ \bibinfo {author} {\bibfnamefont {S.}~\bibnamefont {Yoshizawa.}},\
  }\href@noop {} {\bibfield  {journal} {\bibinfo  {journal} {Proc. IRE}\
  }\textbf {\bibinfo {volume} {50}},\ \bibinfo {pages} {2061} (\bibinfo {year}
  {1962})}\BibitemShut {NoStop}%
\bibitem [{\citenamefont {Chernihovskyi}\ \emph {et~al.}(2005)\citenamefont
  {Chernihovskyi}, \citenamefont {Mormann}, \citenamefont {M{\"u}ller},
  \citenamefont {Elger}, \citenamefont {Baier},\ and\ \citenamefont
  {Lehnertz}}]{CHE05e}%
  \BibitemOpen
  \bibfield  {author} {\bibinfo {author} {\bibfnamefont {A.}~\bibnamefont
  {Chernihovskyi}}, \bibinfo {author} {\bibfnamefont {F.}~\bibnamefont
  {Mormann}}, \bibinfo {author} {\bibfnamefont {M.}~\bibnamefont {M{\"u}ller}},
  \bibinfo {author} {\bibfnamefont {C.~E.}\ \bibnamefont {Elger}}, \bibinfo
  {author} {\bibfnamefont {G.}~\bibnamefont {Baier}}, \ and\ \bibinfo {author}
  {\bibfnamefont {K.}~\bibnamefont {Lehnertz}},\ }\href@noop {} {\bibfield
  {journal} {\bibinfo  {journal} {J. Clin. Neurophysiol.}\ }\textbf {\bibinfo
  {volume} {22}},\ \bibinfo {pages} {314} (\bibinfo {year} {2005})}\BibitemShut
  {NoStop}%
\bibitem [{\citenamefont {Chernihovskyi}\ and\ \citenamefont
  {Lehnertz}(2007)}]{CHE07a}%
  \BibitemOpen
  \bibfield  {author} {\bibinfo {author} {\bibfnamefont {A.}~\bibnamefont
  {Chernihovskyi}}\ and\ \bibinfo {author} {\bibfnamefont {K.}~\bibnamefont
  {Lehnertz}},\ }\href@noop {} {\bibfield  {journal} {\bibinfo  {journal} {Int.
  J. Bifurcat. Chaos}\ }\textbf {\bibinfo {volume} {17}},\ \bibinfo {pages}
  {3425} (\bibinfo {year} {2007})}\BibitemShut {NoStop}%
\bibitem [{\citenamefont {Tzourio-Mazoyer}\ \emph {et~al.}(2002)\citenamefont
  {Tzourio-Mazoyer}, \citenamefont {Landeau}, \citenamefont {Papathanassiou},
  \citenamefont {Crivello}, \citenamefont {Etard}, \citenamefont {Delcroix},
  \citenamefont {Mazoyer},\ and\ \citenamefont {Joliot}}]{TZO02}%
  \BibitemOpen
  \bibfield  {author} {\bibinfo {author} {\bibfnamefont {N.}~\bibnamefont
  {Tzourio-Mazoyer}}, \bibinfo {author} {\bibfnamefont {B.}~\bibnamefont
  {Landeau}}, \bibinfo {author} {\bibfnamefont {D.}~\bibnamefont
  {Papathanassiou}}, \bibinfo {author} {\bibfnamefont {F.}~\bibnamefont
  {Crivello}}, \bibinfo {author} {\bibfnamefont {O.}~\bibnamefont {Etard}},
  \bibinfo {author} {\bibfnamefont {N.}~\bibnamefont {Delcroix}}, \bibinfo
  {author} {\bibfnamefont {B.}~\bibnamefont {Mazoyer}}, \ and\ \bibinfo
  {author} {\bibfnamefont {M.}~\bibnamefont {Joliot}},\ }\href@noop {}
  {\bibfield  {journal} {\bibinfo  {journal} {Neuroimage}\ }\textbf {\bibinfo
  {volume} {15}},\ \bibinfo {pages} {273} (\bibinfo {year} {2002})}\BibitemShut
  {NoStop}%
\bibitem [{\citenamefont {Omelchenko}\ \emph {et~al.}(2013)\citenamefont
  {Omelchenko}, \citenamefont {Omel'chenko}, \citenamefont {H{\"o}vel},\ and\
  \citenamefont {Sch{\"o}ll}}]{OME13}%
  \BibitemOpen
  \bibfield  {author} {\bibinfo {author} {\bibfnamefont {I.}~\bibnamefont
  {Omelchenko}}, \bibinfo {author} {\bibfnamefont {O.~E.}\ \bibnamefont
  {Omel'chenko}}, \bibinfo {author} {\bibfnamefont {P.}~\bibnamefont
  {H{\"o}vel}}, \ and\ \bibinfo {author} {\bibfnamefont {E.}~\bibnamefont
  {Sch{\"o}ll}},\ }\href {\doibase 10.1103/physrevlett.110.224101} {\bibfield
  {journal} {\bibinfo  {journal} {Phys. Rev. Lett.}\ }\textbf {\bibinfo
  {volume} {110}},\ \bibinfo {pages} {224101} (\bibinfo {year}
  {2013})}\BibitemShut {NoStop}%
\bibitem [{\citenamefont {Kiskowski}\ \emph {et~al.}(2004)\citenamefont
  {Kiskowski}, \citenamefont {Alber}, \citenamefont {Thomas}, \citenamefont
  {Glazier}, \citenamefont {Bronstein}, \citenamefont {Pu},\ and\ \citenamefont
  {Newman}}]{KIS04}%
  \BibitemOpen
  \bibfield  {author} {\bibinfo {author} {\bibfnamefont {M.~A.}\ \bibnamefont
  {Kiskowski}}, \bibinfo {author} {\bibfnamefont {M.~S.}\ \bibnamefont
  {Alber}}, \bibinfo {author} {\bibfnamefont {G.~L.}\ \bibnamefont {Thomas}},
  \bibinfo {author} {\bibfnamefont {J.~A.}\ \bibnamefont {Glazier}}, \bibinfo
  {author} {\bibfnamefont {N.~B.}\ \bibnamefont {Bronstein}}, \bibinfo {author}
  {\bibfnamefont {J.}~\bibnamefont {Pu}}, \ and\ \bibinfo {author}
  {\bibfnamefont {S.~A.}\ \bibnamefont {Newman}},\ }\href@noop {} {\bibfield
  {journal} {\bibinfo  {journal} {Dev. Biol.}\ }\textbf {\bibinfo {volume}
  {271}},\ \bibinfo {pages} {372} (\bibinfo {year} {2004})}\BibitemShut
  {NoStop}%
\bibitem [{\citenamefont {Pereda}(2014)}]{PER14a}%
  \BibitemOpen
  \bibfield  {author} {\bibinfo {author} {\bibfnamefont {A.~E.}\ \bibnamefont
  {Pereda}},\ }\href {\doibase 10.1038/nrn3708} {\bibfield  {journal} {\bibinfo
   {journal} {Nat. Rev. Neurosci.}\ }\textbf {\bibinfo {volume} {15}},\
  \bibinfo {pages} {250} (\bibinfo {year} {2014})}\BibitemShut {NoStop}%
\bibitem [{\citenamefont {Massobrio}\ \emph {et~al.}(2015)\citenamefont
  {Massobrio}, \citenamefont {de~Arcangelis}, \citenamefont {Pasquale},
  \citenamefont {Jensen},\ and\ \citenamefont {Plenz}}]{MAS15a}%
  \BibitemOpen
  \bibfield  {author} {\bibinfo {author} {\bibfnamefont {P.}~\bibnamefont
  {Massobrio}}, \bibinfo {author} {\bibfnamefont {L.}~\bibnamefont
  {de~Arcangelis}}, \bibinfo {author} {\bibfnamefont {V.}~\bibnamefont
  {Pasquale}}, \bibinfo {author} {\bibfnamefont {H.~J.}\ \bibnamefont
  {Jensen}}, \ and\ \bibinfo {author} {\bibfnamefont {D.}~\bibnamefont
  {Plenz}},\ }\href@noop {} {\bibfield  {journal} {\bibinfo  {journal} {Front.
  Syst. Neurosci.}\ }\textbf {\bibinfo {volume} {9}},\ \bibinfo {pages} {22}
  (\bibinfo {year} {2015})}\BibitemShut {NoStop}%
\bibitem [{\citenamefont {Omel'chenko}\ \emph {et~al.}(2010)\citenamefont
  {Omel'chenko}, \citenamefont {Wolfrum},\ and\ \citenamefont
  {Maistrenko}}]{OME10a}%
  \BibitemOpen
  \bibfield  {author} {\bibinfo {author} {\bibfnamefont {O.~E.}\ \bibnamefont
  {Omel'chenko}}, \bibinfo {author} {\bibfnamefont {M.}~\bibnamefont
  {Wolfrum}}, \ and\ \bibinfo {author} {\bibfnamefont {Y.}~\bibnamefont
  {Maistrenko}},\ }\href {\doibase 10.1103/physreve.81.065201} {\bibfield
  {journal} {\bibinfo  {journal} {Phys. Rev. E}\ }\textbf {\bibinfo {volume}
  {81}},\ \bibinfo {pages} {065201(R)} (\bibinfo {year} {2010})}\BibitemShut
  {NoStop}%
\bibitem [{\citenamefont {Melicher}\ \emph {et~al.}(2015)\citenamefont
  {Melicher}, \citenamefont {Horacek}, \citenamefont {Hlinka}, \citenamefont
  {Spaniel}, \citenamefont {Tintera}, \citenamefont {Ibrahim}, \citenamefont
  {Mikolas}, \citenamefont {Novak}, \citenamefont {Mohr},\ and\ \citenamefont
  {Hoschl}}]{MEL15}%
  \BibitemOpen
  \bibfield  {author} {\bibinfo {author} {\bibfnamefont {T.}~\bibnamefont
  {Melicher}}, \bibinfo {author} {\bibfnamefont {J.}~\bibnamefont {Horacek}},
  \bibinfo {author} {\bibfnamefont {J.}~\bibnamefont {Hlinka}}, \bibinfo
  {author} {\bibfnamefont {F.}~\bibnamefont {Spaniel}}, \bibinfo {author}
  {\bibfnamefont {J.}~\bibnamefont {Tintera}}, \bibinfo {author} {\bibfnamefont
  {I.}~\bibnamefont {Ibrahim}}, \bibinfo {author} {\bibfnamefont
  {P.}~\bibnamefont {Mikolas}}, \bibinfo {author} {\bibfnamefont
  {T.}~\bibnamefont {Novak}}, \bibinfo {author} {\bibfnamefont
  {P.}~\bibnamefont {Mohr}}, \ and\ \bibinfo {author} {\bibfnamefont
  {C.}~\bibnamefont {Hoschl}},\ }\href@noop {} {\bibfield  {journal} {\bibinfo
  {journal} {Schizophr. Res.}\ }\textbf {\bibinfo {volume} {162}},\ \bibinfo
  {pages} {22} (\bibinfo {year} {2015})}\BibitemShut {NoStop}%
\bibitem [{\citenamefont {Ramlow}\ \emph {et~al.}(2019)\citenamefont {Ramlow},
  \citenamefont {Sawicki}, \citenamefont {Zakharova}, \citenamefont {Hlinka},
  \citenamefont {Claussen},\ and\ \citenamefont {Sch{\"o}ll}}]{RAM19}%
  \BibitemOpen
  \bibfield  {author} {\bibinfo {author} {\bibfnamefont {L.}~\bibnamefont
  {Ramlow}}, \bibinfo {author} {\bibfnamefont {J.}~\bibnamefont {Sawicki}},
  \bibinfo {author} {\bibfnamefont {A.}~\bibnamefont {Zakharova}}, \bibinfo
  {author} {\bibfnamefont {J.}~\bibnamefont {Hlinka}}, \bibinfo {author}
  {\bibfnamefont {J.~C.}\ \bibnamefont {Claussen}}, \ and\ \bibinfo {author}
  {\bibfnamefont {E.}~\bibnamefont {Sch{\"o}ll}},\ }\href@noop {} {\bibfield
  {journal} {\bibinfo  {journal} {EPL}\ }\textbf {\bibinfo {volume} {126}},\
  \bibinfo {pages} {50007} (\bibinfo {year} {2019})},\ \bibinfo {note}
  {highlighted in phys.org
  https://phys.org/news/2019-07-unihemispheric-humans.html and in Europhys.
  News 50, no. 5-6 (2019)}\BibitemShut {NoStop}%
\bibitem [{\citenamefont {Bal}\ \emph {et~al.}(2000)\citenamefont {Bal},
  \citenamefont {Debay},\ and\ \citenamefont {Destexhe}}]{BAL00a}%
  \BibitemOpen
  \bibfield  {author} {\bibinfo {author} {\bibfnamefont {T.}~\bibnamefont
  {Bal}}, \bibinfo {author} {\bibfnamefont {D.}~\bibnamefont {Debay}}, \ and\
  \bibinfo {author} {\bibfnamefont {A.}~\bibnamefont {Destexhe}},\ }\href@noop
  {} {\bibfield  {journal} {\bibinfo  {journal} {J. Neurosci.}\ }\textbf
  {\bibinfo {volume} {20}},\ \bibinfo {pages} {7478} (\bibinfo {year}
  {2000})}\BibitemShut {NoStop}%
\bibitem [{\citenamefont {Blumenfeld}\ and\ \citenamefont
  {McCormick}(2000)}]{BLU00}%
  \BibitemOpen
  \bibfield  {author} {\bibinfo {author} {\bibfnamefont {H.}~\bibnamefont
  {Blumenfeld}}\ and\ \bibinfo {author} {\bibfnamefont {D.~A.}\ \bibnamefont
  {McCormick}},\ }\href@noop {} {\bibfield  {journal} {\bibinfo  {journal} {J.
  Neurosci.}\ }\textbf {\bibinfo {volume} {20}},\ \bibinfo {pages} {5153}
  (\bibinfo {year} {2000})}\BibitemShut {NoStop}%
\bibitem [{\citenamefont {Sadleir}\ \emph {et~al.}(2006)\citenamefont
  {Sadleir}, \citenamefont {Farrell}, \citenamefont {Smith}, \citenamefont
  {Connolly},\ and\ \citenamefont {Scheffer}}]{SAD06}%
  \BibitemOpen
  \bibfield  {author} {\bibinfo {author} {\bibfnamefont {L.}~\bibnamefont
  {Sadleir}}, \bibinfo {author} {\bibfnamefont {K.}~\bibnamefont {Farrell}},
  \bibinfo {author} {\bibfnamefont {S.}~\bibnamefont {Smith}}, \bibinfo
  {author} {\bibfnamefont {M.}~\bibnamefont {Connolly}}, \ and\ \bibinfo
  {author} {\bibfnamefont {I.}~\bibnamefont {Scheffer}},\ }\href@noop {}
  {\bibfield  {journal} {\bibinfo  {journal} {Neurology}\ }\textbf {\bibinfo
  {volume} {67}},\ \bibinfo {pages} {413} (\bibinfo {year} {2006})}\BibitemShut
  {NoStop}%
\bibitem [{\citenamefont {Mormann}\ \emph {et~al.}(2000)\citenamefont
  {Mormann}, \citenamefont {Lehnertz}, \citenamefont {David},\ and\
  \citenamefont {Elger}}]{MOR00}%
  \BibitemOpen
  \bibfield  {author} {\bibinfo {author} {\bibfnamefont {F.}~\bibnamefont
  {Mormann}}, \bibinfo {author} {\bibfnamefont {K.}~\bibnamefont {Lehnertz}},
  \bibinfo {author} {\bibfnamefont {P.}~\bibnamefont {David}}, \ and\ \bibinfo
  {author} {\bibfnamefont {C.~E.}\ \bibnamefont {Elger}},\ }\href@noop {}
  {\bibfield  {journal} {\bibinfo  {journal} {Physica D}\ }\textbf {\bibinfo
  {volume} {144}},\ \bibinfo {pages} {358} (\bibinfo {year}
  {2000})}\BibitemShut {NoStop}%
\bibitem [{\citenamefont {Mormann}\ \emph {et~al.}(2003)\citenamefont
  {Mormann}, \citenamefont {Kreuz}, \citenamefont {Andrzejak}, \citenamefont
  {David}, \citenamefont {Lehnertz},\ and\ \citenamefont {Elger}}]{MOR03a}%
  \BibitemOpen
  \bibfield  {author} {\bibinfo {author} {\bibfnamefont {F.}~\bibnamefont
  {Mormann}}, \bibinfo {author} {\bibfnamefont {T.}~\bibnamefont {Kreuz}},
  \bibinfo {author} {\bibfnamefont {R.~G.}\ \bibnamefont {Andrzejak}}, \bibinfo
  {author} {\bibfnamefont {P.}~\bibnamefont {David}}, \bibinfo {author}
  {\bibfnamefont {K.}~\bibnamefont {Lehnertz}}, \ and\ \bibinfo {author}
  {\bibfnamefont {C.~E.}\ \bibnamefont {Elger}},\ }\href {\doibase
  10.1016/s0920-1211(03)00002-0} {\bibfield  {journal} {\bibinfo  {journal}
  {Epilepsy Res.}\ }\textbf {\bibinfo {volume} {53}},\ \bibinfo {pages} {173}
  (\bibinfo {year} {2003})}\BibitemShut {NoStop}%
\bibitem [{\citenamefont {Feldt}\ \emph {et~al.}(2007)\citenamefont {Feldt},
  \citenamefont {Osterhage}, \citenamefont {Mormann}, \citenamefont
  {Lehnertz},\ and\ \citenamefont {Zochowski}}]{FEL07}%
  \BibitemOpen
  \bibfield  {author} {\bibinfo {author} {\bibfnamefont {S.}~\bibnamefont
  {Feldt}}, \bibinfo {author} {\bibfnamefont {H.}~\bibnamefont {Osterhage}},
  \bibinfo {author} {\bibfnamefont {F.}~\bibnamefont {Mormann}}, \bibinfo
  {author} {\bibfnamefont {K.}~\bibnamefont {Lehnertz}}, \ and\ \bibinfo
  {author} {\bibfnamefont {M.}~\bibnamefont {Zochowski}},\ }\href {\doibase
  10.1103/physreve.76.021920} {\bibfield  {journal} {\bibinfo  {journal} {Phys.
  Rev. E}\ }\textbf {\bibinfo {volume} {76}},\ \bibinfo {eid} {021920}
  (\bibinfo {year} {2007})}\BibitemShut {NoStop}%
\bibitem [{\citenamefont {Barrat}\ \emph {et~al.}(2004)\citenamefont {Barrat},
  \citenamefont {Barth{\'e}lemy}, \citenamefont {Pastor-Satorras},\ and\
  \citenamefont {Vespignani}}]{BAR04b}%
  \BibitemOpen
  \bibfield  {author} {\bibinfo {author} {\bibfnamefont {A.}~\bibnamefont
  {Barrat}}, \bibinfo {author} {\bibfnamefont {M.}~\bibnamefont
  {Barth{\'e}lemy}}, \bibinfo {author} {\bibfnamefont {R.}~\bibnamefont
  {Pastor-Satorras}}, \ and\ \bibinfo {author} {\bibfnamefont {A.}~\bibnamefont
  {Vespignani}},\ }\href {\doibase 10.1073/pnas.0400087101} {\bibfield
  {journal} {\bibinfo  {journal} {Proc. Natl. Acad. Sci. U.S.A.}\ }\textbf
  {\bibinfo {volume} {101}},\ \bibinfo {pages} {3747} (\bibinfo {year}
  {2004})}\BibitemShut {NoStop}%
\bibitem [{\citenamefont {Boccaletti}\ and\ \citenamefont
  {Bragard}(2006)}]{BOC06}%
  \BibitemOpen
  \bibfield  {author} {\bibinfo {author} {\bibfnamefont {S.}~\bibnamefont
  {Boccaletti}}\ and\ \bibinfo {author} {\bibfnamefont {J.}~\bibnamefont
  {Bragard}},\ }\href@noop {} {\bibfield  {journal} {\bibinfo  {journal}
  {Philos. Trans. R. Soc. London, Ser. A}\ }\textbf {\bibinfo {volume} {364}},\
  \bibinfo {pages} {2383} (\bibinfo {year} {2006})}\BibitemShut {NoStop}%
\bibitem [{\citenamefont {Enright}\ and\ \citenamefont
  {Leitner}(2005)}]{ENR05}%
  \BibitemOpen
  \bibfield  {author} {\bibinfo {author} {\bibfnamefont {M.~B.}\ \bibnamefont
  {Enright}}\ and\ \bibinfo {author} {\bibfnamefont {D.~M.}\ \bibnamefont
  {Leitner}},\ }\href@noop {} {\bibfield  {journal} {\bibinfo  {journal} {Phys.
  Rev. E}\ }\textbf {\bibinfo {volume} {71}},\ \bibinfo {pages} {011912}
  (\bibinfo {year} {2005})}\BibitemShut {NoStop}%
\bibitem [{\citenamefont {Hahn}\ \emph {et~al.}(2005)\citenamefont {Hahn},
  \citenamefont {Georg},\ and\ \citenamefont {Peitgen}}]{HAH05}%
  \BibitemOpen
  \bibfield  {author} {\bibinfo {author} {\bibfnamefont {H.~K.}\ \bibnamefont
  {Hahn}}, \bibinfo {author} {\bibfnamefont {M.}~\bibnamefont {Georg}}, \ and\
  \bibinfo {author} {\bibfnamefont {H.~O.}\ \bibnamefont {Peitgen}},\
  }\href@noop {} {\emph {\bibinfo {title} {Fractals in Biology and
  Medicine}}},\ Vol.~\bibinfo {volume} {4}\ (\bibinfo  {publisher} {Springer},\
  \bibinfo {year} {2005})\BibitemShut {NoStop}%
\bibitem [{\citenamefont {Katsaloulis}\ \emph {et~al.}(2009)\citenamefont
  {Katsaloulis}, \citenamefont {Verganelakis},\ and\ \citenamefont
  {Provata}}]{KAT09}%
  \BibitemOpen
  \bibfield  {author} {\bibinfo {author} {\bibfnamefont {P.}~\bibnamefont
  {Katsaloulis}}, \bibinfo {author} {\bibfnamefont {D.~A.}\ \bibnamefont
  {Verganelakis}}, \ and\ \bibinfo {author} {\bibfnamefont {A.}~\bibnamefont
  {Provata}},\ }\href@noop {} {\bibfield  {journal} {\bibinfo  {journal}
  {Fractals}\ }\textbf {\bibinfo {volume} {17}},\ \bibinfo {pages} {181}
  (\bibinfo {year} {2009})}\BibitemShut {NoStop}%
\bibitem [{\citenamefont {Katsaloulis}\ \emph {et~al.}(2012)\citenamefont
  {Katsaloulis}, \citenamefont {Ghosh}, \citenamefont {Philippe}, \citenamefont
  {Provata},\ and\ \citenamefont {Deriche}}]{KAT12}%
  \BibitemOpen
  \bibfield  {author} {\bibinfo {author} {\bibfnamefont {P.}~\bibnamefont
  {Katsaloulis}}, \bibinfo {author} {\bibfnamefont {A.}~\bibnamefont {Ghosh}},
  \bibinfo {author} {\bibfnamefont {A.~C.}\ \bibnamefont {Philippe}}, \bibinfo
  {author} {\bibfnamefont {A.}~\bibnamefont {Provata}}, \ and\ \bibinfo
  {author} {\bibfnamefont {R.}~\bibnamefont {Deriche}},\ }\href@noop {}
  {\bibfield  {journal} {\bibinfo  {journal} {Eur. Phys. J. B}\ }\textbf
  {\bibinfo {volume} {85}},\ \bibinfo {pages} {1} (\bibinfo {year}
  {2012})}\BibitemShut {NoStop}%
\bibitem [{\citenamefont {Omelchenko}\ \emph {et~al.}(2015)\citenamefont
  {Omelchenko}, \citenamefont {Provata}, \citenamefont {Hizanidis},
  \citenamefont {Sch{\"o}ll},\ and\ \citenamefont {H{\"o}vel}}]{OME15}%
  \BibitemOpen
  \bibfield  {author} {\bibinfo {author} {\bibfnamefont {I.}~\bibnamefont
  {Omelchenko}}, \bibinfo {author} {\bibfnamefont {A.}~\bibnamefont {Provata}},
  \bibinfo {author} {\bibfnamefont {J.}~\bibnamefont {Hizanidis}}, \bibinfo
  {author} {\bibfnamefont {E.}~\bibnamefont {Sch{\"o}ll}}, \ and\ \bibinfo
  {author} {\bibfnamefont {P.}~\bibnamefont {H{\"o}vel}},\ }\href {\doibase
  10.1103/physreve.91.022917} {\bibfield  {journal} {\bibinfo  {journal} {Phys.
  Rev. E}\ }\textbf {\bibinfo {volume} {91}},\ \bibinfo {pages} {022917}
  (\bibinfo {year} {2015})}\BibitemShut {NoStop}%
\bibitem [{\citenamefont {Plotnikov}\ \emph {et~al.}(2016)\citenamefont
  {Plotnikov}, \citenamefont {Lehnert}, \citenamefont {Fradkov},\ and\
  \citenamefont {Sch{\"o}ll}}]{PLO16a}%
  \BibitemOpen
  \bibfield  {author} {\bibinfo {author} {\bibfnamefont {S.~A.}\ \bibnamefont
  {Plotnikov}}, \bibinfo {author} {\bibfnamefont {J.}~\bibnamefont {Lehnert}},
  \bibinfo {author} {\bibfnamefont {A.~L.}\ \bibnamefont {Fradkov}}, \ and\
  \bibinfo {author} {\bibfnamefont {E.}~\bibnamefont {Sch{\"o}ll}},\
  }\href@noop {} {\bibfield  {journal} {\bibinfo  {journal} {Phys. Rev. E}\
  }\textbf {\bibinfo {volume} {94}},\ \bibinfo {pages} {012203} (\bibinfo
  {year} {2016})}\BibitemShut {NoStop}%
\bibitem [{\citenamefont {Sawicki}\ \emph {et~al.}(2019)\citenamefont
  {Sawicki}, \citenamefont {Omelchenko}, \citenamefont {Zakharova},\ and\
  \citenamefont {Sch{\"o}ll}}]{SAW19}%
  \BibitemOpen
  \bibfield  {author} {\bibinfo {author} {\bibfnamefont {J.}~\bibnamefont
  {Sawicki}}, \bibinfo {author} {\bibfnamefont {I.}~\bibnamefont {Omelchenko}},
  \bibinfo {author} {\bibfnamefont {A.}~\bibnamefont {Zakharova}}, \ and\
  \bibinfo {author} {\bibfnamefont {E.}~\bibnamefont {Sch{\"o}ll}},\
  }\href@noop {} {\bibfield  {journal} {\bibinfo  {journal} {Eur. Phys. J. B}\
  }\textbf {\bibinfo {volume} {92}},\ \bibinfo {pages} {54} (\bibinfo {year}
  {2019})}\BibitemShut {NoStop}%
\bibitem [{\citenamefont {Nikitin}\ \emph {et~al.}(2019)\citenamefont
  {Nikitin}, \citenamefont {Omelchenko}, \citenamefont {Zakharova},
  \citenamefont {Avetyan}, \citenamefont {Fradkov},\ and\ \citenamefont
  {Sch{\"o}ll}}]{NIK19}%
  \BibitemOpen
  \bibfield  {author} {\bibinfo {author} {\bibfnamefont {D.}~\bibnamefont
  {Nikitin}}, \bibinfo {author} {\bibfnamefont {I.}~\bibnamefont {Omelchenko}},
  \bibinfo {author} {\bibfnamefont {A.}~\bibnamefont {Zakharova}}, \bibinfo
  {author} {\bibfnamefont {M.}~\bibnamefont {Avetyan}}, \bibinfo {author}
  {\bibfnamefont {A.~L.}\ \bibnamefont {Fradkov}}, \ and\ \bibinfo {author}
  {\bibfnamefont {E.}~\bibnamefont {Sch{\"o}ll}},\ }\href@noop {} {\bibfield
  {journal} {\bibinfo  {journal} {Phil. Trans. R. Soc. A}\ }\textbf {\bibinfo
  {volume} {377}},\ \bibinfo {pages} {20180128} (\bibinfo {year}
  {2019})}\BibitemShut {NoStop}%
\bibitem [{\citenamefont {Krishnagopal}\ \emph {et~al.}(2017)\citenamefont
  {Krishnagopal}, \citenamefont {Lehnert}, \citenamefont {Poel}, \citenamefont
  {Zakharova},\ and\ \citenamefont {Sch{\"o}ll}}]{KRI17}%
  \BibitemOpen
  \bibfield  {author} {\bibinfo {author} {\bibfnamefont {S.}~\bibnamefont
  {Krishnagopal}}, \bibinfo {author} {\bibfnamefont {J.}~\bibnamefont
  {Lehnert}}, \bibinfo {author} {\bibfnamefont {W.}~\bibnamefont {Poel}},
  \bibinfo {author} {\bibfnamefont {A.}~\bibnamefont {Zakharova}}, \ and\
  \bibinfo {author} {\bibfnamefont {E.}~\bibnamefont {Sch{\"o}ll}},\ }\href
  {\doibase 10.1098/rsta.2016.0216} {\bibfield  {journal} {\bibinfo  {journal}
  {Phil. Trans. R. Soc. A}\ }\textbf {\bibinfo {volume} {375}},\ \bibinfo
  {pages} {20160216} (\bibinfo {year} {2017})}\BibitemShut {NoStop}%
\bibitem [{\citenamefont {Hizanidis}\ \emph {et~al.}(2015)\citenamefont
  {Hizanidis}, \citenamefont {Panagakou}, \citenamefont {Omelchenko},
  \citenamefont {Sch{\"o}ll}, \citenamefont {H{\"o}vel},\ and\ \citenamefont
  {Provata}}]{HIZ15}%
  \BibitemOpen
  \bibfield  {author} {\bibinfo {author} {\bibfnamefont {J.}~\bibnamefont
  {Hizanidis}}, \bibinfo {author} {\bibfnamefont {E.}~\bibnamefont
  {Panagakou}}, \bibinfo {author} {\bibfnamefont {I.}~\bibnamefont
  {Omelchenko}}, \bibinfo {author} {\bibfnamefont {E.}~\bibnamefont
  {Sch{\"o}ll}}, \bibinfo {author} {\bibfnamefont {P.}~\bibnamefont
  {H{\"o}vel}}, \ and\ \bibinfo {author} {\bibfnamefont {A.}~\bibnamefont
  {Provata}},\ }\href {\doibase 10.1103/physreve.92.012915} {\bibfield
  {journal} {\bibinfo  {journal} {Phys. Rev. E}\ }\textbf {\bibinfo {volume}
  {92}},\ \bibinfo {pages} {012915} (\bibinfo {year} {2015})}\BibitemShut
  {NoStop}%
\bibitem [{\citenamefont {Ulonska}\ \emph {et~al.}(2016)\citenamefont
  {Ulonska}, \citenamefont {Omelchenko}, \citenamefont {Zakharova},\ and\
  \citenamefont {Sch{\"o}ll}}]{ULO16}%
  \BibitemOpen
  \bibfield  {author} {\bibinfo {author} {\bibfnamefont {S.}~\bibnamefont
  {Ulonska}}, \bibinfo {author} {\bibfnamefont {I.}~\bibnamefont {Omelchenko}},
  \bibinfo {author} {\bibfnamefont {A.}~\bibnamefont {Zakharova}}, \ and\
  \bibinfo {author} {\bibfnamefont {E.}~\bibnamefont {Sch{\"o}ll}},\
  }\href@noop {} {\bibfield  {journal} {\bibinfo  {journal} {Chaos}\ }\textbf
  {\bibinfo {volume} {26}},\ \bibinfo {pages} {094825} (\bibinfo {year}
  {2016})}\BibitemShut {NoStop}%
\bibitem [{\citenamefont {Tsigkri-DeSmedt}\ \emph {et~al.}(2016)\citenamefont
  {Tsigkri-DeSmedt}, \citenamefont {Hizanidis}, \citenamefont {H{\"o}vel},\
  and\ \citenamefont {Provata}}]{TSI16}%
  \BibitemOpen
  \bibfield  {author} {\bibinfo {author} {\bibfnamefont {N.~D.}\ \bibnamefont
  {Tsigkri-DeSmedt}}, \bibinfo {author} {\bibfnamefont {J.}~\bibnamefont
  {Hizanidis}}, \bibinfo {author} {\bibfnamefont {P.}~\bibnamefont
  {H{\"o}vel}}, \ and\ \bibinfo {author} {\bibfnamefont {A.}~\bibnamefont
  {Provata}},\ }\href {\doibase 10.1140/epjst/e2016-02661-4} {\bibfield
  {journal} {\bibinfo  {journal} {Eur. Phys. J. ST}\ }\textbf {\bibinfo
  {volume} {225}},\ \bibinfo {pages} {1149} (\bibinfo {year}
  {2016})}\BibitemShut {NoStop}%
\bibitem [{\citenamefont {Tsigkri-DeSmedt}\ \emph {et~al.}(2017)\citenamefont
  {Tsigkri-DeSmedt}, \citenamefont {Hizanidis}, \citenamefont {Sch{\"o}ll},
  \citenamefont {H{\"o}vel},\ and\ \citenamefont {Provata}}]{TSI17}%
  \BibitemOpen
  \bibfield  {author} {\bibinfo {author} {\bibfnamefont {N.~D.}\ \bibnamefont
  {Tsigkri-DeSmedt}}, \bibinfo {author} {\bibfnamefont {J.}~\bibnamefont
  {Hizanidis}}, \bibinfo {author} {\bibfnamefont {E.}~\bibnamefont
  {Sch{\"o}ll}}, \bibinfo {author} {\bibfnamefont {P.}~\bibnamefont
  {H{\"o}vel}}, \ and\ \bibinfo {author} {\bibfnamefont {A.}~\bibnamefont
  {Provata}},\ }\href {\doibase 10.1140/epjb/e2017-80162-0} {\bibfield
  {journal} {\bibinfo  {journal} {Eur. Phys. J. B}\ }\textbf {\bibinfo {volume}
  {90}},\ \bibinfo {pages} {139} (\bibinfo {year} {2017})}\BibitemShut
  {NoStop}%
\bibitem [{\citenamefont {Sawicki}\ \emph {et~al.}(2017)\citenamefont
  {Sawicki}, \citenamefont {Omelchenko}, \citenamefont {Zakharova},\ and\
  \citenamefont {Sch{\"o}ll}}]{SAW17}%
  \BibitemOpen
  \bibfield  {author} {\bibinfo {author} {\bibfnamefont {J.}~\bibnamefont
  {Sawicki}}, \bibinfo {author} {\bibfnamefont {I.}~\bibnamefont {Omelchenko}},
  \bibinfo {author} {\bibfnamefont {A.}~\bibnamefont {Zakharova}}, \ and\
  \bibinfo {author} {\bibfnamefont {E.}~\bibnamefont {Sch{\"o}ll}},\ }\href
  {\doibase 10.1140/epjst/e2017-70036-8} {\bibfield  {journal} {\bibinfo
  {journal} {Eur. Phys. J. Spec. Top.}\ }\textbf {\bibinfo {volume} {226}},\
  \bibinfo {pages} {1883} (\bibinfo {year} {2017})}\BibitemShut {NoStop}%
\bibitem [{\citenamefont {zur Bonsen}\ \emph {et~al.}(2018)\citenamefont {zur
  Bonsen}, \citenamefont {Omelchenko}, \citenamefont {Zakharova},\ and\
  \citenamefont {Sch{\"o}ll}}]{BON18}%
  \BibitemOpen
  \bibfield  {author} {\bibinfo {author} {\bibfnamefont {A.}~\bibnamefont {zur
  Bonsen}}, \bibinfo {author} {\bibfnamefont {I.}~\bibnamefont {Omelchenko}},
  \bibinfo {author} {\bibfnamefont {A.}~\bibnamefont {Zakharova}}, \ and\
  \bibinfo {author} {\bibfnamefont {E.}~\bibnamefont {Sch{\"o}ll}},\
  }\href@noop {} {\bibfield  {journal} {\bibinfo  {journal} {Eur. Phys. J. B}\
  }\textbf {\bibinfo {volume} {91}},\ \bibinfo {pages} {65} (\bibinfo {year}
  {2018})}\BibitemShut {NoStop}%
\bibitem [{\citenamefont {Watts}\ and\ \citenamefont {Strogatz}(1998)}]{WAT98}%
  \BibitemOpen
  \bibfield  {author} {\bibinfo {author} {\bibfnamefont {D.~J.}\ \bibnamefont
  {Watts}}\ and\ \bibinfo {author} {\bibfnamefont {S.~H.}\ \bibnamefont
  {Strogatz}},\ }\href@noop {} {\bibfield  {journal} {\bibinfo  {journal}
  {Nature}\ }\textbf {\bibinfo {volume} {393}},\ \bibinfo {pages} {440}
  (\bibinfo {year} {1998})}\BibitemShut {NoStop}%
\bibitem [{\citenamefont {Erd{\"o}s}\ and\ \citenamefont
  {R\'{e}nyi}(1960)}]{ERD60}%
  \BibitemOpen
  \bibfield  {author} {\bibinfo {author} {\bibfnamefont {P.}~\bibnamefont
  {Erd{\"o}s}}\ and\ \bibinfo {author} {\bibfnamefont {A.}~\bibnamefont
  {R\'{e}nyi}},\ }\href@noop {} {\bibfield  {journal} {\bibinfo  {journal}
  {Publ. Math. Inst. Hung. Acad. Sci}\ }\textbf {\bibinfo {volume} {5}},\
  \bibinfo {pages} {17} (\bibinfo {year} {1960})}\BibitemShut {NoStop}%
\bibitem [{\citenamefont {Karnatak}\ \emph {et~al.}(2014)\citenamefont
  {Karnatak}, \citenamefont {Ansmann}, \citenamefont {Feudel},\ and\
  \citenamefont {Lehnertz}}]{KAR14b}%
  \BibitemOpen
  \bibfield  {author} {\bibinfo {author} {\bibfnamefont {R.}~\bibnamefont
  {Karnatak}}, \bibinfo {author} {\bibfnamefont {G.}~\bibnamefont {Ansmann}},
  \bibinfo {author} {\bibfnamefont {U.}~\bibnamefont {Feudel}}, \ and\ \bibinfo
  {author} {\bibfnamefont {K.}~\bibnamefont {Lehnertz}},\ }\href {\doibase
  10.1103/physreve.90.022917} {\bibfield  {journal} {\bibinfo  {journal} {Phys.
  Rev. E}\ }\textbf {\bibinfo {volume} {90}},\ \bibinfo {pages} {022917}
  (\bibinfo {year} {2014})}\BibitemShut {NoStop}%
\bibitem [{\citenamefont {Ansmann}\ \emph {et~al.}(2016)\citenamefont
  {Ansmann}, \citenamefont {Lehnertz},\ and\ \citenamefont {Feudel}}]{ANS16}%
  \BibitemOpen
  \bibfield  {author} {\bibinfo {author} {\bibfnamefont {G.}~\bibnamefont
  {Ansmann}}, \bibinfo {author} {\bibfnamefont {K.}~\bibnamefont {Lehnertz}}, \
  and\ \bibinfo {author} {\bibfnamefont {U.}~\bibnamefont {Feudel}},\ }\href
  {\doibase 10.1103/physrevx.6.011030} {\bibfield  {journal} {\bibinfo
  {journal} {Phys. Rev. X}\ }\textbf {\bibinfo {volume} {6}},\ \bibinfo {pages}
  {011030} (\bibinfo {year} {2016})}\BibitemShut {NoStop}%
\bibitem [{\citenamefont {Hilgetag}\ and\ \citenamefont
  {Goulas}(2016)}]{HIL16}%
  \BibitemOpen
  \bibfield  {author} {\bibinfo {author} {\bibfnamefont {C.~C.}\ \bibnamefont
  {Hilgetag}}\ and\ \bibinfo {author} {\bibfnamefont {A.}~\bibnamefont
  {Goulas}},\ }\href@noop {} {\bibfield  {journal} {\bibinfo  {journal} {Brain
  Struct. Func.}\ }\textbf {\bibinfo {volume} {221}},\ \bibinfo {pages} {2361}
  (\bibinfo {year} {2016})}\BibitemShut {NoStop}%
\bibitem [{\citenamefont {Papo}\ \emph {et~al.}(2016)\citenamefont {Papo},
  \citenamefont {Zanin}, \citenamefont {Mart{\'i}nez},\ and\ \citenamefont
  {Buld{\'u}}}]{PAP16}%
  \BibitemOpen
  \bibfield  {author} {\bibinfo {author} {\bibfnamefont {D.}~\bibnamefont
  {Papo}}, \bibinfo {author} {\bibfnamefont {M.}~\bibnamefont {Zanin}},
  \bibinfo {author} {\bibfnamefont {J.~H.}\ \bibnamefont {Mart{\'i}nez}}, \
  and\ \bibinfo {author} {\bibfnamefont {J.~M.}\ \bibnamefont {Buld{\'u}}},\
  }\href@noop {} {\bibfield  {journal} {\bibinfo  {journal} {Front. Hum.
  Neurosci.}\ }\textbf {\bibinfo {volume} {10}},\ \bibinfo {pages} {96}
  (\bibinfo {year} {2016})}\BibitemShut {NoStop}%
\bibitem [{\citenamefont {Gastner}\ and\ \citenamefont
  {{\'O}dor}(2016)}]{GAS16}%
  \BibitemOpen
  \bibfield  {author} {\bibinfo {author} {\bibfnamefont {M.~T.}\ \bibnamefont
  {Gastner}}\ and\ \bibinfo {author} {\bibfnamefont {G.}~\bibnamefont
  {{\'O}dor}},\ }\href@noop {} {\bibfield  {journal} {\bibinfo  {journal} {Sci.
  Rep.}\ }\textbf {\bibinfo {volume} {6}},\ \bibinfo {pages} {27249} (\bibinfo
  {year} {2016})}\BibitemShut {NoStop}%
\bibitem [{\citenamefont {Bialonski}\ \emph {et~al.}(2010)\citenamefont
  {Bialonski}, \citenamefont {Horstmann},\ and\ \citenamefont
  {Lehnertz}}]{BIA10}%
  \BibitemOpen
  \bibfield  {author} {\bibinfo {author} {\bibfnamefont {S.}~\bibnamefont
  {Bialonski}}, \bibinfo {author} {\bibfnamefont {M.-T.}\ \bibnamefont
  {Horstmann}}, \ and\ \bibinfo {author} {\bibfnamefont {K.}~\bibnamefont
  {Lehnertz}},\ }\href@noop {} {\bibfield  {journal} {\bibinfo  {journal}
  {Chaos}\ }\textbf {\bibinfo {volume} {20}},\ \bibinfo {pages} {013134}
  (\bibinfo {year} {2010})}\BibitemShut {NoStop}%
\bibitem [{\citenamefont {Hlinka}\ \emph {et~al.}(2012)\citenamefont {Hlinka},
  \citenamefont {Hartman},\ and\ \citenamefont {Palu{\v{s}}}}]{HLI12a}%
  \BibitemOpen
  \bibfield  {author} {\bibinfo {author} {\bibfnamefont {J.}~\bibnamefont
  {Hlinka}}, \bibinfo {author} {\bibfnamefont {D.~H.}\ \bibnamefont {Hartman}},
  \ and\ \bibinfo {author} {\bibfnamefont {M.}~\bibnamefont {Palu{\v{s}}}},\
  }\href@noop {} {\bibfield  {journal} {\bibinfo  {journal} {Chaos}\ }\textbf
  {\bibinfo {volume} {22}},\ \bibinfo {pages} {033107} (\bibinfo {year}
  {2012})}\BibitemShut {NoStop}%
\bibitem [{\citenamefont {Hlinka}\ \emph {et~al.}(2017)\citenamefont {Hlinka},
  \citenamefont {Hartman}, \citenamefont {Jajcay}, \citenamefont
  {Tome{\v{c}}ek}, \citenamefont {Tint{\v{e}}ra},\ and\ \citenamefont
  {Palu{\v{s}}}}]{HLI17}%
  \BibitemOpen
  \bibfield  {author} {\bibinfo {author} {\bibfnamefont {J.}~\bibnamefont
  {Hlinka}}, \bibinfo {author} {\bibfnamefont {D.~H.}\ \bibnamefont {Hartman}},
  \bibinfo {author} {\bibfnamefont {N.}~\bibnamefont {Jajcay}}, \bibinfo
  {author} {\bibfnamefont {D.}~\bibnamefont {Tome{\v{c}}ek}}, \bibinfo {author}
  {\bibfnamefont {J.}~\bibnamefont {Tint{\v{e}}ra}}, \ and\ \bibinfo {author}
  {\bibfnamefont {M.}~\bibnamefont {Palu{\v{s}}}},\ }\href@noop {} {\bibfield
  {journal} {\bibinfo  {journal} {Chaos}\ }\textbf {\bibinfo {volume} {27}},\
  \bibinfo {pages} {035812} (\bibinfo {year} {2017})}\BibitemShut {NoStop}%
\bibitem [{\citenamefont {Horstmann}\ \emph {et~al.}(2010)\citenamefont
  {Horstmann}, \citenamefont {Bialonski}, \citenamefont {Noennig},
  \citenamefont {Mai}, \citenamefont {Prusseit}, \citenamefont {Wellmer},
  \citenamefont {Hinrichs},\ and\ \citenamefont {Lehnertz}}]{HOR10}%
  \BibitemOpen
  \bibfield  {author} {\bibinfo {author} {\bibfnamefont {M.-T.}\ \bibnamefont
  {Horstmann}}, \bibinfo {author} {\bibfnamefont {S.}~\bibnamefont
  {Bialonski}}, \bibinfo {author} {\bibfnamefont {N.}~\bibnamefont {Noennig}},
  \bibinfo {author} {\bibfnamefont {H.}~\bibnamefont {Mai}}, \bibinfo {author}
  {\bibfnamefont {J.}~\bibnamefont {Prusseit}}, \bibinfo {author}
  {\bibfnamefont {J.}~\bibnamefont {Wellmer}}, \bibinfo {author} {\bibfnamefont
  {H.}~\bibnamefont {Hinrichs}}, \ and\ \bibinfo {author} {\bibfnamefont
  {K.}~\bibnamefont {Lehnertz}},\ }\href@noop {} {\bibfield  {journal}
  {\bibinfo  {journal} {Clin. Neurophysiol.}\ }\textbf {\bibinfo {volume}
  {121}},\ \bibinfo {pages} {172} (\bibinfo {year} {2010})}\BibitemShut
  {NoStop}%
\bibitem [{\citenamefont {Ansmann}\ and\ \citenamefont
  {Lehnertz}(2012)}]{ANS12}%
  \BibitemOpen
  \bibfield  {author} {\bibinfo {author} {\bibfnamefont {G.}~\bibnamefont
  {Ansmann}}\ and\ \bibinfo {author} {\bibfnamefont {K.}~\bibnamefont
  {Lehnertz}},\ }\href@noop {} {\bibfield  {journal} {\bibinfo  {journal} {J.
  Neurosci. Methods}\ }\textbf {\bibinfo {volume} {208}},\ \bibinfo {pages}
  {165} (\bibinfo {year} {2012})}\BibitemShut {NoStop}%
\bibitem [{\citenamefont {Xia}\ \emph {et~al.}(2013)\citenamefont {Xia},
  \citenamefont {Wang},\ and\ \citenamefont {He}}]{XIA13a}%
  \BibitemOpen
  \bibfield  {author} {\bibinfo {author} {\bibfnamefont {M.}~\bibnamefont
  {Xia}}, \bibinfo {author} {\bibfnamefont {J.}~\bibnamefont {Wang}}, \ and\
  \bibinfo {author} {\bibfnamefont {Y.}~\bibnamefont {He}},\ }\href {\doibase
  10.1371/journal.pone.0068910} {\bibfield  {journal} {\bibinfo  {journal}
  {PLoS ONE}\ }\textbf {\bibinfo {volume} {8}},\ \bibinfo {pages} {1} (\bibinfo
  {year} {2013})}\BibitemShut {NoStop}%
\bibitem [{\citenamefont {Sharbrough}(1991)}]{SHA91}%
  \BibitemOpen
  \bibfield  {author} {\bibinfo {author} {\bibfnamefont {F.}~\bibnamefont
  {Sharbrough}},\ }\href@noop {} {\bibfield  {journal} {\bibinfo  {journal} {J.
  Clin. Neurophysiol.}\ }\textbf {\bibinfo {volume} {8}},\ \bibinfo {pages}
  {200} (\bibinfo {year} {1991})}\BibitemShut {NoStop}%
\bibitem [{\citenamefont {Okamoto}\ \emph {et~al.}(2004)\citenamefont
  {Okamoto}, \citenamefont {Dan}, \citenamefont {Sakamoto}, \citenamefont
  {Takeo}, \citenamefont {Shimizu}, \citenamefont {Kohno}, \citenamefont {Oda},
  \citenamefont {Isobe}, \citenamefont {Suzuki}, \citenamefont {Kohyama},\ and\
  \citenamefont {Dan}}]{OKA04}%
  \BibitemOpen
  \bibfield  {author} {\bibinfo {author} {\bibfnamefont {M.}~\bibnamefont
  {Okamoto}}, \bibinfo {author} {\bibfnamefont {H.}~\bibnamefont {Dan}},
  \bibinfo {author} {\bibfnamefont {K.}~\bibnamefont {Sakamoto}}, \bibinfo
  {author} {\bibfnamefont {K.}~\bibnamefont {Takeo}}, \bibinfo {author}
  {\bibfnamefont {H.}~\bibnamefont {Shimizu}}, \bibinfo {author} {\bibfnamefont
  {S.}~\bibnamefont {Kohno}}, \bibinfo {author} {\bibfnamefont
  {I.}~\bibnamefont {Oda}}, \bibinfo {author} {\bibfnamefont {S.}~\bibnamefont
  {Isobe}}, \bibinfo {author} {\bibfnamefont {T.}~\bibnamefont {Suzuki}},
  \bibinfo {author} {\bibfnamefont {K.}~\bibnamefont {Kohyama}}, \ and\
  \bibinfo {author} {\bibfnamefont {I.}~\bibnamefont {Dan}},\ }\href@noop {}
  {\bibfield  {journal} {\bibinfo  {journal} {Neuroimage}\ }\textbf {\bibinfo
  {volume} {21}},\ \bibinfo {pages} {99} (\bibinfo {year} {2004})}\BibitemShut
  {NoStop}%
\bibitem [{\citenamefont {Hoke}\ \emph {et~al.}(1989)\citenamefont {Hoke},
  \citenamefont {Lehnertz}, \citenamefont {Pantev},\ and\ \citenamefont
  {L{\"u}tkenh{\"o}ner}}]{HOK89}%
  \BibitemOpen
  \bibfield  {author} {\bibinfo {author} {\bibfnamefont {M.}~\bibnamefont
  {Hoke}}, \bibinfo {author} {\bibfnamefont {K.}~\bibnamefont {Lehnertz}},
  \bibinfo {author} {\bibfnamefont {C.}~\bibnamefont {Pantev}}, \ and\ \bibinfo
  {author} {\bibfnamefont {B.}~\bibnamefont {L{\"u}tkenh{\"o}ner}},\ }in\
  \href@noop {} {\emph {\bibinfo {booktitle} {Brain Dynamics}}}\ (\bibinfo
  {publisher} {Springer},\ \bibinfo {year} {1989})\ pp.\ \bibinfo {pages}
  {84--105}\BibitemShut {NoStop}%
\bibitem [{\citenamefont {Chernihovskyi}\ \emph {et~al.}(2009)\citenamefont
  {Chernihovskyi}, \citenamefont {Elger},\ and\ \citenamefont
  {Lehnertz}}]{CHE09a}%
  \BibitemOpen
  \bibfield  {author} {\bibinfo {author} {\bibfnamefont {A.}~\bibnamefont
  {Chernihovskyi}}, \bibinfo {author} {\bibfnamefont {C.~E.}\ \bibnamefont
  {Elger}}, \ and\ \bibinfo {author} {\bibfnamefont {K.}~\bibnamefont
  {Lehnertz}},\ }\href@noop {} {\bibfield  {journal} {\bibinfo  {journal}
  {EURASIP J. Adv. Sig. Pr.}\ ,\ \bibinfo {pages} {1}} (\bibinfo {year}
  {2009})}\BibitemShut {NoStop}%
\bibitem [{\citenamefont {Kasatkin}\ \emph {et~al.}(2017)\citenamefont
  {Kasatkin}, \citenamefont {Yanchuk}, \citenamefont {Sch{\"o}ll},\ and\
  \citenamefont {Nekorkin}}]{KAS17}%
  \BibitemOpen
  \bibfield  {author} {\bibinfo {author} {\bibfnamefont {D.~V.}\ \bibnamefont
  {Kasatkin}}, \bibinfo {author} {\bibfnamefont {S.}~\bibnamefont {Yanchuk}},
  \bibinfo {author} {\bibfnamefont {E.}~\bibnamefont {Sch{\"o}ll}}, \ and\
  \bibinfo {author} {\bibfnamefont {V.~I.}\ \bibnamefont {Nekorkin}},\ }\href
  {\doibase 10.1103/physreve.96.062211} {\bibfield  {journal} {\bibinfo
  {journal} {Phys. Rev. E}\ }\textbf {\bibinfo {volume} {96}},\ \bibinfo
  {pages} {062211} (\bibinfo {year} {2017})}\BibitemShut {NoStop}%
\bibitem [{\citenamefont {Kilpatrick}\ \emph {et~al.}(1990)\citenamefont
  {Kilpatrick}, \citenamefont {Davis}, \citenamefont {Tress}, \citenamefont
  {Rossiter}, \citenamefont {Hopper},\ and\ \citenamefont
  {Vandendriesen}}]{KIL90}%
  \BibitemOpen
  \bibfield  {author} {\bibinfo {author} {\bibfnamefont {C.~J.}\ \bibnamefont
  {Kilpatrick}}, \bibinfo {author} {\bibfnamefont {S.~M.}\ \bibnamefont
  {Davis}}, \bibinfo {author} {\bibfnamefont {B.~M.}\ \bibnamefont {Tress}},
  \bibinfo {author} {\bibfnamefont {S.~C.}\ \bibnamefont {Rossiter}}, \bibinfo
  {author} {\bibfnamefont {J.~L.}\ \bibnamefont {Hopper}}, \ and\ \bibinfo
  {author} {\bibfnamefont {M.~L.}\ \bibnamefont {Vandendriesen}},\ }\href@noop
  {} {\bibfield  {journal} {\bibinfo  {journal} {Arch. Neurol.}\ }\textbf
  {\bibinfo {volume} {47}},\ \bibinfo {pages} {157} (\bibinfo {year}
  {1990})}\BibitemShut {NoStop}%
\bibitem [{\citenamefont {Goodfellow}\ \emph {et~al.}(2016)\citenamefont
  {Goodfellow}, \citenamefont {Rummel}, \citenamefont {Abela}, \citenamefont
  {Richardson}, \citenamefont {Schindler},\ and\ \citenamefont
  {Terry}}]{GOO16}%
  \BibitemOpen
  \bibfield  {author} {\bibinfo {author} {\bibfnamefont {M.}~\bibnamefont
  {Goodfellow}}, \bibinfo {author} {\bibfnamefont {C.}~\bibnamefont {Rummel}},
  \bibinfo {author} {\bibfnamefont {E.}~\bibnamefont {Abela}}, \bibinfo
  {author} {\bibfnamefont {M.~P.}\ \bibnamefont {Richardson}}, \bibinfo
  {author} {\bibfnamefont {K.}~\bibnamefont {Schindler}}, \ and\ \bibinfo
  {author} {\bibfnamefont {J.~R.}\ \bibnamefont {Terry}},\ }\href {\doibase
  10.1038/srep29215} {\bibfield  {journal} {\bibinfo  {journal} {Sci. Rep.}\
  }\textbf {\bibinfo {volume} {6}},\ \bibinfo {pages} {29215} (\bibinfo {year}
  {2016})}\BibitemShut {NoStop}%
\bibitem [{\citenamefont {Sinha}\ \emph {et~al.}(2016)\citenamefont {Sinha},
  \citenamefont {Dauwels}, \citenamefont {Kaiser}, \citenamefont {Cash},
  \citenamefont {Brandon~Westover}, \citenamefont {Wang},\ and\ \citenamefont
  {Taylor}}]{SIN16a}%
  \BibitemOpen
  \bibfield  {author} {\bibinfo {author} {\bibfnamefont {N.}~\bibnamefont
  {Sinha}}, \bibinfo {author} {\bibfnamefont {J.}~\bibnamefont {Dauwels}},
  \bibinfo {author} {\bibfnamefont {M.}~\bibnamefont {Kaiser}}, \bibinfo
  {author} {\bibfnamefont {S.~S.}\ \bibnamefont {Cash}}, \bibinfo {author}
  {\bibfnamefont {M.}~\bibnamefont {Brandon~Westover}}, \bibinfo {author}
  {\bibfnamefont {Y.}~\bibnamefont {Wang}}, \ and\ \bibinfo {author}
  {\bibfnamefont {P.~N.}\ \bibnamefont {Taylor}},\ }\href {\doibase
  10.1093/brain/aww299} {\bibfield  {journal} {\bibinfo  {journal} {Brain}\
  }\textbf {\bibinfo {volume} {140}},\ \bibinfo {pages} {319} (\bibinfo {year}
  {2016})}\BibitemShut {NoStop}%
\bibitem [{\citenamefont {Olmi}\ \emph {et~al.}(2019)\citenamefont {Olmi},
  \citenamefont {Petkoski}, \citenamefont {Guye}, \citenamefont {Bartolomei},\
  and\ \citenamefont {Jirsa}}]{OLM19}%
  \BibitemOpen
  \bibfield  {author} {\bibinfo {author} {\bibfnamefont {S.}~\bibnamefont
  {Olmi}}, \bibinfo {author} {\bibfnamefont {S.}~\bibnamefont {Petkoski}},
  \bibinfo {author} {\bibfnamefont {M.}~\bibnamefont {Guye}}, \bibinfo {author}
  {\bibfnamefont {F.}~\bibnamefont {Bartolomei}}, \ and\ \bibinfo {author}
  {\bibfnamefont {V.}~\bibnamefont {Jirsa}},\ }\href@noop {} {\bibfield
  {journal} {\bibinfo  {journal} {PLoS Comp. Biol.}\ }\textbf {\bibinfo
  {volume} {15}},\ \bibinfo {pages} {e1006805} (\bibinfo {year}
  {2019})}\BibitemShut {NoStop}%
\bibitem [{\citenamefont {Cabral}\ \emph {et~al.}(2013)\citenamefont {Cabral},
  \citenamefont {Fernandes}, \citenamefont {Van~Hartevelt}, \citenamefont
  {James},\ and\ \citenamefont {Kringelbach}}]{CAB13b}%
  \BibitemOpen
  \bibfield  {author} {\bibinfo {author} {\bibfnamefont {J.}~\bibnamefont
  {Cabral}}, \bibinfo {author} {\bibfnamefont {H.~M.}\ \bibnamefont
  {Fernandes}}, \bibinfo {author} {\bibfnamefont {T.~J.}\ \bibnamefont
  {Van~Hartevelt}}, \bibinfo {author} {\bibfnamefont {A.~C.}\ \bibnamefont
  {James}}, \ and\ \bibinfo {author} {\bibfnamefont {M.~L.}\ \bibnamefont
  {Kringelbach}},\ }\href@noop {} {\bibfield  {journal} {\bibinfo  {journal}
  {Chaos}\ }\textbf {\bibinfo {volume} {23}},\ \bibinfo {pages} {046111}
  (\bibinfo {year} {2013})}\BibitemShut {NoStop}%
\bibitem [{\citenamefont {Schilling}\ \emph {et~al.}(2019)\citenamefont
  {Schilling}, \citenamefont {Daducci}, \citenamefont {Maier-Hein},
  \citenamefont {Poupon}, \citenamefont {Houde}, \citenamefont {Nath},
  \citenamefont {Anderson}, \citenamefont {Landman},\ and\ \citenamefont
  {Descoteaux}}]{SCH19d}%
  \BibitemOpen
  \bibfield  {author} {\bibinfo {author} {\bibfnamefont {K.~G.}\ \bibnamefont
  {Schilling}}, \bibinfo {author} {\bibfnamefont {A.}~\bibnamefont {Daducci}},
  \bibinfo {author} {\bibfnamefont {K.}~\bibnamefont {Maier-Hein}}, \bibinfo
  {author} {\bibfnamefont {C.}~\bibnamefont {Poupon}}, \bibinfo {author}
  {\bibfnamefont {J.-C.}\ \bibnamefont {Houde}}, \bibinfo {author}
  {\bibfnamefont {V.}~\bibnamefont {Nath}}, \bibinfo {author} {\bibfnamefont
  {A.~W.}\ \bibnamefont {Anderson}}, \bibinfo {author} {\bibfnamefont {B.~A.}\
  \bibnamefont {Landman}}, \ and\ \bibinfo {author} {\bibfnamefont
  {M.}~\bibnamefont {Descoteaux}},\ }\href@noop {} {\bibfield  {journal}
  {\bibinfo  {journal} {Magn. Res. Imaging}\ }\textbf {\bibinfo {volume}
  {57}},\ \bibinfo {pages} {194} (\bibinfo {year} {2019})}\BibitemShut
  {NoStop}%
\bibitem [{\citenamefont {Hlinka}\ and\ \citenamefont {Coombes}(2012)}]{HLI12}%
  \BibitemOpen
  \bibfield  {author} {\bibinfo {author} {\bibfnamefont {J.}~\bibnamefont
  {Hlinka}}\ and\ \bibinfo {author} {\bibfnamefont {S.}~\bibnamefont
  {Coombes}},\ }\href@noop {} {\bibfield  {journal} {\bibinfo  {journal}
  {European Journal of Neuroscience}\ }\textbf {\bibinfo {volume} {36}},\
  \bibinfo {pages} {2137} (\bibinfo {year} {2012})}\BibitemShut {NoStop}%
\end{thebibliography}




\begin{table}[h!]
	\begin{tabular}{|c|c|c|} 
		\hline
		Label L/R & Region &Lobe \\
		\hline
		\hline
		1/46 & Precentral &  Central region\\
		2/47 & Frontal Sup &  Frontal lobe\\
		3/48 & Frontal Sup Orb & Frontal lobe\\
		4/49 & Frontal Mid &  Frontal lobe\\
		5/50 & Frontal Mid Orb & Frontal lobe\\
		6/51 & Frontal Inf Oper &Frontal lobe \\
		7/52& Frontal Inf Tri  &  Frontal lobe\\
		8/53 & Frontal Inf Orb &  Frontal lobe\\
		9/54 & Rolandic Oper &  Central Region \\
		10/55 & Supp Motor Area &  Frontal lobe\\
		11/56 & Olfactory & Frontal lobe\\
		12/57 & Frontal Sup Medial &  Frontal lobe\\
		13/58 & Frontal Med Orb  &  Frontal lobe\\
		14/59 & Rectus & Frontal lobe\\
		15/60 & Insula  & Insula\\
		16/61 & Cingulum Ant  & Limbic lobe\\
		17/62 & Cingulum Mid & Limbic lobe\\
		18/63 & Cingulum Post & Limbic lobe\\
		19/64 & Hippocampus  & Limbic lobe \\
		20/65 & ParaHippocampal & Limbic lobe\\
		21/66 & Amygdala & Sub cort. gray nuc.\\
		22/67 & Calcarine  & Occipital lobe\\
		23/68 & Cuneus  & Occipital lobe\\
		24/69 & Lingual  & Occipital lobe\\
		25/70 & Occipital Sup & Occipital lobe \\
		26/71 & Occipital Mid  & Occipital lobe\\
		27/72 & Occipital Inf  & Occipital lobe \\
		28/73 & Fusiform   & Occipital lobe\\
		29/74& Postcentral & Central region\\
		30/75 & Parietal Sup & Parietal lobe \\
		31/76 & Parietal Inf  & Parietal lobe\\
		32/77 & Supramarginal & Parietal lobe \\
		33/78 & Angular  & Parietal lobe\\
		34/79 & Precuneus & Parietal lobe\\
		35/80 & Paracentral Lobule & Frontal lobe\\
		36/81 & Caudate & Sub cort. gray nuc.\\
		37/82 & Putamen &Sub cort. gray nuc.\\
		38/83 & Pallidum &  Sub cort. gray nuc.\\
		39/84 & Thalamus & Sub cort. gray nuc.\\
		40/85 & Heschl &  Temporal lobe\\
		41/86 & Temporal Sup & Temporal lobe\\
		42/87 & Temporal Pole Sup & Limbic lobe\\
		43/88 & Temporal Mid & Temporal lobe\\
		44/89 & Temporal Pole Mid & Limbic lobe\\
		45/90 & Temporal Inf & Temporal lobe \\ \hline
	\end{tabular}
	\caption{Cortical and subcortical regions, according to the Automated Anatomical Labeling atlas (AAL) [N. Tzourio-Mazoyer, B. Landeau, D. Papathanassiou, F. Crivello, O. Etard, N. Delcroix, B. Mazoyer, and M. Joliot, Neuroimage 15, 273 (2002)]. Note that the numbering of the brain regions is different from the original numbering, as described in the first paragraph of Sec.~II.}
	\label{tab:appA_aal}
	\label{tab:appA_aal}
\end{table}

\end{document}